\pdfoutput=1

\documentclass[sigplan,dvipsnames]{acmart}

\acmJournal{PACMPL}
\acmVolume{1}
\acmNumber{ICFP} 
\acmArticle{1}
\acmYear{2022}
\acmMonth{5}
\acmDOI{} 
\startPage{1}

\setcopyright{none}

\usepackage{booktabs}   
\usepackage{subcaption} 
\usepackage[utf8]{inputenc}
\usepackage{microtype}
\usepackage{amsmath}
\usepackage{amsthm}
\usepackage{framed}
\usepackage{proof}
\usepackage{braket}
\usepackage{fancyvrb}
\usepackage{listings}
\usepackage[inline,shortlabels]{enumitem}
\usepackage[capitalize]{cleveref}
\usepackage{xcolor}
\usepackage{pgffor}
\usepackage{ragged2e}
\usepackage{upgreek} 
\usepackage{multirow}
\usepackage{diagbox}
\usepackage{csquotes}
\usepackage{comment}  

\usepackage{pgfplots}
\usepackage{pgfplotstable}
\pgfplotsset{compat=1.16}

\VerbatimFootnotes

\let\restriction\relax

\theoremstyle{theorem}

\theoremstyle{definition}

\newtheorem{restriction}{Restriction}

\theoremstyle{remark}

\crefname{figure}{Fig.}{Figs.}
\Crefname{figure}{Fig.}{Figs.}
\crefname{restriction}{Restriction}{Restrictions}

\floatstyle{plaintop}
\restylefloat{table}

\definecolor{dkcyan}{rgb}{0.1, 0.3, 0.3}
\definecolor{dkgreen}{rgb}{0,0.3,0}
\definecolor{olive}{rgb}{0.5, 0.5, 0.0}
\definecolor{dkblue}{rgb}{0,0.1,0.5}

\definecolor{col:ln}{rgb}  {0.1, 0.1, 0.7}
\definecolor{col:str}{rgb} {0.8, 0.0, 0.0}
\definecolor{col:db}{rgb}  {0.9, 0.5, 0.0}
\definecolor{col:ours}{rgb}{0.0, 0.7, 0.0}

\definecolor{lightgreen}{RGB}{170, 255, 220}
\definecolor{darkbrown}{RGB}{121,37,0}

\colorlet{listing-comment}{gray}
\colorlet{operator-color}{darkbrown}

\lstdefinestyle{default}{
    basicstyle=\ttfamily\fontsize{8.7}{9.5}\selectfont,
    columns=fullflexible,
    commentstyle=\sffamily\color{black!50!white},
    escapechar=\#,
    framexleftmargin=1em,
    framexrightmargin=1ex,
    keepspaces=true,
    keywordstyle=\color{dkblue},
    mathescape,
    numbers=none,
    numberblanklines=false,
    numbersep=1.25em,
    numberstyle=\relscale{0.8}\color{gray}\ttfamily,
    showstringspaces=true,
    stepnumber=1,
    xleftmargin=1em
}

\lstdefinelanguage{custom-haskell}{
    language=Haskell,
    deletekeywords={lookup, delete, map, mapMaybe, Ord, Maybe, String, Just, Nothing, Int, Bool},
    keywordstyle=[2]\color{dkgreen},
    morekeywords=[2]{String, Map, Ord, Maybe, Int, Bool},
    morekeywords=[2]{Name, Expression, ESummary, PosTree, Structure, HashCode, VarMap},
    keywordstyle=[3]\color{dkcyan},
    mathescape=false, 
    escapechar=\%,    
    literate=%
        {=}{{{\color{operator-color}=}}}1
        {|}{{{\color{operator-color}|}}}1
        {\\}{{{\color{operator-color}\textbackslash$\,\!$}}}1
        {.}{{{\color{operator-color}.}}}1
        {=>}{{{\color{operator-color}=>}}}1
        {->}{{{\color{operator-color}->}}}1
        {<-}{{{\color{operator-color}<-}}}1
        {::}{{{\color{operator-color}::}}}1
}

\lstnewenvironment{code}[1][]
  {\small\lstset{language=custom-haskell,#1}}
  {}

\lstnewenvironment{expression}[1][]
  {\small\lstset{#1}}
  {}

\let\Bbbk\relax
\makeatletter
\@ifundefined{lhs2tex.lhs2tex.sty.read}%
  {\@namedef{lhs2tex.lhs2tex.sty.read}{}%
   \newcommand\SkipToFmtEnd{}%
   \newcommand\EndFmtInput{}%
   \long\def\SkipToFmtEnd#1\EndFmtInput{}%
  }\SkipToFmtEnd

\newcommand\ReadOnlyOnce[1]{\@ifundefined{#1}{\@namedef{#1}{}}\SkipToFmtEnd}
\usepackage{newtxmath}
\usepackage{stmaryrd}
\DeclareFontFamily{OT1}{cmtex}{}
\DeclareFontShape{OT1}{cmtex}{m}{n}
  {<5><6><7><8>cmtex8
   <9>cmtex9
   <10><10.95><12><14.4><17.28><20.74><24.88>cmtex10}{}
\DeclareFontShape{OT1}{cmtex}{m}{it}
  {<-> ssub * cmtt/m/it}{}

\DeclareFontShape{OT1}{cmtt}{bx}{n}
  {<5><6><7><8>cmtt8
   <9>cmbtt9
   <10><10.95><12><14.4><17.28><20.74><24.88>cmbtt10}{}
\DeclareFontShape{OT1}{cmtex}{bx}{n}
  {<-> ssub * cmtt/bx/n}{}

\newcommand{\anonymous}{\kern0.06em \vbox{\hrule\@width.5em}}

\usepackage{polytable}

\@ifundefined{mathindent}%
  {\newdimen\mathindent\mathindent\leftmargini}%
  {}%

\def\resethooks{%
  \global\let\SaveRestoreHook\empty
  \global\let\ColumnHook\empty}
\newcommand*{\savecolumns}[1][default]%
  {\g@addto@macro\SaveRestoreHook{\savecolumns[#1]}}
\newcommand*{\restorecolumns}[1][default]%
  {\g@addto@macro\SaveRestoreHook{\restorecolumns[#1]}}
\newcommand*{\aligncolumn}[2]%
  {\g@addto@macro\ColumnHook{\column{#1}{#2}}}

\resethooks

\newcommand{\onelinecommentchars}{\quad-{}- }
\newcommand{\commentbeginchars}{\enskip\{-}
\newcommand{\commentendchars}{-\}\enskip}

\newcommand{\visiblecomments}{%
  \let\onelinecomment=\onelinecommentchars
  \let\commentbegin=\commentbeginchars
  \let\commentend=\commentendchars}

\newcommand{\invisiblecomments}{%
  \let\onelinecomment=\empty
  \let\commentbegin=\empty
  \let\commentend=\empty}

\visiblecomments

\newlength{\blanklineskip}
\newcommand{\hsindent}[1]{\quad}
\let\hspre\empty
\let\hspost\empty

\EndFmtInput
\makeatother
\ReadOnlyOnce{polycode.fmt}%
\makeatletter

\newcommand{\hsnewpar}[1]%
  {{\parskip=0pt\parindent=0pt\par\vskip #1\noindent}}

\newcommand{\hscodestyle}{}

\newcommand{\sethscode}[1]%
  {\expandafter\let\expandafter\hscode\csname #1\endcsname
   \expandafter\let\expandafter\endhscode\csname end#1\endcsname}

  {\par\noindent
   \advance\leftskip\mathindent
   \hscodestyle
   \let\\=\@normalcr
   \let\hspre\(\let\hspost\)%
   \pboxed}%
  {\endpboxed\)%
   \par\noindent
   \ignorespacesafterend}

  {\hsnewpar\abovedisplayskip
   \advance\leftskip\mathindent
   \hscodestyle
   \let\hspre\(\let\hspost\)%
   \pboxed}%
  {\endpboxed%
   \hsnewpar\belowdisplayskip
   \ignorespacesafterend}

  {\hsnewpar\abovedisplayskip
   \advance\leftskip\mathindent
   \hscodestyle
   \let\\=\@normalcr
   \(\pboxed}%
  {\endpboxed\)%
   \hsnewpar\belowdisplayskip
   \ignorespacesafterend}

\newcommand{\plainhs}{\sethscode{plainhscode}}

\plainhs

  {\hsnewpar\abovedisplayskip
   \advance\leftskip\mathindent
   \hscodestyle
   \let\\=\@normalcr
   \(\parray}%
  {\endparray\)%
   \hsnewpar\belowdisplayskip
   \ignorespacesafterend}

  {\parray}{\endparray}

  {\(\parray}{\endparray\)}

\def\codeframewidth{\arrayrulewidth}
\RequirePackage{calc}

  {\parskip=\abovedisplayskip\par\noindent
   \hscodestyle
   \arrayrulewidth=\codeframewidth
   \tabular{@{}|p{\linewidth-2\arraycolsep-2\arrayrulewidth-2pt}|@{}}%
   \hline\framedhslinecorrect\\{-1.5ex}%
   \let\endoflinesave=\\
   \let\\=\@normalcr
   \(\pboxed}%
  {\endpboxed\)%
   \framedhslinecorrect\endoflinesave{.5ex}\hline
   \endtabular
   \parskip=\belowdisplayskip\par\noindent
   \ignorespacesafterend}

\newcommand{\framedhslinecorrect}[2]%
  {#1[#2]}

  {\(\def\column##1##2{}%
   \let\>\undefined\let\<\undefined\let\\\undefined
   \newcommand\>[1][]{}\newcommand\<[1][]{}\newcommand\\[1][]{}%
   \def\fromto##1##2##3{##3}%
   }{\) }%

  {\let\orighscode=\hscode
   \let\origendhscode=\endhscode
   \def\endhscode{\def\hscode{\endgroup\def\@currenvir{hscode}\\}\begingroup}
   \orighscode\def\hscode{\endgroup\def\@currenvir{hscode}}}%
  {\origendhscode
   \global\let\hscode=\orighscode
   \global\let\endhscode=\origendhscode}%

\makeatother
\EndFmtInput
\ReadOnlyOnce{forall.fmt}%
\makeatletter

\let\HaskellResetHook\empty
\newcommand*{\AtHaskellReset}[1]{%
  \g@addto@macro\HaskellResetHook{#1}}
\newcommand*{\HaskellReset}{\HaskellResetHook}

\newcommand\hsforall{\global\let\hsdot=\hsperiodonce}
\newcommand\hsexists{\global\let\hsdot=\hsperiodonce}
\newcommand*\hsperiodonce[2]{#2\global\let\hsdot=\hscompose}
\newcommand*\hscompose[2]{#1}

\AtHaskellReset{\global\let\hsdot=\hscompose}

\HaskellReset

\makeatother
\EndFmtInput
\newcommand{\keyword}[1]{\textcolor{BlueViolet}{\textbf{#1}}}
\newcommand{\id}[1]{\textsf{\textsl{#1}}}
\newcommand{\varid}[1]{\textcolor{Sepia}{\id{#1}}}
\newcommand{\conid}[1]{\textcolor{OliveGreen}{\id{#1}}}
\newcommand{\tick}{\text{\textquoteright}}

\renewcommand{\hscodestyle}{\small}

\begin{document}

\setlength{\pdfpageheight}{\paperheight}
\setlength{\pdfpagewidth}{\paperwidth}

\newcommand{\simon}[1]{}
\newcommand{\js}[1]{}
\newcommand{\rae}[1]{}
\newcommand{\sg}[1]{}

\newcommand{\bv}[1]{\#_{#1}}    
\newcommand{\pv}[1]{\$_{#1}}    
\newcommand{\pvo}[1]{\%_{#1}}   

\newcommand{\benchname}[1]{\texttt{#1}}
\newcommand{\insigdig}[1]{\ensuremath{\tilde{\text{#1}}}} 
\newcommand{\hackage}[1]{\varid{#1}\footnote{\url{https://hackage.haskell.org/package/#1}}}

\title
{Triemaps that match}         
\subtitle{Technical Report}               


\author{Simon Peyton Jones}
\affiliation{
  \institution{Epic Games}
  \city{Cambridge}
  \country{UK}
}
\email{simon.peytonjones@gmail.com}

\author{Sebastian Graf}
\affiliation{%
  \institution{Karlsruhe Institute of Technology}
  \city{Karlsruhe}
  \country{Germany}
}
\email{sebastian.graf@kit.edu}


\begin{abstract}
  The \emph{trie} data structure is a good choice for finite maps whose
  keys are data structures (trees) rather than atomic values. But what if we want
  the keys to be \emph{patterns}, each of which matches many lookup keys?
  Efficient matching of this kind is well studied in the theorem prover
  community, but much less so in the context of statically typed functional programming.
  Doing so yields an interesting new viewpoint --- and a practically useful design
  pattern, with good runtime performance.
\end{abstract}

\maketitle

\section{Introduction} \label{sec:intro}

Many functional languages provide \emph{finite maps} either as a
built-in data type, or as a mature, well-optimised library.  Generally the keys
of such a map will be small: an integer, a string, or perhaps a pair of integers.
But in some applications the key is large: an entire tree structure.  For example,
consider the Haskell expression
\begin{hscode}\SaveRestoreHook
\column{B}{@{}>{\hspre}l<{\hspost}@{}}%
\column{3}{@{}>{\hspre}l<{\hspost}@{}}%
\column{E}{@{}>{\hspre}l<{\hspost}@{}}%
\>[3]{}\keyword{let}\;\varid{x}\mathrel{=}\varid{a}\mathbin{+}\varid{b}\;\keyword{in}\mathbin{...}(\keyword{let}\;\varid{y}\mathrel{=}\varid{a}\mathbin{+}\varid{b}\;\keyword{in}\;\varid{x}\mathbin{+}\varid{y})\mathbin{....}{}\<[E]%
\ColumnHook
\end{hscode}\resethooks
We might hope that the compiler will recognise the repeated sub-expression
\ensuremath{(\varid{a}\mathbin{+}\varid{b})} and transform to
\begin{hscode}\SaveRestoreHook
\column{B}{@{}>{\hspre}l<{\hspost}@{}}%
\column{3}{@{}>{\hspre}l<{\hspost}@{}}%
\column{E}{@{}>{\hspre}l<{\hspost}@{}}%
\>[3]{}\keyword{let}\;\varid{x}\mathrel{=}\varid{a}\mathbin{+}\varid{b}\;\keyword{in}\mathbin{...}(\varid{x}\mathbin{+}\varid{x})\mathbin{....}{}\<[E]%
\ColumnHook
\end{hscode}\resethooks
An easy way to do so is to build a finite map that maps the expression \ensuremath{(\varid{a}\mathbin{+}\varid{b})} to \ensuremath{\varid{x}}.
Then, when encountering the inner \ensuremath{\keyword{let}}, we can look up the right hand side in the map,
and replace \ensuremath{\varid{y}} by \ensuremath{\varid{x}}.  All we need is a \emph{finite map keyed by syntax trees}.

Another similar challenge is this: we often want a finite map keyed
by \emph{patterns} rather than expressions.
GHC (a Haskell compiler) supports user-written rewrite rules. For example:
\begin{hscode}\SaveRestoreHook
\column{B}{@{}>{\hspre}l<{\hspost}@{}}%
\column{29}{@{}>{\hspre}l<{\hspost}@{}}%
\column{45}{@{}>{\hspre}l<{\hspost}@{}}%
\column{65}{@{}>{\hspre}l<{\hspost}@{}}%
\column{E}{@{}>{\hspre}l<{\hspost}@{}}%
\>[B]{}\{-{\#}\;\keyword{RULES}\;\text{\ttfamily \char34 map/map\char34}\;{}\<[29]%
\>[29]{}\keyword{$\forall$}\!\! \hsforall \;\varid{f}\;\varid{g}\;\varid{xs}\hsdot{\circ }{.\,}{}\<[45]%
\>[45]{}\varid{map}\;\varid{f}\;(\varid{map}\;\varid{g}\;\varid{xs})\mathrel{=}{}\<[65]%
\>[65]{}\varid{map}\;(\varid{f}\hsdot{\circ }{.\,}\varid{g})\;\varid{xs}\;{\#}{-}\}{}\<[E]%
\ColumnHook
\end{hscode}\resethooks
Given a target expression to optimise, the compiler must decide if any
rewrite rule matches the target, after intantiating
the forall-bound variables of the rule.  Another application, which also appears in GHC,
is type class instance lookup.
In both cases we need a fast way to see if a target expression matches one of
the patterns in a set of (\varid{pattern},~\varid{rhs}) pairs.
We could do this by testing the patterns one by one, but if there are many patterns it would
be much better to consult some kind of efficient \emph{finite map keyed by patterns}.

When the key is a perhaps-large data structure, traditional finite-map
implementations can behave badly, because they are often based on balanced trees
and make the assumption that comparing two keys is a fast, constant-time
operation.  That assumption is false when the key is large.  For the lookup task,
a good solution is well known: use a \emph{trie} (e.g. \cite{hinze:generalized}).
The matching task is also well studied but, surprisingly, only in the automated
reasoning community: they use so-called \emph{discrimination trees}, as we discuss in \Cref{sec:discrim-trees}.
In this paper we apply these trie-based ideas in the context of a statically-typed functional
programming language, Haskell.
This shift of context is surprisingly fruitful, and we make the following contributions:
\begin{itemize}
\item Following \citet{hinze:generalized}, we develop a standard pattern for
  a \emph{statically-typed triemap} for an arbitrary algebraic data type (\Cref{sec:basic}). In
  contrast, most of the literature describes untyped tries for a
  fixed, generic tree type.
  In particular:
  \begin{itemize}
    \item Supported by type classes, we can make good use of polymorphism to build triemaps
      for polymorphic data types, such as lists (\Cref{sec:class}).

    \item We cover the full range of operations expected for finite maps:
      not only \ensuremath{\varid{insert}}ion and \ensuremath{\varid{lookup}}, but \ensuremath{\varid{alter}}, \ensuremath{\varid{union}}, \ensuremath{\varid{foldr}}, \ensuremath{\varid{map}} and \ensuremath{\varid{filter}}
      (\Cref{sec:basic}).

    \item We develop a generic optimisation for singleton maps that
      compresses leaf paths. Intriguingly, the resulting triemap
      \emph{transformer} can be easily mixed into arbitrary triemap definitions
      (\Cref{sec:singleton}).

   \end{itemize}
\item We show how to make our triemaps insensitive to \emph{$\alpha$-renamings} in
       keys that include binding forms (\Cref{sec:binders}).
       Accounting for $\alpha$-equivalence is not hard, but it is crucial for
       the applications in compilers.

\item We extend our triemaps to support \emph{matching} lookups
  (\Cref{sec:matching}).  This is an important step, because the only
  readily-available alternative is linear lookup. Our main
  contribution is to extend the established idea of tries keyed by
  arbitrary data types, so that it can handle matching too.

\item We present measurements that compare the performance of our triemaps (ignoring
  their matching capability) with traditional finite-map implementations in
  Haskell (\Cref{sec:eval}).
\end{itemize}
All the code in this paper is available online ancillary to the extended version
of this paper~\cite{triemaps-extended} or in the accompanying GitHub
repository~\cite{triemaps-github}.
It is written in Haskell and makes crucial use of some distinctive features of
Haskell, including type classes, associated types, higher kinded type variables,
and polymorphic recursion.

We discuss related work in \Cref{sec:related}.
Our contribution is not so much a clever new idea as an exposition of
some old ideas in a new context.  Nevertheless, we found it
surprisingly tricky to develop the ``right'' abstractions, such as
the \ensuremath{\conid{TrieMap}} and \ensuremath{\conid{Matchable}} classes, the singleton-and-empty map
data type, and the combinators we use in our instances.
These abstractions have been through \emph{many} iterations, and we
hope that by laying them out here we may shorten the path for others.

\section{The problem we address} \label{sec:problem}

Our general task is as follows: \emph{implement an efficient finite mapping
from keys to values, in which the key is a tree}.
Semantically, such a finite map is just a set of \emph{(key,value)}
pairs; we query the map by looking up a \emph{target}.

\subsection{Expressions as keys}

As a concrete example of the general problem, suppose a compiler is maniuplating
syntax trees representing expressions in the language being compiled.
An expression might be defined like this:
\begin{hscode}\SaveRestoreHook
\column{B}{@{}>{\hspre}l<{\hspost}@{}}%
\column{34}{@{}>{\hspre}l<{\hspost}@{}}%
\column{E}{@{}>{\hspre}l<{\hspost}@{}}%
\>[B]{}\keyword{type}\;\conid{Var}\mathrel{=}\conid{String}{}\<[E]%
\\
\>[B]{}\keyword{data}\;\conid{Expr}\mathrel{=}\conid{App}\;\conid{Expr}\;\conid{Expr}\mid \conid{Lam}\;{}\<[34]%
\>[34]{}\conid{Var}\;\conid{Expr}\mid \conid{Var}\;\conid{Var}{}\<[E]%
\ColumnHook
\end{hscode}\resethooks
Here the type \ensuremath{\conid{Var}} represents variables; these can be compared for
equality and used as the key of a finite map.  The definition of \ensuremath{\conid{Var}} is not important
for this paper, but for the sake of concreteness
we have defined \ensuremath{\conid{Var}} to be \ensuremath{\conid{String}}.  In a real compiler, a variable would
have a unique identifier (a number, say) that supports constant-time comparison, so
despite the \ensuremath{\conid{String}} we will assume that variables can be compared in constant time.

The data type \ensuremath{\conid{Expr}} is capable of representing expressions like \ensuremath{\varid{add}\;\varid{x}\;\varid{y}} and
\ensuremath{\lambda \varid{x}\to \varid{add}\;\varid{x}\;\varid{y}}. We will use this data type throughout the paper, because it
has all the features that occur in real expression data types: free variables like \ensuremath{\varid{add}},
represented by a \ensuremath{\conid{Var}} node;
lambdas which can bind variables (\ensuremath{\conid{Lam}}), and occurrences of those bound variables (\ensuremath{\conid{Var}});
and nodes with multiple children (\ensuremath{\conid{App}}).  A real-world expression type would have
many more constructors, including literals, let-expressions and suchlike.


\subsection{Alpha-renaming} \label{sec:alpha-renaming}

In the context of a compiler, where the keys are expressions or types,
the keys may contain internal \emph{binders}, such as the binder \ensuremath{\varid{x}} in
\ensuremath{\lambda \varid{x}\to \varid{x}}. If so, we would expect insertion and lookup to be insensitive
to $\alpha$-renaming, so we could, for example, insert with key \ensuremath{\lambda \varid{x}\to \varid{x}}
and look up with key \ensuremath{\lambda \varid{y}\to \varid{y}}, to find the inserted value.

\subsection{Lookup modulo matching} \label{sec:matching-intro}

Beyond just the basic finite maps we have described, our practical setting
in GHC demands more: we want to do a lookup that does \emph{matching}.  As mentioned
above, GHC supports
so-called \emph{rewrite rules}~\cite{rewrite-rules}, which the user can specify
in their source program, like this:
\begin{hscode}\SaveRestoreHook
\column{B}{@{}>{\hspre}l<{\hspost}@{}}%
\column{29}{@{}>{\hspre}l<{\hspost}@{}}%
\column{45}{@{}>{\hspre}l<{\hspost}@{}}%
\column{65}{@{}>{\hspre}l<{\hspost}@{}}%
\column{E}{@{}>{\hspre}l<{\hspost}@{}}%
\>[B]{}\{-{\#}\;\keyword{RULES}\;\text{\ttfamily \char34 map/map\char34}\;{}\<[29]%
\>[29]{}\keyword{$\forall$}\!\! \hsforall \;\varid{f}\;\varid{g}\;\varid{xs}\hsdot{\circ }{.\,}{}\<[45]%
\>[45]{}\varid{map}\;\varid{f}\;(\varid{map}\;\varid{g}\;\varid{xs})\mathrel{=}{}\<[65]%
\>[65]{}\varid{map}\;(\varid{f}\hsdot{\circ }{.\,}\varid{g})\;\varid{xs}\;{\#}{-}\}{}\<[E]%
\ColumnHook
\end{hscode}\resethooks
This rule asks the compiler to rewrite any target expression that matches the shape
of the left-hand side (LHS) of the rule into the right-hand side
(RHS).  We use the term \emph{pattern} to describe the LHS, and \emph{target} to describe
the expression we are looking up in the map.
The pattern is explicitly quantified over the \emph{pattern variables}
(here \ensuremath{\varid{f}}, \ensuremath{\varid{g}}, and \ensuremath{\varid{xs}}) that
can be bound during the matching process.  In other words, we seek a substitution
for the pattern variables that makes the pattern equal to the target expression.
For example, if the program we are compiling contains the target expression
\ensuremath{\varid{map}\;\varid{double}\;(\varid{map}\;\varid{square}\;\varid{nums})}, we would like to produce a substitution
\ensuremath{[\mskip1.5mu \varid{f}\mapsto\varid{double},\varid{g}\mapsto\varid{square},\varid{xs}\mapsto\varid{nums}\mskip1.5mu]} so that the substituted RHS
becomes \ensuremath{\varid{map}\;(\varid{double}\hsdot{\circ }{.\,}\varid{square})\;\varid{nums}}; we would replace the former expression
with the latter in the code under consideration.

Of course, the pattern might itself have bound variables, and we would
like to be insensitive to $\alpha$-conversion for those. For example:
\begin{hscode}\SaveRestoreHook
\column{B}{@{}>{\hspre}l<{\hspost}@{}}%
\column{28}{@{}>{\hspre}l<{\hspost}@{}}%
\column{E}{@{}>{\hspre}l<{\hspost}@{}}%
\>[B]{}\{-{\#}\;\keyword{RULES}\;\text{\ttfamily \char34 map/id\char34}\;{}\<[28]%
\>[28]{}\varid{map}\;(\lambda \varid{x}\to \varid{x})\mathrel{=}\lambda \varid{y}\to \varid{y}\;{\#}{-}\}{}\<[E]%
\ColumnHook
\end{hscode}\resethooks
We want to find a successful match if we see a call \ensuremath{\varid{map}\;(\lambda \varid{z}\to \varid{z})},
even though the bound variable has a different name.

A similar application is GHC's lookup for type class and type family instances.
Suppose we have
\begin{hscode}\SaveRestoreHook
\column{B}{@{}>{\hspre}l<{\hspost}@{}}%
\column{3}{@{}>{\hspre}l<{\hspost}@{}}%
\column{E}{@{}>{\hspre}l<{\hspost}@{}}%
\>[3]{}\keyword{instance}\;\conid{C}\;[\mskip1.5mu \varid{a}\mskip1.5mu]\;\conid{Int}\;\keyword{where}\mathbin{...}{}\<[E]%
\\
\>[3]{}\keyword{instance}\;\conid{C}\;\conid{Bool}\;(\conid{Maybe}\;\varid{b})\;\keyword{where}\mathbin{...}{}\<[E]%
\\
\>[3]{}\mathbin{...}\varid{and}\;\varid{so}\;\varid{on}{}\<[E]%
\ColumnHook
\end{hscode}\resethooks
When solving an instance constraint \ensuremath{\conid{C}\;[\mskip1.5mu \conid{Char}\mskip1.5mu]\;\conid{Int}}, say, GHC must look through all
the type class instances to find one that matches.  In this case, the first instance
matches, by binding \ensuremath{[\mskip1.5mu \varid{a}\mapsto\conid{Char}\mskip1.5mu]}.

There may occasionally be thousands of rules, or thousands of type class or
type family instances.  We would like to find a matching candidate more
efficiently than by linear search.

\subsection{Non-solutions} \label{sec:ord}

At first sight, our task can be done easily: define a total order on \ensuremath{\conid{Expr}}
and use a standard finite map library.
Indeed that works, but it can be terribly slow.  A finite map is
implemented as a binary search tree; at every node of this tree,
we compare the key (an \ensuremath{\conid{Expr}}, remember) with
the key stored at the node; if it is smaller, go left; if larger, go right. Each lookup
thus must perform a (logarithmic) number of
potentially-full-depth comparisons of two expressions.

Another possibility might be to hash the \ensuremath{\conid{Expr}} and use the
hash-code as the lookup key.  That would make lookup much faster, but
it requires at least two full traversals of the key for every lookup:
one to compute its hash code for every lookup, and a full equality
comparison on a ``hit'' because hash-codes can collide.

But the killer is this: \emph{neither binary search trees nor hashing is compatible
with matching lookup}.  For our purposes they are non-starters.

What other standard solutions to matching lookup are there, apart from linear search?
The theorem proving and automated reasoning community has been working with huge sets
of rewrite rules, just as we describe, for many years.
They have developed term indexing techniques for the job \cite[Chapter 26]{handbook:2001},
which attack the same problem from a rather different angle, as we discuss in \Cref{sec:discrim-trees}.

\section{Tries} \label{sec:Expr}

A standard approach to a finite map in which the key has internal structure
is to use a \emph{trie}~\cite{Knuth1973}.
Generalising tries to handle an arbitrary algebraic data type as the key is
a well established, albeit under-used, idea \cite{connelly-morris,hinze:generalized}.
We review these ideas in this section.
Let us consider a simplified
form of expression:
\begin{hscode}\SaveRestoreHook
\column{B}{@{}>{\hspre}l<{\hspost}@{}}%
\column{E}{@{}>{\hspre}l<{\hspost}@{}}%
\>[B]{}\keyword{data}\;\conid{Expr}\mathrel{=}\conid{Var}\;\conid{Var}\mid \conid{App}\;\conid{Expr}\;\conid{Expr}{}\<[E]%
\ColumnHook
\end{hscode}\resethooks
We omit lambdas for now,
so that all \ensuremath{\conid{Var}} nodes represent free variables, which are treated as constants.
We will return to lambdas in \Cref{sec:binders}.

\subsection{The interface of a finite map} \label{sec:interface}

\begin{figure}
\begin{hscode}\SaveRestoreHook
\column{B}{@{}>{\hspre}l<{\hspost}@{}}%
\column{7}{@{}>{\hspre}l<{\hspost}@{}}%
\column{8}{@{}>{\hspre}l<{\hspost}@{}}%
\column{19}{@{}>{\hspre}l<{\hspost}@{}}%
\column{23}{@{}>{\hspre}l<{\hspost}@{}}%
\column{26}{@{}>{\hspre}l<{\hspost}@{}}%
\column{30}{@{}>{\hspre}l<{\hspost}@{}}%
\column{36}{@{}>{\hspre}l<{\hspost}@{}}%
\column{37}{@{}>{\hspre}l<{\hspost}@{}}%
\column{E}{@{}>{\hspre}l<{\hspost}@{}}%
\>[B]{}\keyword{type}\;\conid{TF}\;\varid{v}\mathrel{=}\conid{Maybe}\;\varid{v}\to \conid{Maybe}\;\varid{v}{}\<[E]%
\\[\blanklineskip]%
\>[B]{}\keyword{data}\;\conid{Map}\;\varid{k}\;\varid{v}\mathrel{=}\ldots{}\<[23]%
\>[23]{}\mbox{\onelinecomment  Keys k, values v}{}\<[E]%
\\
\>[B]{}\varid{\conid{Map}.empty}{}\<[26]%
\>[26]{}\mathbin{::}\conid{Map}\;\varid{k}\;\varid{v}{}\<[E]%
\\
\>[B]{}\varid{\conid{Map}.insert}{}\<[26]%
\>[26]{}\mathbin{::}\conid{Ord}\;\varid{k}{}\<[36]%
\>[36]{}\Rightarrow \varid{k}\to \varid{v}\to \conid{Map}\;\varid{k}\;\varid{v}\to \conid{Map}\;\varid{k}\;\varid{v}{}\<[E]%
\\
\>[B]{}\varid{\conid{Map}.lookup}{}\<[26]%
\>[26]{}\mathbin{::}\conid{Ord}\;\varid{k}{}\<[36]%
\>[36]{}\Rightarrow \varid{k}\to \conid{Map}\;\varid{k}\;\varid{v}\to \conid{Maybe}\;\varid{v}{}\<[E]%
\\
\>[B]{}\varid{\conid{Map}.alter}{}\<[26]%
\>[26]{}\mathbin{::}{}\<[30]%
\>[30]{}\conid{Ord}\;\varid{k}{}\<[37]%
\>[37]{}\Rightarrow \conid{TF}\;\varid{v}\to \varid{k}{}\<[E]%
\\
\>[30]{}\to \conid{Map}\;\varid{k}\;\varid{v}\to \conid{Map}\;\varid{k}\;\varid{v}{}\<[E]%
\\
\>[B]{}\varid{\conid{Map}.foldr}{}\<[26]%
\>[26]{}\mathbin{::}(\varid{v}\to \varid{r}\to \varid{r})\to \varid{r}\to \conid{Map}\;\varid{k}\;\varid{v}\to \varid{r}{}\<[E]%
\\
\>[B]{}\varid{\conid{Map}.map}{}\<[26]%
\>[26]{}\mathbin{::}(\varid{v}\to \varid{w})\to \conid{Map}\;\varid{k}\;\varid{v}\to \conid{Map}\;\varid{k}\;\varid{w}{}\<[E]%
\\
\>[B]{}\varid{\conid{Map}.unionWith}{}\<[26]%
\>[26]{}\mathbin{::}{}\<[30]%
\>[30]{}\conid{Ord}\;\varid{k}{}\<[37]%
\>[37]{}\Rightarrow (\varid{v}\to \varid{v}\to \varid{v}){}\<[E]%
\\
\>[30]{}\to \conid{Map}\;\varid{k}\;\varid{v}\to \conid{Map}\;\varid{k}\;\varid{v}\to \conid{Map}\;\varid{k}\;\varid{v}{}\<[E]%
\\
\>[B]{}\varid{\conid{Map}.size}{}\<[26]%
\>[26]{}\mathbin{::}\conid{Map}\;\varid{k}\;\varid{v}\to \conid{Int}{}\<[E]%
\\
\>[B]{}\varid{\conid{Map}.compose}{}\<[26]%
\>[26]{}\mathbin{::}{}\<[30]%
\>[30]{}\conid{Ord}\;\varid{b}{}\<[37]%
\>[37]{}\Rightarrow \conid{Map}\;\varid{b}\;\varid{c}\to \conid{Map}\;\varid{a}\;\varid{b}\to \conid{Map}\;\varid{a}\;\varid{c}{}\<[E]%
\\[\blanklineskip]%
\>[B]{}\keyword{infixr}\;\mathrm{1}\mathrel{{>}\hspace{-0.32em}{=}\hspace{-0.32em}{>}}{}\<[23]%
\>[23]{}\mbox{\onelinecomment  Kleisli composition}{}\<[E]%
\\
\>[B]{}(\mathrel{{>}\hspace{-0.32em}{=}\hspace{-0.32em}{>}})\mathbin{::}\conid{Monad}\;\varid{m}{}\<[19]%
\>[19]{}\Rightarrow (\varid{a}\to \varid{m}\;\varid{b})\to (\varid{b}\to \varid{m}\;\varid{c}){}\<[E]%
\\
\>[19]{}\to \varid{a}\to \varid{m}\;\varid{c}{}\<[E]%
\\[\blanklineskip]%
\>[B]{}\keyword{infixr}\;\mathrm{1}\mathrel{{>}\hspace{-0.4em}{>}\hspace{-0.4em}{>}}{}\<[23]%
\>[23]{}\mbox{\onelinecomment  Forward composition}{}\<[E]%
\\
\>[B]{}(\mathrel{{>}\hspace{-0.4em}{>}\hspace{-0.4em}{>}}){}\<[8]%
\>[8]{}\mathbin{::}(\varid{a}\to \varid{b})\to (\varid{b}\to \varid{c})\to \varid{a}\to \varid{c}{}\<[E]%
\\[\blanklineskip]%
\>[B]{}\keyword{infixr}\;\mathrm{0}\triangleright{}\<[23]%
\>[23]{}\mbox{\onelinecomment  Reverse function application}{}\<[E]%
\\
\>[B]{}(\triangleright){}\<[7]%
\>[7]{}\mathbin{::}\varid{a}\to (\varid{a}\to \varid{b})\to \varid{b}{}\<[E]%
\ColumnHook
\end{hscode}\resethooks
\caption{API for various library functions used in this work}
\label{fig:containers} \label{fig:library}
\end{figure}

Building on the design of widely
used functions in Haskell (see \cref{fig:containers}), we
seek these basic operations:
\begin{hscode}\SaveRestoreHook
\column{B}{@{}>{\hspre}l<{\hspost}@{}}%
\column{11}{@{}>{\hspre}l<{\hspost}@{}}%
\column{E}{@{}>{\hspre}l<{\hspost}@{}}%
\>[B]{}\varid{emptyEM}{}\<[11]%
\>[11]{}\mathbin{::}\conid{ExprMap}\;\varid{v}{}\<[E]%
\\
\>[B]{}\varid{lkEM}{}\<[11]%
\>[11]{}\mathbin{::}\conid{Expr}\to \conid{ExprMap}\;\varid{v}\to \conid{Maybe}\;\varid{v}{}\<[E]%
\\
\>[B]{}\varid{atEM}{}\<[11]%
\>[11]{}\mathbin{::}\conid{Expr}\to \conid{TF}\;\varid{v}\to \conid{ExprMap}\;\varid{v}\to \conid{ExprMap}\;\varid{v}{}\<[E]%
\ColumnHook
\end{hscode}\resethooks
The lookup function \ensuremath{\varid{lkEM}}\footnote{We use short names \ensuremath{\varid{lkEM}} and \ensuremath{\varid{atEM}}
  consistently in this paper for compact code listings.
}
has a type that is familiar from every finite map.
The update function \ensuremath{\varid{atEM}}, typically called \ensuremath{\varid{alter}} in Haskell libraries,
changes the value stored at a particular key.
The caller provides a \emph{value transformation function} \ensuremath{\conid{TF}\;\varid{v}}, an
abbreviation for \ensuremath{\conid{Maybe}\;\varid{v}\to \conid{Maybe}\;\varid{v}} (see \Cref{fig:library}). This function
transforms the existing value associated with the key, if any (hence the input
\ensuremath{\conid{Maybe}}), to a new value, if any (hence the output \ensuremath{\conid{Maybe}}).
We can easily define \ensuremath{\varid{insertEM}} and \ensuremath{\varid{deleteEM}} from \ensuremath{\varid{atEM}}:
\begin{hscode}\SaveRestoreHook
\column{B}{@{}>{\hspre}l<{\hspost}@{}}%
\column{E}{@{}>{\hspre}l<{\hspost}@{}}%
\>[B]{}\varid{insertEM}\mathbin{::}\conid{Expr}\to \varid{v}\to \conid{ExprMap}\;\varid{v}\to \conid{ExprMap}\;\varid{v}{}\<[E]%
\\
\>[B]{}\varid{insertEM}\;\varid{e}\;\varid{v}\mathrel{=}\varid{atEM}\;\varid{e}\;(\mathbin{\char92 \char95 }\to \conid{Just}\;\varid{v}){}\<[E]%
\\[\blanklineskip]%
\>[B]{}\varid{deleteEM}\mathbin{::}\conid{Expr}\to \conid{ExprMap}\;\varid{v}\to \conid{ExprMap}\;\varid{v}{}\<[E]%
\\
\>[B]{}\varid{deleteEM}\;\varid{e}\mathrel{=}\varid{atEM}\;\varid{e}\;(\mathbin{\char92 \char95 }\to \conid{Nothing}){}\<[E]%
\ColumnHook
\end{hscode}\resethooks
You might wonder whether, for the purposes of this paper, we could just define \ensuremath{\varid{insertEM}},
leaving \ensuremath{\varid{atEM}} to the supplemental material of the extended version of this
paper~\cite{triemaps-extended,triemaps-github}, but as we will see in \Cref{sec:alter}, our
approach using tries requires the generality of \ensuremath{\varid{atEM}}.

We also support other standard operations on finite maps,
with types analogous to those in \Cref{fig:library}, including \ensuremath{\varid{unionEM}}, \ensuremath{\varid{mapEM}}, and \ensuremath{\varid{foldrEM}}.
%

\subsection{Tries: the basic idea} \label{sec:basic}

Here is a trie-based implementation for \ensuremath{\conid{Expr}}:
\begin{hscode}\SaveRestoreHook
\column{B}{@{}>{\hspre}l<{\hspost}@{}}%
\column{3}{@{}>{\hspre}l<{\hspost}@{}}%
\column{18}{@{}>{\hspre}l<{\hspost}@{}}%
\column{40}{@{}>{\hspre}l<{\hspost}@{}}%
\column{E}{@{}>{\hspre}l<{\hspost}@{}}%
\>[B]{}\keyword{data}\;\conid{ExprMap}\;\varid{v}{}\<[E]%
\\
\>[B]{}\hsindent{3}{}\<[3]%
\>[3]{}\mathrel{=}\conid{EM}\;\{\mskip1.5mu \varid{em\char95 var}{}\<[18]%
\>[18]{}\mathbin{::}\conid{Map}\;\conid{Var}\;\varid{v},\varid{em\char95 app}{}\<[40]%
\>[40]{}\mathbin{::}\conid{ExprMap}\;(\conid{ExprMap}\;\varid{v})\mskip1.5mu\}{}\<[E]%
\ColumnHook
\end{hscode}\resethooks
Here \ensuremath{\conid{Map}\;\conid{Var}\;\varid{v}} is any standard finite map (e.g. in \hackage{containers})
keyed by \ensuremath{\conid{Var}}, with values \ensuremath{\varid{v}}.
One way to understand this slightly odd data type is to study its lookup function:
\begin{hscode}\SaveRestoreHook
\column{B}{@{}>{\hspre}l<{\hspost}@{}}%
\column{3}{@{}>{\hspre}l<{\hspost}@{}}%
\column{6}{@{}>{\hspre}l<{\hspost}@{}}%
\column{14}{@{}>{\hspre}l<{\hspost}@{}}%
\column{15}{@{}>{\hspre}l<{\hspost}@{}}%
\column{18}{@{}>{\hspre}l<{\hspost}@{}}%
\column{E}{@{}>{\hspre}l<{\hspost}@{}}%
\>[B]{}\varid{lkEM}\mathbin{::}\conid{Expr}\to \conid{ExprMap}\;\varid{v}\to \conid{Maybe}\;\varid{v}{}\<[E]%
\\
\>[B]{}\varid{lkEM}\;\varid{e}\;(\conid{EM}\;\{\mskip1.5mu \varid{em\char95 var}\mathrel{=}\varid{var},\varid{em\char95 app}\mathrel{=}\varid{app}\mskip1.5mu\})\mathrel{=}\keyword{case}\;\varid{e}\;\keyword{of}{}\<[E]%
\\
\>[B]{}\hsindent{3}{}\<[3]%
\>[3]{}\conid{Var}\;\varid{x}{}\<[14]%
\>[14]{}\to \varid{\conid{Map}.lookup}\;\varid{x}\;\varid{var}{}\<[E]%
\\
\>[B]{}\hsindent{3}{}\<[3]%
\>[3]{}\conid{App}\;\varid{e}_{1}\;\varid{e}_{2}{}\<[14]%
\>[14]{}\to {}\<[18]%
\>[18]{}\keyword{case}\;\varid{lkEM}\;\varid{e}_{1}\;\varid{app}\;\keyword{of}{}\<[E]%
\\
\>[3]{}\hsindent{3}{}\<[6]%
\>[6]{}\conid{Nothing}{}\<[15]%
\>[15]{}\to \conid{Nothing}{}\<[E]%
\\
\>[3]{}\hsindent{3}{}\<[6]%
\>[6]{}\conid{Just}\;\varid{m}_{1}{}\<[15]%
\>[15]{}\to \varid{lkEM}\;\varid{e}_{2}\;\varid{m}_{1}{}\<[E]%
\ColumnHook
\end{hscode}\resethooks
This function pattern-matches on the target \ensuremath{\varid{e}}.  The \ensuremath{\conid{Var}} alternative
says that to look up a variable occurrence, just look that variable up in the
\ensuremath{\varid{em\char95 var}} field.
But if the expression is an \ensuremath{\conid{App}\;\varid{e}_{1}\;\varid{e}_{2}} node, we first look up \ensuremath{\varid{e}_{1}}
in the \ensuremath{\varid{em\char95 app}} field, \emph{which returns an \ensuremath{\conid{ExprMap}}}.  We then look up \ensuremath{\varid{e}_{2}}
in that map.  Each distinct \ensuremath{\varid{e}_{1}} yields a different \ensuremath{\conid{ExprMap}} in which to look up \ensuremath{\varid{e}_{2}}.

We can substantially abbreviate this code, at the expense of making it more cryptic, thus:
\begin{hscode}\SaveRestoreHook
\column{B}{@{}>{\hspre}l<{\hspost}@{}}%
\column{23}{@{}>{\hspre}l<{\hspost}@{}}%
\column{33}{@{}>{\hspre}l<{\hspost}@{}}%
\column{E}{@{}>{\hspre}l<{\hspost}@{}}%
\>[B]{}\varid{lkEM}\;(\conid{Var}\;\varid{x}){}\<[23]%
\>[23]{}\mathrel{=}\varid{em\char95 var}{}\<[33]%
\>[33]{}\mathrel{{>}\hspace{-0.4em}{>}\hspace{-0.4em}{>}}\varid{\conid{Map}.lookup}\;\varid{x}{}\<[E]%
\\
\>[B]{}\varid{lkEM}\;(\conid{App}\;\varid{e}_{1}\;\varid{e}_{2}){}\<[23]%
\>[23]{}\mathrel{=}\varid{em\char95 app}{}\<[33]%
\>[33]{}\mathrel{{>}\hspace{-0.4em}{>}\hspace{-0.4em}{>}}\varid{lkEM}\;\varid{e}_{1}\mathrel{{>}\hspace{-0.32em}{=}\hspace{-0.32em}{>}}\varid{lkEM}\;\varid{e}_{2}{}\<[E]%
\ColumnHook
\end{hscode}\resethooks
The function \ensuremath{\varid{em\char95 var}\mathbin{::}\conid{ExprMap}\;\varid{v}\to \conid{Map}\;\conid{Var}\;\varid{v}}
is the auto-generated selector that picks the \ensuremath{\varid{em\char95 var}} field from an \ensuremath{\conid{EM}} record, and similarly \ensuremath{\varid{em\char95 app}}.
The functions \ensuremath{(\mathrel{{>}\hspace{-0.32em}{=}\hspace{-0.32em}{>}})} and \ensuremath{(\mathrel{{>}\hspace{-0.4em}{>}\hspace{-0.4em}{>}})} are right-associative forward composition
operators, respectively monadic and non-monadic,
that chain the individual operations together (see \Cref{fig:library}).
Finally, we have $\eta$-reduced the definition, by omitting the \ensuremath{\varid{m}} parameter.
These abbreviations become quite worthwhile when we add more constructors, each with more fields,
to the key data type.

Notice that in contrast to the approach of \Cref{sec:ord}, \emph{we never compare two expressions
for equality or ordering}.  We simply walk down the \ensuremath{\conid{ExprMap}} structure, guided
at each step by the next node in the target.

This definition is extremely short and natural. But it embodies a hidden
complexity: \emph{it requires polymorphic recursion} \cite{mycroft-poly}. The recursive call to \ensuremath{\varid{lkEM}\;\varid{e}_{1}}
instantiates \ensuremath{\varid{v}} to a different type than the parent function definition.
Haskell supports polymorphic recursion readily, provided you give a type signature to
\ensuremath{\varid{lkEM}}, but not all languages do.

\subsection{Modifying tries} \label{sec:alter} \label{sec:empty-infinite}

It is not enough to look up in a trie -- we need to \emph{build} them too.
First, we need an empty trie. Here is one way to define it:
\begin{hscode}\SaveRestoreHook
\column{B}{@{}>{\hspre}l<{\hspost}@{}}%
\column{E}{@{}>{\hspre}l<{\hspost}@{}}%
\>[B]{}\varid{emptyEM}\mathbin{::}\conid{ExprMap}\;\varid{v}{}\<[E]%
\\
\>[B]{}\varid{emptyEM}\mathrel{=}\conid{EM}\;\{\mskip1.5mu \varid{em\char95 var}\mathrel{=}\varid{\conid{Map}.empty},\varid{em\char95 app}\mathrel{=}\varid{emptyEM}\mskip1.5mu\}{}\<[E]%
\ColumnHook
\end{hscode}\resethooks
It is interesting to note that \ensuremath{\varid{emptyEM}} is an infinite, recursive structure:
the \ensuremath{\varid{em\char95 app}} field refers back to \ensuremath{\varid{emptyEM}}.  We will change this
definition in \Cref{sec:empty}, but it works perfectly well for now.
Next, we need to \emph{alter} a triemap \emph{at} some key, implemented as
\ensuremath{\varid{atEM}}:%
\footnote{The name \ensuremath{\varid{atEM}} is borrowed from the \ensuremath{\conid{At}} type class of the
\hackage{lens} package. Indeed, \ensuremath{\conid{ExprMap}} could have an \ensuremath{\conid{At}} instance if \ensuremath{\varid{atEM}}
were generalised to an applicative traversal.}
\begin{hscode}\SaveRestoreHook
\column{B}{@{}>{\hspre}l<{\hspost}@{}}%
\column{3}{@{}>{\hspre}l<{\hspost}@{}}%
\column{14}{@{}>{\hspre}l<{\hspost}@{}}%
\column{21}{@{}>{\hspre}l<{\hspost}@{}}%
\column{29}{@{}>{\hspre}l<{\hspost}@{}}%
\column{E}{@{}>{\hspre}l<{\hspost}@{}}%
\>[B]{}\varid{atEM}\mathbin{::}\conid{Expr}\to \conid{TF}\;\varid{v}\to \conid{ExprMap}\;\varid{v}\to \conid{ExprMap}\;\varid{v}{}\<[E]%
\\
\>[B]{}\varid{atEM}\;\varid{e}\;\varid{tf}\;\varid{m}\mathord{@}(\conid{EM}\;\{\mskip1.5mu \varid{em\char95 var}\mathrel{=}\varid{var},\varid{em\char95 app}\mathrel{=}\varid{app}\mskip1.5mu\})\mathrel{=}\keyword{case}\;\varid{e}\;\keyword{of}{}\<[E]%
\\
\>[B]{}\hsindent{3}{}\<[3]%
\>[3]{}\conid{Var}\;\varid{x}{}\<[14]%
\>[14]{}\to \varid{m}\;\{\mskip1.5mu \varid{em\char95 var}{}\<[29]%
\>[29]{}\mathrel{=}\varid{\conid{Map}.alter}\;\varid{tf}\;\varid{x}\;\varid{var}\mskip1.5mu\}{}\<[E]%
\\
\>[B]{}\hsindent{3}{}\<[3]%
\>[3]{}\conid{App}\;\varid{e}_{1}\;\varid{e}_{2}{}\<[14]%
\>[14]{}\to \varid{m}\;\{\mskip1.5mu \varid{em\char95 app}{}\<[29]%
\>[29]{}\mathrel{=}\varid{atEM}\;\varid{e}_{1}\;(\varid{liftTF}\;(\varid{atEM}\;\varid{e}_{2}\;\varid{tf}))\;\varid{app}\mskip1.5mu\}{}\<[E]%
\\[\blanklineskip]%
\>[B]{}\varid{liftTF}\mathbin{::}(\conid{ExprMap}\;\varid{v}\to \conid{ExprMap}\;\varid{v})\to \conid{TF}\;(\conid{ExprMap}\;\varid{v}){}\<[E]%
\\
\>[B]{}\varid{liftTF}\;\varid{f}\;\conid{Nothing}{}\<[21]%
\>[21]{}\mathrel{=}\conid{Just}\;(\varid{f}\;\varid{emptyEM}){}\<[E]%
\\
\>[B]{}\varid{liftTF}\;\varid{f}\;(\conid{Just}\;\varid{m}){}\<[21]%
\>[21]{}\mathrel{=}\conid{Just}\;(\varid{f}\;\varid{m}){}\<[E]%
\ColumnHook
\end{hscode}\resethooks
In the \ensuremath{\conid{Var}} case, we must just update the map stored in the \ensuremath{\varid{em\char95 var}} field,
using the \ensuremath{\varid{\conid{Map}.alter}} function from \Cref{fig:containers}.
In the \ensuremath{\conid{App}} case we look up \ensuremath{\varid{e}_{1}} in \ensuremath{\varid{app}};
we should find a \ensuremath{\conid{ExprMap}} there, which we want to alter with \ensuremath{\varid{tf}}.
We can do that with a recursive call to \ensuremath{\varid{atEM}}, using \ensuremath{\varid{liftTF}}
for impedance-matching.

The \ensuremath{\conid{App}} case shows why we need the generality of \ensuremath{\varid{atEM}}.
Suppose we attempted to define an apparently-simpler \ensuremath{\varid{insertEM}} operation.
Its equation for \ensuremath{(\conid{App}\;\varid{e}_{1}\;\varid{e}_{2})} would look up \ensuremath{\varid{e}_{1}} --- and would then
need to alter that entry (an \ensuremath{\conid{ExprMap}}, remember) with the result of
inserting \ensuremath{(\varid{e}_{2},\varid{v})}.  So we are forced to define \ensuremath{\varid{atEM}} anyway.

We can abbreviate the code for \ensuremath{\varid{atEM}} using combinators, as we did in the case of
lookup, and doing so pays dividends when the key is a data type with
many constructors, each with many fields.  However, the details are
fiddly and not illuminating, so we omit them here.  Indeed, for the
same reason, in the rest of this paper we will typically omit the code
for \ensuremath{\varid{atEM}}, though the full code is available in the
supplement~\cite{triemaps-extended,triemaps-github}.

\subsection{Unions of maps}

A common operation on finite maps is to take their union:
\begin{hscode}\SaveRestoreHook
\column{B}{@{}>{\hspre}l<{\hspost}@{}}%
\column{E}{@{}>{\hspre}l<{\hspost}@{}}%
\>[B]{}\varid{unionEM}\mathbin{::}\conid{ExprMap}\;\varid{v}\to \conid{ExprMap}\;\varid{v}\to \conid{ExprMap}\;\varid{v}{}\<[E]%
\ColumnHook
\end{hscode}\resethooks
In tree-based implementations of finite maps, such union operations can be tricky.
The two trees, which have been built independently, might not have the same
left-subtree/right-subtree structure, so some careful rebalancing may be required.
But for tries there are no such worries --
their structure is identical, and we can simply zip them together.  There is one
wrinkle: just as we had to generalise \ensuremath{\varid{insertEM}} to \ensuremath{\varid{atEM}},
to accommodate the nested map in \ensuremath{\varid{em\char95 app}}, so we need to generalise \ensuremath{\varid{unionEM}} to \ensuremath{\varid{unionWithEM}}:
\begin{hscode}\SaveRestoreHook
\column{B}{@{}>{\hspre}l<{\hspost}@{}}%
\column{14}{@{}>{\hspre}c<{\hspost}@{}}%
\column{14E}{@{}l@{}}%
\column{18}{@{}>{\hspre}l<{\hspost}@{}}%
\column{E}{@{}>{\hspre}l<{\hspost}@{}}%
\>[B]{}\varid{unionWithEM}{}\<[14]%
\>[14]{}\mathbin{::}{}\<[14E]%
\>[18]{}(\varid{v}\to \varid{v}\to \varid{v}){}\<[E]%
\\
\>[14]{}\to {}\<[14E]%
\>[18]{}\conid{ExprMap}\;\varid{v}\to \conid{ExprMap}\;\varid{v}\to \conid{ExprMap}\;\varid{v}{}\<[E]%
\ColumnHook
\end{hscode}\resethooks
When a key appears on both maps, the combining function is used to
combine the two corresponding values.
With that generalisation, the code is as follows:
\begin{hscode}\SaveRestoreHook
\column{B}{@{}>{\hspre}l<{\hspost}@{}}%
\column{3}{@{}>{\hspre}l<{\hspost}@{}}%
\column{9}{@{}>{\hspre}l<{\hspost}@{}}%
\column{16}{@{}>{\hspre}l<{\hspost}@{}}%
\column{E}{@{}>{\hspre}l<{\hspost}@{}}%
\>[B]{}\varid{unionWithEM}\;\varid{f}\;{}\<[16]%
\>[16]{}(\conid{EM}\;\{\mskip1.5mu \varid{em\char95 var}\mathrel{=}\varid{var}_{1},\varid{em\char95 app}\mathrel{=}\varid{app}_{1}\mskip1.5mu\})\;{}\<[E]%
\\
\>[16]{}(\conid{EM}\;\{\mskip1.5mu \varid{em\char95 var}\mathrel{=}\varid{var}_{2},\varid{em\char95 app}\mathrel{=}\varid{app}_{2}\mskip1.5mu\}){}\<[E]%
\\
\>[B]{}\hsindent{3}{}\<[3]%
\>[3]{}\mathrel{=}\conid{EM}\;{}\<[9]%
\>[9]{}\{\mskip1.5mu \varid{em\char95 var}\mathrel{=}\varid{\conid{Map}.unionWith}\;\varid{f}\;\varid{var}_{1}\;\varid{var}_{2}{}\<[E]%
\\
\>[9]{},\varid{em\char95 app}\mathrel{=}\varid{unionWithEM}\;(\varid{unionWithEM}\;\varid{f})\;\varid{app}_{1}\;\varid{app}_{2}\mskip1.5mu\}{}\<[E]%
\ColumnHook
\end{hscode}\resethooks
It could hardly be simpler.

\subsection{Folds and the empty map} \label{sec:fold} \label{sec:empty}

The strange, infinite definition of \ensuremath{\varid{emptyEM}} given in \Cref{sec:empty-infinite}
works fine (in a lazy language at least) for lookup, alteration, and union, but it fails
fundamentally when we want to \emph{iterate} over the elements of the trie.
For example, suppose we wanted to count the number of elements in the finite map; in \ensuremath{\varid{containers}}
this is the function \ensuremath{\varid{\conid{Map}.size}} (\Cref{fig:containers}).  We might attempt:
\begin{hscode}\SaveRestoreHook
\column{B}{@{}>{\hspre}l<{\hspost}@{}}%
\column{3}{@{}>{\hspre}l<{\hspost}@{}}%
\column{E}{@{}>{\hspre}l<{\hspost}@{}}%
\>[B]{}\varid{sizeEM}\mathbin{::}\conid{ExprMap}\;\varid{v}\to \conid{Int}{}\<[E]%
\\
\>[B]{}\varid{sizeEM}\;(\conid{EM}\;\{\mskip1.5mu \varid{em\char95 var}\mathrel{=}\varid{var},\varid{em\char95 app}\mathrel{=}\varid{app}\mskip1.5mu\}){}\<[E]%
\\
\>[B]{}\hsindent{3}{}\<[3]%
\>[3]{}\mathrel{=}\varid{\conid{Map}.size}\;\varid{var} \; \mathbin{+} \; ???{}\<[E]%
\ColumnHook
\end{hscode}\resethooks
We seem stuck because the size of the \ensuremath{\varid{app}} map is not what we want: rather,
we want to add up the sizes of its \emph{elements}, and we don't have a way to do that yet.
The right thing to do is to generalise to a fold:
\begin{hscode}\SaveRestoreHook
\column{B}{@{}>{\hspre}l<{\hspost}@{}}%
\column{3}{@{}>{\hspre}l<{\hspost}@{}}%
\column{5}{@{}>{\hspre}l<{\hspost}@{}}%
\column{E}{@{}>{\hspre}l<{\hspost}@{}}%
\>[B]{}\varid{foldrEM}\mathbin{::}\keyword{$\forall$}\!\! \hsforall \;\varid{v}\hsdot{\circ }{.\,}(\varid{v}\to \varid{r}\to \varid{r})\to \varid{r}\to \conid{ExprMap}\;\varid{v}\to \varid{r}{}\<[E]%
\\
\>[B]{}\varid{foldrEM}\;\varid{k}\;\varid{z}\;(\conid{EM}\;\{\mskip1.5mu \varid{em\char95 var}\mathrel{=}\varid{var},\varid{em\char95 app}\mathrel{=}\varid{app}\mskip1.5mu\}){}\<[E]%
\\
\>[B]{}\hsindent{3}{}\<[3]%
\>[3]{}\mathrel{=}\varid{\conid{Map}.foldr}\;\varid{k}\;\varid{z}_{1}\;\varid{var}{}\<[E]%
\\
\>[B]{}\hsindent{3}{}\<[3]%
\>[3]{}\keyword{where}{}\<[E]%
\\
\>[3]{}\hsindent{2}{}\<[5]%
\>[5]{}\varid{z}_{1}\mathrel{=}\varid{foldrEM}\;\varid{kapp}\;\varid{z}\;(\varid{app}\mathbin{::}\conid{ExprMap}\;(\conid{ExprMap}\;\varid{v})){}\<[E]%
\\
\>[3]{}\hsindent{2}{}\<[5]%
\>[5]{}\varid{kapp}\;\varid{m}_{1}\;\varid{r}\mathrel{=}\varid{foldrEM}\;\varid{k}\;\varid{r}\;\varid{m}_{1}{}\<[E]%
\ColumnHook
\end{hscode}\resethooks
In the binding for \ensuremath{\varid{z}_{1}} we fold over \ensuremath{\varid{app}}, using
\ensuremath{\varid{kapp}} to combine the map we find with the accumulator, by again
folding over the map with \ensuremath{\varid{foldrEM}}.

But alas, \ensuremath{\varid{foldrEM}} will never terminate!  It always invokes itself immediately
(in \ensuremath{\varid{z}_{1}}) on \ensuremath{\varid{app}}; but that invocation will again recursively invoke
\ensuremath{\varid{foldrEM}}; and so on forever.
The solution is simple: we just need an explicit representation of the empty map.
Here is one way to do it (we will see another in \Cref{sec:singleton}):
\begin{hscode}\SaveRestoreHook
\column{B}{@{}>{\hspre}l<{\hspost}@{}}%
\column{3}{@{}>{\hspre}l<{\hspost}@{}}%
\column{5}{@{}>{\hspre}l<{\hspost}@{}}%
\column{17}{@{}>{\hspre}l<{\hspost}@{}}%
\column{E}{@{}>{\hspre}l<{\hspost}@{}}%
\>[B]{}\keyword{data}\;\conid{ExprMap}\;\varid{v}{}\<[17]%
\>[17]{}\mathrel{=}\conid{EmptyEM}\mid \conid{EM}\;\{\mskip1.5mu \varid{em\char95 var}\mathbin{::}\mathbin{...},\varid{em\char95 app}\mathbin{::}\mathbin{...}\mskip1.5mu\}{}\<[E]%
\\[\blanklineskip]%
\>[B]{}\varid{emptyEM}\mathbin{::}\conid{ExprMap}\;\varid{v}{}\<[E]%
\\
\>[B]{}\varid{emptyEM}\mathrel{=}\conid{EmptyEM}{}\<[E]%
\\[\blanklineskip]%
\>[B]{}\varid{foldrEM}\mathbin{::}(\varid{v}\to \varid{r}\to \varid{r})\to \varid{r}\to \conid{ExprMap}\;\varid{v}\to \varid{r}{}\<[E]%
\\
\>[B]{}\varid{foldrEM}\;\varid{k}\;\varid{z}\;\conid{EmptyEM}\mathrel{=}\varid{z}{}\<[E]%
\\
\>[B]{}\varid{foldrEM}\;\varid{k}\;\varid{z}\;(\conid{EM}\;\{\mskip1.5mu \varid{em\char95 var}\mathrel{=}\varid{var},\varid{em\char95 app}\mathrel{=}\varid{app}\mskip1.5mu\}){}\<[E]%
\\
\>[B]{}\hsindent{3}{}\<[3]%
\>[3]{}\mathrel{=}\varid{\conid{Map}.foldr}\;\varid{k}\;\varid{z}_{1}\;\varid{var}{}\<[E]%
\\
\>[B]{}\hsindent{3}{}\<[3]%
\>[3]{}\keyword{where}{}\<[E]%
\\
\>[3]{}\hsindent{2}{}\<[5]%
\>[5]{}\varid{z}_{1}\mathrel{=}\varid{foldrEM}\;\varid{kapp}\;\varid{z}\;\varid{app}{}\<[E]%
\\
\>[3]{}\hsindent{2}{}\<[5]%
\>[5]{}\varid{kapp}\;\varid{m}_{1}\;\varid{r}\mathrel{=}\varid{foldrEM}\;\varid{k}\;\varid{r}\;\varid{m}_{1}{}\<[E]%
\ColumnHook
\end{hscode}\resethooks
Equipped with a fold, we can easily define the size function, and another
that returns the range of the map:
\begin{hscode}\SaveRestoreHook
\column{B}{@{}>{\hspre}l<{\hspost}@{}}%
\column{E}{@{}>{\hspre}l<{\hspost}@{}}%
\>[B]{}\varid{sizeEM}\mathbin{::}\conid{ExprMap}\;\varid{v}\to \conid{Int}{}\<[E]%
\\
\>[B]{}\varid{sizeEM}\mathrel{=}\varid{foldrEM}\;(\lambda \anonymous \;\varid{n}\to \varid{n}\mathbin{+}\mathrm{1})\;\mathrm{0}{}\<[E]%
\\[\blanklineskip]%
\>[B]{}\varid{elemsEM}\mathbin{::}\conid{ExprMap}\;\varid{v}\to [\mskip1.5mu \varid{v}\mskip1.5mu]{}\<[E]%
\\
\>[B]{}\varid{elemsEM}\mathrel{=}\varid{foldrEM}\;(\mathbin{:})\;[\mskip1.5mu \mskip1.5mu]{}\<[E]%
\ColumnHook
\end{hscode}\resethooks

\subsection{A type class for triemaps} \label{sec:generalised} \label{sec:class}

Since all our triemaps share a common interface, it is useful to define
a type class for them:
\begin{hscode}\SaveRestoreHook
\column{B}{@{}>{\hspre}l<{\hspost}@{}}%
\column{4}{@{}>{\hspre}l<{\hspost}@{}}%
\column{17}{@{}>{\hspre}l<{\hspost}@{}}%
\column{E}{@{}>{\hspre}l<{\hspost}@{}}%
\>[B]{}\keyword{class}\;\conid{Eq}\;(\conid{Key}\;\varid{tm})\Rightarrow \conid{TrieMap}\;\varid{tm}\;\keyword{where}{}\<[E]%
\\
\>[B]{}\hsindent{4}{}\<[4]%
\>[4]{}\keyword{type}\;\conid{Key}\;\varid{tm}\mathbin{::}\conid{Type}{}\<[E]%
\\
\>[B]{}\hsindent{4}{}\<[4]%
\>[4]{}\varid{emptyTM}{}\<[17]%
\>[17]{}\mathbin{::}\varid{tm}\;\varid{a}{}\<[E]%
\\
\>[B]{}\hsindent{4}{}\<[4]%
\>[4]{}\varid{lkTM}{}\<[17]%
\>[17]{}\mathbin{::}\conid{Key}\;\varid{tm}\to \varid{tm}\;\varid{a}\to \conid{Maybe}\;\varid{a}{}\<[E]%
\\
\>[B]{}\hsindent{4}{}\<[4]%
\>[4]{}\varid{atTM}{}\<[17]%
\>[17]{}\mathbin{::}\conid{Key}\;\varid{tm}\to \conid{TF}\;\varid{a}\to \varid{tm}\;\varid{a}\to \varid{tm}\;\varid{a}{}\<[E]%
\\
\>[B]{}\hsindent{4}{}\<[4]%
\>[4]{}\varid{foldrTM}{}\<[17]%
\>[17]{}\mathbin{::}(\varid{a}\to \varid{b}\to \varid{b})\to \varid{tm}\;\varid{a}\to \varid{b}\to \varid{b}{}\<[E]%
\\
\>[B]{}\hsindent{4}{}\<[4]%
\>[4]{}\varid{unionWithTM}{}\<[17]%
\>[17]{}\mathbin{::}(\varid{a}\to \varid{a}\to \varid{a})\to \varid{tm}\;\varid{a}\to \varid{tm}\;\varid{a}\to \varid{tm}\;\varid{a}{}\<[E]%
\\
\>[B]{}\hsindent{4}{}\<[4]%
\>[4]{}\ldots{}\<[E]%
\ColumnHook
\end{hscode}\resethooks
The class constraint \ensuremath{\conid{TrieMap}\;\varid{tm}} says that the type \ensuremath{\varid{tm}} is a triemap, with operations
\ensuremath{\varid{emptyTM}}, \ensuremath{\varid{lkTM}} etc.
The class has an \emph{associated type} \cite{associated-types}, \ensuremath{\conid{Key}\;\varid{tm}},
a type-level function that transforms the type of the triemap into
the type of \emph{keys} of that triemap.

Now we can witness the fact that \ensuremath{\conid{ExprMap}} is a \ensuremath{\conid{TrieMap}}, like this:
\begin{hscode}\SaveRestoreHook
\column{B}{@{}>{\hspre}l<{\hspost}@{}}%
\column{3}{@{}>{\hspre}l<{\hspost}@{}}%
\column{13}{@{}>{\hspre}l<{\hspost}@{}}%
\column{E}{@{}>{\hspre}l<{\hspost}@{}}%
\>[B]{}\keyword{instance}\;\conid{TrieMap}\;\conid{ExprMap}\;\keyword{where}{}\<[E]%
\\
\>[B]{}\hsindent{3}{}\<[3]%
\>[3]{}\keyword{type}\;\conid{Key}\;\conid{ExprMap}\mathrel{=}\conid{Expr}{}\<[E]%
\\
\>[B]{}\hsindent{3}{}\<[3]%
\>[3]{}\varid{emptyTM}{}\<[13]%
\>[13]{}\mathrel{=}\varid{emptyEM}{}\<[E]%
\\
\>[B]{}\hsindent{3}{}\<[3]%
\>[3]{}\varid{lkTM}{}\<[13]%
\>[13]{}\mathrel{=}\varid{lkEM}{}\<[E]%
\\
\>[B]{}\hsindent{3}{}\<[3]%
\>[3]{}\varid{atTM}{}\<[13]%
\>[13]{}\mathrel{=}\varid{atEM}{}\<[E]%
\\
\>[B]{}\hsindent{3}{}\<[3]%
\>[3]{}\ldots{}\<[E]%
\ColumnHook
\end{hscode}\resethooks
Having a class allows us to write helper functions that work for any triemap
implementation, such as
\begin{hscode}\SaveRestoreHook
\column{B}{@{}>{\hspre}l<{\hspost}@{}}%
\column{E}{@{}>{\hspre}l<{\hspost}@{}}%
\>[B]{}\varid{insertTM}\mathbin{::}\conid{TrieMap}\;\varid{tm}\Rightarrow \conid{Key}\;\varid{tm}\to \varid{v}\to \varid{tm}\;\varid{v}\to \varid{tm}\;\varid{v}{}\<[E]%
\\
\>[B]{}\varid{insertTM}\;\varid{k}\;\varid{v}\mathrel{=}\varid{atTM}\;\varid{k}\;(\mathbin{\char92 \char95 }\to \conid{Just}\;\varid{v}){}\<[E]%
\\[\blanklineskip]%
\>[B]{}\varid{deleteTM}\mathbin{::}\conid{TrieMap}\;\varid{tm}\Rightarrow \conid{Key}\;\varid{tm}\to \varid{tm}\;\varid{v}\to \varid{tm}\;\varid{v}{}\<[E]%
\\
\>[B]{}\varid{deleteTM}\;\varid{k}\mathrel{=}\varid{atTM}\;\varid{k}\;(\mathbin{\char92 \char95 }\to \conid{Nothing}){}\<[E]%
\ColumnHook
\end{hscode}\resethooks
But that is not all.
Suppose our expressions had multi-argument apply nodes, \ensuremath{\conid{AppV}}, thus
\begin{hscode}\SaveRestoreHook
\column{B}{@{}>{\hspre}l<{\hspost}@{}}%
\column{E}{@{}>{\hspre}l<{\hspost}@{}}%
\>[B]{}\keyword{data}\;\conid{Expr}\mathrel{=}\ldots\mid \conid{AppV}\;\conid{Expr}\;[\mskip1.5mu \conid{Expr}\mskip1.5mu]{}\<[E]%
\ColumnHook
\end{hscode}\resethooks
Then we would need to build a trie keyed by a \emph{list} of \ensuremath{\conid{Expr}}.
A list is just another algebraic data type, built with nil and cons,
so we \emph{could} use exactly the same approach, thus
\begin{hscode}\SaveRestoreHook
\column{B}{@{}>{\hspre}l<{\hspost}@{}}%
\column{E}{@{}>{\hspre}l<{\hspost}@{}}%
\>[B]{}\varid{lkLEM}\mathbin{::}[\mskip1.5mu \conid{Expr}\mskip1.5mu]\to \conid{ListExprMap}\;\varid{v}\to \conid{Maybe}\;\varid{v}{}\<[E]%
\ColumnHook
\end{hscode}\resethooks
But rather than define a \ensuremath{\conid{ListExprMap}} for keys of type \ensuremath{[\mskip1.5mu \conid{Expr}\mskip1.5mu]},
and a \ensuremath{\conid{ListDeclMap}} for keys of type \ensuremath{[\mskip1.5mu \conid{Decl}\mskip1.5mu]}, etc, we would obviously prefer
to build a trie for lists of \emph{any type}, like this \cite{hinze:generalized}:
\begin{hscode}\SaveRestoreHook
\column{B}{@{}>{\hspre}l<{\hspost}@{}}%
\column{4}{@{}>{\hspre}l<{\hspost}@{}}%
\column{14}{@{}>{\hspre}l<{\hspost}@{}}%
\column{18}{@{}>{\hspre}l<{\hspost}@{}}%
\column{25}{@{}>{\hspre}l<{\hspost}@{}}%
\column{35}{@{}>{\hspre}l<{\hspost}@{}}%
\column{54}{@{}>{\hspre}l<{\hspost}@{}}%
\column{E}{@{}>{\hspre}l<{\hspost}@{}}%
\>[B]{}\keyword{instance}\;\conid{TrieMap}\;\varid{tm}\Rightarrow \conid{TrieMap}\;(\conid{ListMap}\;\varid{tm})\;\keyword{where}{}\<[E]%
\\
\>[B]{}\hsindent{4}{}\<[4]%
\>[4]{}\keyword{type}\;\conid{Key}\;(\conid{ListMap}\;\varid{tm})\mathrel{=}[\mskip1.5mu \conid{Key}\;\varid{tm}\mskip1.5mu]{}\<[E]%
\\
\>[B]{}\hsindent{4}{}\<[4]%
\>[4]{}\varid{emptyTM}{}\<[14]%
\>[14]{}\mathrel{=}\varid{emptyLM}{}\<[E]%
\\
\>[B]{}\hsindent{4}{}\<[4]%
\>[4]{}\varid{lkTM}{}\<[14]%
\>[14]{}\mathrel{=}\varid{lkLM}{}\<[E]%
\\
\>[B]{}\hsindent{4}{}\<[4]%
\>[4]{}\mathbin{...}{}\<[E]%
\\[\blanklineskip]%
\>[B]{}\keyword{data}\;\conid{ListMap}\;\varid{tm}\;\varid{v}\mathrel{=}\conid{LM}\;{}\<[25]%
\>[25]{}\{\mskip1.5mu \varid{lm\char95 nil}{}\<[35]%
\>[35]{}\mathbin{::}\conid{Maybe}\;\varid{v}{}\<[E]%
\\
\>[25]{},\varid{lm\char95 cons}\mathbin{::}\varid{tm}\;(\conid{ListMap}\;\varid{tm}\;{}\<[54]%
\>[54]{}\varid{v})\mskip1.5mu\}{}\<[E]%
\\[\blanklineskip]%
\>[B]{}\varid{emptyLM}\mathbin{::}\conid{TrieMap}\;\varid{tm}\Rightarrow \conid{ListMap}\;\varid{tm}{}\<[E]%
\\
\>[B]{}\varid{emptyLM}\mathrel{=}\conid{LM}\;\{\mskip1.5mu \varid{lm\char95 nil}\mathrel{=}\conid{Nothing},\varid{lm\char95 cons}\mathrel{=}\varid{emptyTM}\mskip1.5mu\}{}\<[E]%
\\[\blanklineskip]%
\>[B]{}\varid{lkLM}\mathbin{::}\conid{TrieMap}\;\varid{tm}\Rightarrow [\mskip1.5mu \conid{Key}\;\varid{tm}\mskip1.5mu]\to \conid{ListMap}\;\varid{tm}\;\varid{v}\to \conid{Maybe}\;\varid{v}{}\<[E]%
\\
\>[B]{}\varid{lkLM}\;[\mskip1.5mu \mskip1.5mu]{}\<[18]%
\>[18]{}\mathrel{=}\varid{lm\char95 nil}{}\<[E]%
\\
\>[B]{}\varid{lkLM}\;(\varid{k}\mathbin{:}\varid{ks}){}\<[18]%
\>[18]{}\mathrel{=}\varid{lm\char95 cons}\mathrel{{>}\hspace{-0.4em}{>}\hspace{-0.4em}{>}}\varid{lkTM}\;\varid{k}\mathrel{{>}\hspace{-0.32em}{=}\hspace{-0.32em}{>}}\varid{lkLM}\;\varid{ks}{}\<[E]%
\ColumnHook
\end{hscode}\resethooks
The code for \ensuremath{\varid{atLM}} and \ensuremath{\varid{foldrLM}} is routine. Notice that all of
these functions are polymorphic in \ensuremath{\varid{tm}}, the triemap for the list elements.

\subsection{Singleton maps, and empty maps revisited} \label{sec:singleton}

Suppose we start with an empty map, and insert a value
with a key (an \ensuremath{\conid{Expr}}) such as
\begin{hscode}\SaveRestoreHook
\column{B}{@{}>{\hspre}l<{\hspost}@{}}%
\column{3}{@{}>{\hspre}l<{\hspost}@{}}%
\column{E}{@{}>{\hspre}l<{\hspost}@{}}%
\>[3]{}\conid{App}\;(\conid{App}\;(\conid{Var}\;\text{\ttfamily \char34 f\char34})\;(\conid{Var}\;\text{\ttfamily \char34 x\char34}))\;(\conid{Var}\;\text{\ttfamily \char34 y\char34}){}\<[E]%
\ColumnHook
\end{hscode}\resethooks
Looking at the code
for \ensuremath{\varid{atEM}} in \Cref{sec:alter}, you can see that
because there is an \ensuremath{\conid{App}} at the root, we will build an
\ensuremath{\conid{EM}} record with an empty \ensuremath{\varid{em\char95 var}}, and an
\ensuremath{\varid{em\char95 app}} field that is... another \ensuremath{\conid{EM}}
record.  Again the \ensuremath{\varid{em\char95 var}} field will contain an
empty map, while the \ensuremath{\varid{em\char95 app}} field is a further \ensuremath{\conid{EM}} record.

In effect, the key is linearised into a chain of \ensuremath{\conid{EM}} records.
This is great when there are a lot of keys with shared structure, but
once we are in a sub-tree that represents a \emph{single} key-value pair it is
a rather inefficient way to represent the key.  So a simple idea is this:
when an \ensuremath{\conid{ExprMap}} represents a single key-value pair, represent it
directly as a key-value pair, like this:
\begin{hscode}\SaveRestoreHook
\column{B}{@{}>{\hspre}l<{\hspost}@{}}%
\column{17}{@{}>{\hspre}c<{\hspost}@{}}%
\column{17E}{@{}l@{}}%
\column{20}{@{}>{\hspre}l<{\hspost}@{}}%
\column{38}{@{}>{\hspre}l<{\hspost}@{}}%
\column{E}{@{}>{\hspre}l<{\hspost}@{}}%
\>[B]{}\keyword{data}\;\conid{ExprMap}\;\varid{v}{}\<[17]%
\>[17]{}\mathrel{=}{}\<[17E]%
\>[20]{}\conid{EmptyEM}{}\<[E]%
\\
\>[17]{}\mid {}\<[17E]%
\>[20]{}\conid{SingleEM}\;\conid{Expr}\;\varid{v}{}\<[38]%
\>[38]{}\mbox{\onelinecomment  A single key/value pair}{}\<[E]%
\\
\>[17]{}\mid {}\<[17E]%
\>[20]{}\conid{EM}\;\{\mskip1.5mu \varid{em\char95 var}\mathbin{::}\mathbin{...},\varid{em\char95 app}\mathbin{::}\mathbin{...}\mskip1.5mu\}{}\<[E]%
\ColumnHook
\end{hscode}\resethooks
But in the triemap for each new data type \ensuremath{\conid{X}},
we will have to tiresomely repeat these extra data constructors, \ensuremath{\conid{EmptyX}} and \ensuremath{\conid{SingleX}}.
For example we would have to add \ensuremath{\conid{EmptyList}} and \ensuremath{\conid{SingleList}} to the \ensuremath{\conid{ListMap}} data type
of \Cref{sec:class}.
It is better instead to abstract over the enclosed triemap, as follows:%
\footnote{\ensuremath{\conid{SEMap}} stands for \enquote{singleton or empty map}.}
\begin{hscode}\SaveRestoreHook
\column{B}{@{}>{\hspre}l<{\hspost}@{}}%
\column{3}{@{}>{\hspre}l<{\hspost}@{}}%
\column{13}{@{}>{\hspre}l<{\hspost}@{}}%
\column{18}{@{}>{\hspre}c<{\hspost}@{}}%
\column{18E}{@{}l@{}}%
\column{21}{@{}>{\hspre}l<{\hspost}@{}}%
\column{31}{@{}>{\hspre}l<{\hspost}@{}}%
\column{E}{@{}>{\hspre}l<{\hspost}@{}}%
\>[B]{}\keyword{data}\;\conid{SEMap}\;\varid{tm}\;\varid{v}{}\<[18]%
\>[18]{}\mathrel{=}{}\<[18E]%
\>[21]{}\conid{EmptySEM}{}\<[E]%
\\
\>[18]{}\mid {}\<[18E]%
\>[21]{}\conid{SingleSEM}\;(\conid{Key}\;\varid{tm})\;\varid{v}{}\<[E]%
\\
\>[18]{}\mid {}\<[18E]%
\>[21]{}\conid{MultiSEM}\;{}\<[31]%
\>[31]{}(\varid{tm}\;\varid{v}){}\<[E]%
\\[\blanklineskip]%
\>[B]{}\keyword{instance}\;\conid{TrieMap}\;\varid{tm}\Rightarrow \conid{TrieMap}\;(\conid{SEMap}\;\varid{tm})\;\keyword{where}{}\<[E]%
\\
\>[B]{}\hsindent{3}{}\<[3]%
\>[3]{}\keyword{type}\;\conid{Key}\;(\conid{SEMap}\;\varid{tm})\mathrel{=}\conid{Key}\;\varid{tm}{}\<[E]%
\\
\>[B]{}\hsindent{3}{}\<[3]%
\>[3]{}\varid{emptyTM}{}\<[13]%
\>[13]{}\mathrel{=}\conid{EmptySEM}{}\<[E]%
\\
\>[B]{}\hsindent{3}{}\<[3]%
\>[3]{}\varid{lkTM}{}\<[13]%
\>[13]{}\mathrel{=}\varid{lkSEM}{}\<[E]%
\\
\>[B]{}\hsindent{3}{}\<[3]%
\>[3]{}\varid{atTM}{}\<[13]%
\>[13]{}\mathrel{=}\varid{atSEM}{}\<[E]%
\\
\>[B]{}\hsindent{3}{}\<[3]%
\>[3]{}\mathbin{...}{}\<[E]%
\ColumnHook
\end{hscode}\resethooks
The code for lookup practically writes itself. We abstract over \ensuremath{\conid{Maybe}}
with some \ensuremath{\conid{MonadPlus}} combinators to enjoy forward compatibility to
\Cref{sec:matching}:
\begin{hscode}\SaveRestoreHook
\column{B}{@{}>{\hspre}l<{\hspost}@{}}%
\column{3}{@{}>{\hspre}l<{\hspost}@{}}%
\column{19}{@{}>{\hspre}l<{\hspost}@{}}%
\column{E}{@{}>{\hspre}l<{\hspost}@{}}%
\>[B]{}\varid{lkSEM}\mathbin{::}\conid{TrieMap}\;\varid{tm}\Rightarrow \conid{Key}\;\varid{tm}\to \conid{SEMap}\;\varid{tm}\;\varid{v}\to \conid{Maybe}\;\varid{v}{}\<[E]%
\\
\>[B]{}\varid{lkSEM}\;\varid{k}\;\varid{m}\mathrel{=}\keyword{case}\;\varid{m}\;\keyword{of}{}\<[E]%
\\
\>[B]{}\hsindent{3}{}\<[3]%
\>[3]{}\conid{EmptySEM}{}\<[19]%
\>[19]{}\to \varid{mzero}{}\<[E]%
\\
\>[B]{}\hsindent{3}{}\<[3]%
\>[3]{}\conid{SingleSEM}\;\varid{pk}\;\varid{v}{}\<[19]%
\>[19]{}\to \varid{guard}\;(\varid{k}{}\mathop{==}{}\varid{pk})\mathrel{{>}\hspace{-0.4em}{>}}\varid{pure}\;\varid{v}{}\<[E]%
\\
\>[B]{}\hsindent{3}{}\<[3]%
\>[3]{}\conid{MultiSEM}\;\varid{m}{}\<[19]%
\>[19]{}\to \varid{lkTM}\;\varid{k}\;\varid{m}{}\<[E]%
\ColumnHook
\end{hscode}\resethooks
Where \ensuremath{\varid{mzero}} means \ensuremath{\conid{Nothing}} and \ensuremath{\varid{pure}} means \ensuremath{\conid{Just}}. The \ensuremath{\varid{guard}} expression
in the \ensuremath{\conid{SingleSEM}} will return \ensuremath{\conid{Nothing}} when the key expression \ensuremath{\varid{k}} doesn't
equate to the pattern expression \ensuremath{\varid{pk}}.
To test for said equality we require an \ensuremath{\conid{Eq}\;(\conid{Key}\;\varid{tm})} instance, hence it is
a superclass of \ensuremath{\conid{TrieMap}\;\varid{tm}} in the \ensuremath{\keyword{class}} declaration in \Cref{sec:class}.

The code for alter is more interesting, because it governs the shift from
\ensuremath{\conid{EmptySEM}} to \ensuremath{\conid{SingleSEM}} and thence to \ensuremath{\conid{MultiSEM}}:
\begin{hscode}\SaveRestoreHook
\column{B}{@{}>{\hspre}l<{\hspost}@{}}%
\column{3}{@{}>{\hspre}l<{\hspost}@{}}%
\column{7}{@{}>{\hspre}l<{\hspost}@{}}%
\column{11}{@{}>{\hspre}l<{\hspost}@{}}%
\column{16}{@{}>{\hspre}l<{\hspost}@{}}%
\column{46}{@{}>{\hspre}l<{\hspost}@{}}%
\column{55}{@{}>{\hspre}l<{\hspost}@{}}%
\column{E}{@{}>{\hspre}l<{\hspost}@{}}%
\>[B]{}\varid{atSEM}{}\<[11]%
\>[11]{}\mathbin{::}\conid{TrieMap}\;\varid{tm}{}\<[E]%
\\
\>[11]{}\Rightarrow \conid{Key}\;\varid{tm}\to \conid{TF}\;\varid{v}\to \conid{SEMap}\;\varid{tm}\;\varid{v}\to \conid{SEMap}\;\varid{tm}\;\varid{v}{}\<[E]%
\\
\>[B]{}\varid{atSEM}\;\varid{k}\;\varid{tf}\;\conid{EmptySEM}\mathrel{=}\keyword{case}\;\varid{tf}\;\conid{Nothing}\;\keyword{of}\;{}\<[46]%
\>[46]{}\conid{Nothing}{}\<[55]%
\>[55]{}\to \conid{EmptySEM}{}\<[E]%
\\
\>[46]{}\conid{Just}\;\varid{v}{}\<[55]%
\>[55]{}\to \conid{SingleSEM}\;\varid{k}\;\varid{v}{}\<[E]%
\\
\>[B]{}\varid{atSEM}\;\varid{k}_{1}\;\varid{tf}\;(\conid{SingleSEM}\;\varid{k}_{2}\;\varid{v}_{2})\mathrel{=}\keyword{if}\;\varid{k}_{1}{}\mathop{==}{}\varid{k}_{2}{}\<[E]%
\\
\>[B]{}\hsindent{3}{}\<[3]%
\>[3]{}\keyword{then}\;\keyword{case}\;\varid{tf}\;(\conid{Just}\;\varid{v}_{2})\;\keyword{of}{}\<[E]%
\\
\>[3]{}\hsindent{4}{}\<[7]%
\>[7]{}\conid{Nothing}{}\<[16]%
\>[16]{}\to \conid{EmptySEM}{}\<[E]%
\\
\>[3]{}\hsindent{4}{}\<[7]%
\>[7]{}\conid{Just}\;\varid{v'}{}\<[16]%
\>[16]{}\to \conid{SingleSEM}\;\varid{k}_{2}\;\varid{v'}{}\<[E]%
\\
\>[B]{}\hsindent{3}{}\<[3]%
\>[3]{}\keyword{else}\;\keyword{case}\;\varid{tf}\;\conid{Nothing}\;\keyword{of}{}\<[E]%
\\
\>[3]{}\hsindent{4}{}\<[7]%
\>[7]{}\conid{Nothing}{}\<[16]%
\>[16]{}\to \conid{SingleSEM}\;\varid{k}_{2}\;\varid{v}_{2}{}\<[E]%
\\
\>[3]{}\hsindent{4}{}\<[7]%
\>[7]{}\conid{Just}\;\varid{v}_{1}{}\<[16]%
\>[16]{}\to \conid{MultiSEM}\;(\varid{insertTM}\;\varid{k}_{1}\;\varid{v}_{1}\;(\varid{insertTM}\;\varid{k}_{2}\;\varid{v}_{2}\;\varid{emptyTM})){}\<[E]%
\\
\>[B]{}\varid{atSEM}\;\varid{k}\;\varid{tf}\;(\conid{MultiSEM}\;\varid{tm})\mathrel{=}\conid{MultiSEM}\;(\varid{atTM}\;\varid{k}\;\varid{tf}\;\varid{tm}){}\<[E]%
\ColumnHook
\end{hscode}\resethooks
Adding a new item to a triemap can turn \ensuremath{\conid{EmptySEM}} into \ensuremath{\conid{SingleSEM}} and \ensuremath{\conid{SingleSEM}}
into \ensuremath{\conid{MultiSEM}}; and deleting an item from a \ensuremath{\conid{SingleSEM}} turns it back into \ensuremath{\conid{EmptySEM}}.
You might wonder whether we can shrink a \ensuremath{\conid{MultiSEM}} back to a \ensuremath{\conid{SingleSEM}} when it has
only one remaining element?
Yes we can, but it takes quite a bit of code, and it is far from clear
that it is worth doing so.

Finally, we need to re-define \ensuremath{\conid{ExprMap}} and \ensuremath{\conid{ListMap}} using \ensuremath{\conid{SEMap}}:
\begin{hscode}\SaveRestoreHook
\column{B}{@{}>{\hspre}l<{\hspost}@{}}%
\column{3}{@{}>{\hspre}l<{\hspost}@{}}%
\column{22}{@{}>{\hspre}l<{\hspost}@{}}%
\column{23}{@{}>{\hspre}l<{\hspost}@{}}%
\column{37}{@{}>{\hspre}l<{\hspost}@{}}%
\column{38}{@{}>{\hspre}l<{\hspost}@{}}%
\column{53}{@{}>{\hspre}l<{\hspost}@{}}%
\column{E}{@{}>{\hspre}l<{\hspost}@{}}%
\>[3]{}\keyword{type}\;\conid{ExprMap}{}\<[22]%
\>[22]{}\mathrel{=}\conid{SEMap}\;\conid{ExprMap'}{}\<[E]%
\\
\>[3]{}\keyword{data}\;\conid{ExprMap'}\;\varid{v}{}\<[22]%
\>[22]{}\mathrel{=}\conid{EM}\;\{\mskip1.5mu \varid{em\char95 var}{}\<[37]%
\>[37]{}\mathbin{::}\mathbin{...},\varid{em\char95 app}{}\<[53]%
\>[53]{}\mathbin{::}\conid{ExprMap}\;(\conid{ExprMap}\;\varid{v})\mskip1.5mu\}{}\<[E]%
\\[\blanklineskip]%
\>[3]{}\keyword{type}\;\conid{ListMap}{}\<[23]%
\>[23]{}\mathrel{=}\conid{SEMap}\;\conid{ListMap'}{}\<[E]%
\\
\>[3]{}\keyword{data}\;\conid{ListMap'}\;\varid{tm}\;\varid{v}{}\<[23]%
\>[23]{}\mathrel{=}\conid{LM}\;\{\mskip1.5mu \varid{lm\char95 nil}{}\<[38]%
\>[38]{}\mathbin{::}\mathbin{...},\varid{lm\char95 cons}\mathbin{::}\varid{tm}\;(\conid{ListMap}\;\varid{tm}\;\varid{v})\mskip1.5mu\}{}\<[E]%
\ColumnHook
\end{hscode}\resethooks
The auxiliary data types \ensuremath{\conid{ExprMap'}} and \ensuremath{\conid{ListMap'}} have only a single constructor, because
the empty and singleton cases are dealt with by \ensuremath{\conid{SEMap}}.  We reserve the original,
un-primed, names for the user-visible \ensuremath{\conid{ExprMap}} and \ensuremath{\conid{ListMap}} constructors.

\subsection{Generic programming}\label{sec:generic}

We have not described a triemap \emph{library}; rather we have described a \emph{design pattern}.
More precisely, given a new algebraic data type \ensuremath{\conid{X}}, we have described a systematic way
of defining a triemap, \ensuremath{\conid{XMap}}, keyed by values of type \ensuremath{\conid{X}}.
Such a triemap is represented by a record:
\begin{itemize}
\item Each \emph{constructor} \ensuremath{\conid{K}} of \ensuremath{\conid{X}} becomes a \emph{field} \ensuremath{\varid{x\char95 k}} in \ensuremath{\conid{XMap}}.
\item Each \emph{field} of a constructor \ensuremath{\conid{K}} becomes a \emph{nested triemap} in the type of the field \ensuremath{\varid{x\char95 k}}.
\item If \ensuremath{\conid{X}} is polymorphic then \ensuremath{\conid{XMap}} becomes a triemap transformer, like
  \ensuremath{\conid{ListMap}} above.
\end{itemize}
Actually writing out all this boilerplate code is tiresome, and it can of course be automated.
One way to do so would be to
use generic or polytypic programming, and Hinze describes precisely this \cite{hinze:generalized}.
Another approach would be to use Template Haskell.

We do not develop either of these approaches here, because our focus is only the
functionality and expressiveness of the triemaps.  However, everything we do is compatible
with an automated approach to generating boilerplate code.

\section{Keys with binders} \label{sec:binders}
\begin{figure}
\begin{hscode}\SaveRestoreHook
\column{B}{@{}>{\hspre}l<{\hspost}@{}}%
\column{3}{@{}>{\hspre}l<{\hspost}@{}}%
\column{E}{@{}>{\hspre}l<{\hspost}@{}}%
\>[B]{}\keyword{type}\;\conid{DBNum}\mathrel{=}\conid{Int}{}\<[E]%
\\
\>[B]{}\keyword{data}\;\conid{DBEnv}\mathrel{=}\conid{DBE}\;\{\mskip1.5mu \varid{dbe\char95 next}\mathbin{::}\conid{DBNum},\varid{dbe\char95 env}\mathbin{::}\conid{Map}\;\conid{Var}\;\conid{DBNum}\mskip1.5mu\}{}\<[E]%
\\[\blanklineskip]%
\>[B]{}\varid{emptyDBE}\mathbin{::}\conid{DBEnv}{}\<[E]%
\\
\>[B]{}\varid{emptyDBE}\mathrel{=}\conid{DBE}\;\{\mskip1.5mu \varid{dbe\char95 next}\mathrel{=}\mathrm{1},\varid{dbe\char95 env}\mathrel{=}\varid{\conid{Map}.empty}\mskip1.5mu\}{}\<[E]%
\\[\blanklineskip]%
\>[B]{}\varid{extendDBE}\mathbin{::}\conid{Var}\to \conid{DBEnv}\to \conid{DBEnv}{}\<[E]%
\\
\>[B]{}\varid{extendDBE}\;\varid{v}\;(\conid{DBE}\;\{\mskip1.5mu \varid{dbe\char95 next}\mathrel{=}\varid{n},\varid{dbe\char95 env}\mathrel{=}\varid{dbe}\mskip1.5mu\}){}\<[E]%
\\
\>[B]{}\hsindent{3}{}\<[3]%
\>[3]{}\mathrel{=}\conid{DBE}\;\{\mskip1.5mu \varid{dbe\char95 next}\mathrel{=}\varid{n}\mathbin{+}\mathrm{1},\varid{dbe\char95 env}\mathrel{=}\varid{\conid{Map}.insert}\;\varid{v}\;\varid{n}\;\varid{dbe}\mskip1.5mu\}{}\<[E]%
\\[\blanklineskip]%
\>[B]{}\varid{lookupDBE}\mathbin{::}\conid{Var}\to \conid{DBEnv}\to \conid{Maybe}\;\conid{DBNum}{}\<[E]%
\\
\>[B]{}\varid{lookupDBE}\;\varid{v}\;(\conid{DBE}\;\{\mskip1.5mu \varid{dbe\char95 env}\mathrel{=}\varid{dbe}\mskip1.5mu\})\mathrel{=}\varid{\conid{Map}.lookup}\;\varid{v}\;\varid{dbe}{}\<[E]%
\ColumnHook
\end{hscode}\resethooks
\caption{De Bruijn leveling}
\label{fig:debruijn}
\end{figure}

If our keys are expressions (in a compiler, say) they may contain binders,
and we want insert and lookup to be insensitive to $\alpha$-renaming.
That is the challenge we address next. Here is our data type \ensuremath{\conid{Expr}} from
\Cref{sec:alpha-renaming}, which brings back binding semantics through the \ensuremath{\conid{Lam}}
constructor:
\begin{hscode}\SaveRestoreHook
\column{B}{@{}>{\hspre}l<{\hspost}@{}}%
\column{E}{@{}>{\hspre}l<{\hspost}@{}}%
\>[B]{}\keyword{data}\;\conid{Expr}\mathrel{=}\conid{App}\;\conid{Expr}\;\conid{Expr}\mid \conid{Lam}\;\conid{Var}\;\conid{Expr}\mid \conid{Var}\;\conid{Var}{}\<[E]%
\ColumnHook
\end{hscode}\resethooks
The key idea is simple: we perform de Bruijn numbering on the fly,
renaming each binder to a natural number, from outside in.
So, when inserting or looking up a key $(\lambda x.\, \mathit{foo}~ (\lambda y.\, x+y))$ we
behave as if the key was $(\lambda.\, \mathit{foo} ~(\lambda. \bv{1} + \bv{2}))$, where
each $\bv{i}$ stands for an occurrence of the variable bound by the $i$'th
lambda, counting from the root of the expression. In effect, then, we behave as
if the data type was like this:
\begin{hscode}\SaveRestoreHook
\column{B}{@{}>{\hspre}l<{\hspost}@{}}%
\column{E}{@{}>{\hspre}l<{\hspost}@{}}%
\>[B]{}\keyword{data}\;\conid{Expr'}\mathrel{=}\conid{App}\;\conid{Expr}\;\conid{Expr}\mid \conid{Lam}\;\conid{Expr}\mid \conid{FVar}\;\conid{Var}\mid \conid{BVar}\;\conid{BoundKey}{}\<[E]%
\ColumnHook
\end{hscode}\resethooks
Notice (a) the \ensuremath{\conid{Lam}} node no longer has a binder and (b) there are
two sorts of \ensuremath{\conid{Var}} nodes, one for free variables and one for bound
variables, carrying a \ensuremath{\conid{BoundKey}} (see below). We will not actually
build a value of type \ensuremath{\conid{Expr'}} and look that up in a trie keyed by \ensuremath{\conid{Expr'}};
rather, we are going to \emph{behave as if we did}. Here is the code
(which uses \Cref{fig:debruijn}):
\begin{hscode}\SaveRestoreHook
\column{B}{@{}>{\hspre}l<{\hspost}@{}}%
\column{3}{@{}>{\hspre}l<{\hspost}@{}}%
\column{5}{@{}>{\hspre}l<{\hspost}@{}}%
\column{9}{@{}>{\hspre}c<{\hspost}@{}}%
\column{9E}{@{}l@{}}%
\column{12}{@{}>{\hspre}l<{\hspost}@{}}%
\column{14}{@{}>{\hspre}l<{\hspost}@{}}%
\column{16}{@{}>{\hspre}l<{\hspost}@{}}%
\column{21}{@{}>{\hspre}l<{\hspost}@{}}%
\column{26}{@{}>{\hspre}l<{\hspost}@{}}%
\column{42}{@{}>{\hspre}l<{\hspost}@{}}%
\column{E}{@{}>{\hspre}l<{\hspost}@{}}%
\>[B]{}\keyword{data}\;\conid{ModAlpha}\;\varid{a}\mathrel{=}\conid{A}\;\conid{DBEnv}\;\varid{a}{}\<[E]%
\\
\>[B]{}\keyword{type}\;\conid{AlphaExpr}\mathrel{=}\conid{ModAlpha}\;\conid{Expr}{}\<[E]%
\\
\>[B]{}\keyword{instance}\;\conid{Eq}\;\conid{AlphaExpr}\;\keyword{where}\mathbin{...}{}\<[E]%
\\[\blanklineskip]%
\>[B]{}\keyword{type}\;\conid{BoundKey}{}\<[16]%
\>[16]{}\mathrel{=}\conid{DBNum}{}\<[E]%
\\
\>[B]{}\keyword{type}\;\conid{ExprMap}\mathrel{=}\conid{SEMap}\;\conid{ExprMap'}{}\<[E]%
\\
\>[B]{}\keyword{data}\;\conid{ExprMap'}\;\varid{v}{}\<[E]%
\\
\>[B]{}\hsindent{3}{}\<[3]%
\>[3]{}\mathrel{=}\conid{EM}\;{}\<[9]%
\>[9]{}\{\mskip1.5mu {}\<[9E]%
\>[12]{}\varid{em\char95 fvar}{}\<[21]%
\>[21]{}\mathbin{::}\conid{Map}\;\conid{Var}\;\varid{v}{}\<[42]%
\>[42]{}\mbox{\onelinecomment  Free vars}{}\<[E]%
\\
\>[9]{},{}\<[9E]%
\>[12]{}\varid{em\char95 bvar}{}\<[21]%
\>[21]{}\mathbin{::}\conid{Map}\;\conid{BoundKey}\;\varid{v}{}\<[42]%
\>[42]{}\mbox{\onelinecomment  Lambda-bound vars}{}\<[E]%
\\
\>[9]{},{}\<[9E]%
\>[12]{}\varid{em\char95 app}{}\<[21]%
\>[21]{}\mathbin{::}\conid{ExprMap}\;(\conid{ExprMap}\;\varid{v}){}\<[E]%
\\
\>[9]{},{}\<[9E]%
\>[12]{}\varid{em\char95 lam}{}\<[21]%
\>[21]{}\mathbin{::}\conid{ExprMap}\;\varid{v}\mskip1.5mu\}{}\<[E]%
\\[\blanklineskip]%
\>[B]{}\keyword{instance}\;\conid{TrieMap}\;\conid{ExprMap'}\;\keyword{where}{}\<[E]%
\\
\>[B]{}\hsindent{3}{}\<[3]%
\>[3]{}\keyword{type}\;\conid{Key}\;\conid{ExprMap'}\mathrel{=}\conid{AlphaExpr}{}\<[E]%
\\
\>[B]{}\hsindent{3}{}\<[3]%
\>[3]{}\varid{lkTM}\mathrel{=}\varid{lkEM}{}\<[E]%
\\
\>[B]{}\hsindent{3}{}\<[3]%
\>[3]{}\mathbin{...}{}\<[E]%
\\[\blanklineskip]%
\>[B]{}\varid{lkEM}\mathbin{::}\conid{AlphaExpr}\to \conid{ExprMap'}\;\varid{v}\to \conid{Maybe}\;\varid{v}{}\<[E]%
\\
\>[B]{}\varid{lkEM}\;(\conid{A}\;\varid{bve}\;\varid{e})\mathrel{=}\keyword{case}\;\varid{e}\;\keyword{of}{}\<[E]%
\\
\>[B]{}\hsindent{3}{}\<[3]%
\>[3]{}\conid{Var}\;\varid{v}\to \keyword{case}\;\varid{lookupDBE}\;\varid{v}\;\varid{bve}\;\keyword{of}{}\<[E]%
\\
\>[3]{}\hsindent{2}{}\<[5]%
\>[5]{}\conid{Nothing}{}\<[14]%
\>[14]{}\to \varid{em\char95 fvar}{}\<[26]%
\>[26]{}\mathrel{{>}\hspace{-0.4em}{>}\hspace{-0.4em}{>}}\varid{\conid{Map}.lookup}\;\varid{v}{}\<[E]%
\\
\>[3]{}\hsindent{2}{}\<[5]%
\>[5]{}\conid{Just}\;\varid{bv}{}\<[14]%
\>[14]{}\to \varid{em\char95 bvar}{}\<[26]%
\>[26]{}\mathrel{{>}\hspace{-0.4em}{>}\hspace{-0.4em}{>}}\varid{\conid{Map}.lookup}\;\varid{bv}{}\<[E]%
\\
\>[B]{}\hsindent{3}{}\<[3]%
\>[3]{}\conid{App}\;\varid{e}_{1}\;\varid{e}_{2}{}\<[14]%
\>[14]{}\to \varid{em\char95 app}{}\<[26]%
\>[26]{}\mathrel{{>}\hspace{-0.4em}{>}\hspace{-0.4em}{>}}\varid{lkTM}\;(\conid{A}\;\varid{bve}\;\varid{e}_{1})\mathrel{{>}\hspace{-0.32em}{=}\hspace{-0.32em}{>}}\varid{lkTM}\;(\conid{A}\;\varid{bve}\;\varid{e}_{2}){}\<[E]%
\\
\>[B]{}\hsindent{3}{}\<[3]%
\>[3]{}\conid{Lam}\;\varid{v}\;\varid{e}{}\<[14]%
\>[14]{}\to \varid{em\char95 lam}{}\<[26]%
\>[26]{}\mathrel{{>}\hspace{-0.4em}{>}\hspace{-0.4em}{>}}\varid{lkTM}\;(\conid{A}\;(\varid{extendDBE}\;\varid{v}\;\varid{bve})\;\varid{e}){}\<[E]%
\\[\blanklineskip]%
\>[B]{}\varid{lookupClosedExpr}\mathbin{::}\conid{Expr}\to \conid{ExprMap}\;\varid{v}\to \conid{Maybe}\;\varid{v}{}\<[E]%
\\
\>[B]{}\varid{lookupClosedExpr}\;\varid{e}\mathrel{=}\varid{lkEM}\;(\conid{A}\;\varid{emptyDBE}\;\varid{e}){}\<[E]%
\ColumnHook
\end{hscode}\resethooks
We maintain a \ensuremath{\conid{DBEnv}} (cf.~\cref{fig:debruijn}) that
maps each lambda-bound variable to its de Bruijn level \cite{debruijn},%
\footnote{
  The de Bruijn \emph{index} of the occurrence of a variable $v$ counts the number
  of lambdas between the occurrence of $v$ and its binding site.  The de Bruijn \emph{level}
  of $v$ counts the number of lambdas between the root of the expression and $v$'s binding site.
  It is convenient for us to use \emph{levels}.}
its \ensuremath{\conid{BoundKey}}.
The expression we look up --- the first argument of \ensuremath{\varid{lkEM}} --- becomes an
\ensuremath{\conid{AlphaExpr}}, which is a pair of a \ensuremath{\conid{DBEnv}} and an \ensuremath{\conid{Expr}}.
At a \ensuremath{\conid{Lam}}
node we extend the \ensuremath{\conid{DBEnv}}. At a \ensuremath{\conid{Var}} node we
look up the variable in the \ensuremath{\conid{DBEnv}} to decide whether it is
lambda-bound or free, and behave appropriately.%
\footnote{The implementation in the supplement~\cite{triemaps-extended,triemaps-github} uses
more efficient \ensuremath{\conid{IntMap}}s for mapping \ensuremath{\conid{BoundKey}}. \ensuremath{\conid{IntMap}} is itself a trie
data structure, so it could have made for a nice \enquote{Tries all the way
down} argument. But we found it distracting to present here, hence regular
ordered \ensuremath{\conid{Map}}.}

The construction of \Cref{sec:singleton}, to handle empty and singleton maps,
applies without difficulty to this generalised map. To use it
we must define an instance \ensuremath{\conid{Eq}\;\conid{AlphaExpr}} to satisfy the \ensuremath{\conid{Eq}} super class constraint
on the trie key, so that we can instantiate \ensuremath{\conid{TrieMap}\;\conid{ExprMap'}}.
That \ensuremath{\conid{Eq}\;\conid{AlphaExpr}} instance simply equates two
$\alpha$-equivalent expressions in the standard way.
The code for \ensuremath{\varid{atEM}} and \ensuremath{\varid{foldrEM}} holds no new surprises either.

And that is really all there is to it: it is remarkably easy to extend the basic
trie idea to be insensitive to $\alpha$-conversion and even mix in trie
transformers such as \ensuremath{\conid{SEMap}} at no cost other than writing two instance
declarations.

\section{Tries that match} \label{sec:matching}

A key advantage of tries over hash-maps and balanced trees is
that we can support \emph{matching} (\Cref{sec:matching-intro}).

\subsection{What ``matching'' means} \label{sec:matching-spec}

Semantically, a matching trie can be thought of as a set of (pattern, value) pairs.
What is a pattern? It is a pair $(\mathit{vs},p)$ where
\begin{itemize}
\item $\mathit{vs}$ is a set of \emph{pattern variables}, such as $[a,b,c]$.
\item $p$ is a \emph{pattern expression}, such as $f\, a\, (g\, b\, c)$.
\end{itemize}
A pattern may of course contain free variables (not bound by the pattern), such as $f$ and $g$
in the above example, which are regarded as constants by the algorithm.
A pattern $(\mathit{vs}, p)$ \emph{matches} a target expression $e$ iff there is a unique substitution
$S$ whose domain is $\mathit{vs}$, such that $S(p) = e$.

We allow the same variable to occur more than once in the pattern.
For example, the pattern $([x], f~ x~ x)$
should match targets like $(f~ 1~ 1)$ or $(f ~(g~ v)~ (g ~v))$,
but not $(f~ 1~ (g~ v))$.  This ability is important if we are to use matching tries
to implement class or type family lookup in GHC.

%

\subsection{Matching expressions} \label{sec:matching-alphaexpr}

Our matching trie is founded on a function \ensuremath{\varid{matchE}} that matches a target \ensuremath{\conid{AlphaExpr}} against
a \emph{single} pattern:
\begin{hscode}\SaveRestoreHook
\column{B}{@{}>{\hspre}l<{\hspost}@{}}%
\column{3}{@{}>{\hspre}l<{\hspost}@{}}%
\column{E}{@{}>{\hspre}l<{\hspost}@{}}%
\>[3]{}\varid{matchE}\mathbin{::}\conid{PatExpr}\to \conid{AlphaExpr}\to \conid{MatchME}\;(){}\<[E]%
\ColumnHook
\end{hscode}\resethooks
Here \ensuremath{\varid{matchE}} takes a \emph{pattern} of type \ensuremath{\conid{PatExpr}}, and a \emph{target} key of type \ensuremath{\conid{AlphaExpr}}.
It sees whether the pattern matches the target, and returns a value of type \ensuremath{\conid{MatchME}\;()}
to say whether or not it matched, and if so with what substitution of the pattern variables.
But what exactly are \ensuremath{\conid{PatExpr}} and \ensuremath{\conid{MatchME}}?  We look at each in turn.

\subsubsection{Patterns} \label{sec:patterns}

A pattern \ensuremath{\conid{PatExpr}} over \ensuremath{\conid{AlphaExpr}} can be defined like this:
\begin{hscode}\SaveRestoreHook
\column{B}{@{}>{\hspre}l<{\hspost}@{}}%
\column{14}{@{}>{\hspre}l<{\hspost}@{}}%
\column{E}{@{}>{\hspre}l<{\hspost}@{}}%
\>[B]{}\keyword{data}\;\conid{PatExpr}\mathrel{=}\conid{P}\;\conid{PatKeys}\;\conid{AlphaExpr}{}\<[E]%
\\
\>[B]{}\keyword{type}\;\conid{PatKeys}\mathrel{=}\conid{Map}\;\conid{PatVar}\;\conid{PatKey}{}\<[E]%
\\
\>[B]{}\keyword{type}\;\conid{PatVar}{}\<[14]%
\>[14]{}\mathrel{=}\conid{Var}{}\<[E]%
\\
\>[B]{}\keyword{type}\;\conid{PatKey}{}\<[14]%
\>[14]{}\mathrel{=}\conid{DBNum}{}\<[E]%
\ColumnHook
\end{hscode}\resethooks
A pattern \ensuremath{\conid{PatExpr}} is a pair of an \ensuremath{\conid{AlphaExpr}} and a \ensuremath{\conid{PatKeys}} that maps
each of the quantified pattern variables to a canonical de Bruijn \ensuremath{\conid{PatKey}}.
Just as in \Cref{sec:binders}, \ensuremath{\conid{PatKeys}} make the pattern insensitive to
the particular names, and order of quantification, of the pattern variables.
We canonicalise the quantified pattern variables before starting a lookup,
numbering pattern variables in the order they appear in the expression in a
left-to-right scan.
For example
$$
\begin{array}{r@{\hspace{5mm}}l}
\text{Original pattern} & \text{Canonical \ensuremath{\conid{PatExpr}}} \\
\ensuremath{([\mskip1.5mu \varid{a},\varid{b}\mskip1.5mu],\varid{f}\;\varid{a}\;\varid{b}\;\varid{a})}  &  \ensuremath{\conid{P}\;[\mskip1.5mu \varid{a}\mapsto\mathrm{1},\varid{b}\mapsto\mathrm{2}\mskip1.5mu]\;(\varid{f}\;\varid{a}\;\varid{b}\;\varid{a})}\\
\ensuremath{([\mskip1.5mu \varid{x},\varid{g}\mskip1.5mu],\varid{f}\;(\varid{g}\;\varid{x})}  &  \ensuremath{\conid{P}\;[\mskip1.5mu \varid{x}\mapsto\mathrm{2},\varid{g}\mapsto\mathrm{1}\mskip1.5mu]\;(\varid{f}\;(\varid{g}\;\varid{x}))}
\end{array}
$$

\subsubsection{The matching monad} \label{sec:matching-monad}

Since matching must accommodate failure, it turns out to be convenient
for the result of matching, \ensuremath{\conid{MatchME}}, to be a monad.\footnote{``\ensuremath{\conid{MatchME}}'' connotes
  a monadic matcher \ensuremath{\conid{MatchM}} on expressions, hence the ``\ensuremath{\conid{E}}''.
}
Here is one possible implementation, defining \ensuremath{\conid{MatchME}} as a state transfomer over lists:
\begin{hscode}\SaveRestoreHook
\column{B}{@{}>{\hspre}l<{\hspost}@{}}%
\column{E}{@{}>{\hspre}l<{\hspost}@{}}%
\>[B]{}\keyword{type}\;\conid{MatchME}\;\varid{v}\mathrel{=}\conid{StateT}\;\conid{SubstE}\;[\mskip1.5mu \mskip1.5mu]\;\varid{v}{}\<[E]%
\\
\>[B]{}\keyword{type}\;\conid{SubstE}\mathrel{=}\conid{Map}\;\conid{PatKey}\;\conid{Expr}{}\<[E]%
\ColumnHook
\end{hscode}\resethooks
More
concretely, \ensuremath{\conid{MatchME}\;\varid{v}} is a function taking a substitution (of type \ensuremath{\conid{SubstE}})
for pattern variables, and yielding a possibly-empty list of values (of type \ensuremath{\varid{v}}),
each paired with an extended \ensuremath{\conid{SubstE}}.  Notice that:
\begin{itemize}
\item the \emph{domain} of the substitution \ensuremath{\conid{SubstE}} is the canonical pattern keys \ensuremath{\conid{PatKey}}, not \ensuremath{\conid{PatVar}}.
\item the \emph{range} of the substitution is \ensuremath{\conid{Expr}}, not \ensuremath{\conid{AlphaExpr}}, because the pattern
variables cannot mention lambda-bound variables within the target expression.
\end{itemize}
\noindent
Type \ensuremath{\conid{MatchME}} comes with some auxiliary functions that we will need later:
\begin{hscode}\SaveRestoreHook
\column{B}{@{}>{\hspre}l<{\hspost}@{}}%
\column{14}{@{}>{\hspre}l<{\hspost}@{}}%
\column{E}{@{}>{\hspre}l<{\hspost}@{}}%
\>[B]{}\varid{runMatchExpr}\mathbin{::}\conid{MatchME}\;\varid{v}\to [\mskip1.5mu (\conid{SubstE},\varid{v})\mskip1.5mu]{}\<[E]%
\\
\>[B]{}\varid{liftMaybe}{}\<[14]%
\>[14]{}\mathbin{::}\conid{Maybe}\;\varid{v}\to \conid{MatchME}\;\varid{v}{}\<[E]%
\\
\>[B]{}\varid{refineMatch}{}\<[14]%
\>[14]{}\mathbin{::}(\conid{SubstE}\to \conid{Maybe}\;\conid{SubstE})\to \conid{MatchME}\;(){}\<[E]%
\ColumnHook
\end{hscode}\resethooks
Their semantics should be apparent from their types.  For example, \ensuremath{\varid{runMatchExpr}}
runs a \ensuremath{\conid{MatchME}} computation, starting with an empty \ensuremath{\conid{SubstE}}, and
returning a list of all the successful \ensuremath{(\conid{SubstE},\varid{v})} matches.  The function \ensuremath{\varid{refineMatch}\;\varid{f}}
extends the current substitution by applying \ensuremath{\varid{f}} to it; if the result is \ensuremath{\conid{Nothing}} the
match fails; otherwise it turns a single match with the new substitution.


\subsubsection{Matching summary}

The implementation of \ensuremath{\varid{matchE}} is entirely straightforward, using
simultaneous recursive descent over the pattern and target.
The code is given in the supplement~\cite{triemaps-extended,triemaps-github}.

The key point is this: nothing in this section is concerned with
tries.  Here we are simply concerned with the mechanics of matching,
and its underlying monad.  There is ample room for flexibility. For
example, if the key terms had two kinds of variables (say type
variables and term variables) we could easily define \ensuremath{\conid{MatchME}} to carry
two substitutions; or \ensuremath{\conid{MatchME}} could return just the first result
rather than a list of all of them; and so on.

\subsection{Matching tries for \ensuremath{\conid{AlphaExpr}}}

Next, we show how to implement a matching triemap for our running
example, \ensuremath{\conid{AlphaExpr}}.
The data type follows the pattern we developed for \ensuremath{\conid{ExprMap}}
(\Cref{sec:Expr,sec:binders}):\footnote{
To avoid clutter we have, for now, added empty and singleton maps
directly to \ensuremath{\conid{MExprMap}}, via \ensuremath{\conid{EmptyMEM}} and \ensuremath{\conid{SingleMEM}}, much as in \Cref{sec:empty}.
In \Cref{sec:matching-trie-class} we show how to use the type classes
to share the empty/singleton case as we did in \Cref{sec:singleton}.
}
\begin{hscode}\SaveRestoreHook
\column{B}{@{}>{\hspre}l<{\hspost}@{}}%
\column{3}{@{}>{\hspre}l<{\hspost}@{}}%
\column{15}{@{}>{\hspre}c<{\hspost}@{}}%
\column{15E}{@{}l@{}}%
\column{18}{@{}>{\hspre}l<{\hspost}@{}}%
\column{27}{@{}>{\hspre}l<{\hspost}@{}}%
\column{47}{@{}>{\hspre}l<{\hspost}@{}}%
\column{E}{@{}>{\hspre}l<{\hspost}@{}}%
\>[B]{}\keyword{data}\;\conid{MExprMap}\;\varid{v}{}\<[E]%
\\
\>[B]{}\hsindent{3}{}\<[3]%
\>[3]{}\mathrel{=}\conid{EmptyMEM}{}\<[E]%
\\
\>[B]{}\hsindent{3}{}\<[3]%
\>[3]{}\mid \conid{SingleMEM}\;\conid{PatExpr}\;\varid{v}{}\<[E]%
\\
\>[B]{}\hsindent{3}{}\<[3]%
\>[3]{}\mid \conid{MultiMEM}\;{}\<[15]%
\>[15]{}\{\mskip1.5mu {}\<[15E]%
\>[18]{}\varid{mm\char95 fvar}{}\<[27]%
\>[27]{}\mathbin{::}\conid{Map}\;\conid{Var}\;\varid{v}{}\<[47]%
\>[47]{}\mbox{\onelinecomment  Free var}{}\<[E]%
\\
\>[15]{},{}\<[15E]%
\>[18]{}\varid{mm\char95 bvar}{}\<[27]%
\>[27]{}\mathbin{::}\conid{Map}\;\conid{BoundKey}\;\varid{v}{}\<[47]%
\>[47]{}\mbox{\onelinecomment  Bound var}{}\<[E]%
\\
\>[15]{},{}\<[15E]%
\>[18]{}\varid{mm\char95 pvar}{}\<[27]%
\>[27]{}\mathbin{::}\conid{Map}\;\conid{PatKey}\;\varid{v}{}\<[47]%
\>[47]{}\mbox{\onelinecomment  Pattern var}{}\<[E]%
\\
\>[15]{},{}\<[15E]%
\>[18]{}\varid{mm\char95 app}{}\<[27]%
\>[27]{}\mathbin{::}\conid{MExprMap}\;(\conid{MExprMap}\;\varid{v}){}\<[E]%
\\
\>[15]{},{}\<[15E]%
\>[18]{}\varid{mm\char95 lam}{}\<[27]%
\>[27]{}\mathbin{::}\conid{MExprMap}\;\varid{v}\mskip1.5mu\}{}\<[E]%
\ColumnHook
\end{hscode}\resethooks
The main new feature is that we add an extra field \ensuremath{\varid{mm\char95 pvar}} to \ensuremath{\conid{MExprMap}},
for occurrences of a pattern variable.  You can see how this field is used
in the lookup code:
\begin{hscode}\SaveRestoreHook
\column{B}{@{}>{\hspre}l<{\hspost}@{}}%
\column{3}{@{}>{\hspre}l<{\hspost}@{}}%
\column{5}{@{}>{\hspre}l<{\hspost}@{}}%
\column{7}{@{}>{\hspre}l<{\hspost}@{}}%
\column{9}{@{}>{\hspre}l<{\hspost}@{}}%
\column{18}{@{}>{\hspre}l<{\hspost}@{}}%
\column{30}{@{}>{\hspre}c<{\hspost}@{}}%
\column{30E}{@{}l@{}}%
\column{35}{@{}>{\hspre}l<{\hspost}@{}}%
\column{E}{@{}>{\hspre}l<{\hspost}@{}}%
\>[B]{}\varid{lookupPatMM}\mathbin{::}\keyword{$\forall$}\!\! \hsforall \;\varid{v}\hsdot{\circ }{.\,}\conid{AlphaExpr}\to \conid{MExprMap}\;\varid{v}\to \conid{MatchME}\;\varid{v}{}\<[E]%
\\
\>[B]{}\varid{lookupPatMM}\;\varid{ae}\mathord{@}(\conid{A}\;\varid{bve}\;\varid{e})\;\conid{EmptyMEM}{}\<[E]%
\\
\>[B]{}\hsindent{3}{}\<[3]%
\>[3]{}\mathrel{=}\varid{mzero}{}\<[E]%
\\
\>[B]{}\varid{lookupPatMM}\;\varid{ae}\mathord{@}(\conid{A}\;\varid{bve}\;\varid{e})\;(\conid{SingleSEM}\;\varid{pat}\;\varid{val}){}\<[E]%
\\
\>[B]{}\hsindent{3}{}\<[3]%
\>[3]{}\mathrel{=}\varid{matchE}\;\varid{pat}\;\varid{ae}\mathrel{{>}\hspace{-0.4em}{>}}\varid{pure}\;\varid{val}{}\<[E]%
\\
\>[B]{}\varid{lookupPatMM}\;\varid{ae}\mathord{@}(\conid{A}\;\varid{bve}\;\varid{e})\;(\conid{MultiMEM}\;\{\mskip1.5mu \mathinner{\ldotp\ldotp}\mskip1.5mu\}){}\<[E]%
\\
\>[B]{}\hsindent{3}{}\<[3]%
\>[3]{}\mathrel{=}\varid{rigid}\mathbin{`\varid{mplus}`}\varid{flexi}{}\<[E]%
\\
\>[B]{}\hsindent{3}{}\<[3]%
\>[3]{}\keyword{where}{}\<[E]%
\\
\>[3]{}\hsindent{2}{}\<[5]%
\>[5]{}\varid{rigid}\mathrel{=}\keyword{case}\;\varid{e}\;\keyword{of}{}\<[E]%
\\
\>[5]{}\hsindent{2}{}\<[7]%
\>[7]{}\conid{Var}\;\varid{x}{}\<[18]%
\>[18]{}\to \keyword{case}\;\varid{lookupDBE}\;\varid{x}\;\varid{bve}\;\keyword{of}{}\<[E]%
\\
\>[7]{}\hsindent{2}{}\<[9]%
\>[9]{}\conid{Just}\;\varid{bv}{}\<[18]%
\>[18]{}\to \varid{mm\char95 bvar}{}\<[30]%
\>[30]{}\triangleright{}\<[30E]%
\>[35]{}\varid{liftMaybe}\hsdot{\circ }{.\,}\varid{\conid{Map}.lookup}\;\varid{bv}{}\<[E]%
\\
\>[7]{}\hsindent{2}{}\<[9]%
\>[9]{}\conid{Nothing}{}\<[18]%
\>[18]{}\to \varid{mm\char95 fvar}{}\<[30]%
\>[30]{}\triangleright{}\<[30E]%
\>[35]{}\varid{liftMaybe}\hsdot{\circ }{.\,}\varid{\conid{Map}.lookup}\;\varid{x}{}\<[E]%
\\
\>[5]{}\hsindent{2}{}\<[7]%
\>[7]{}\conid{App}\;\varid{e}_{1}\;\varid{e}_{2}{}\<[18]%
\>[18]{}\to \varid{mm\char95 app}{}\<[30]%
\>[30]{}\triangleright{}\<[30E]%
\>[35]{}\varid{lkMTM}\;(\conid{A}\;\varid{bve}\;\varid{e}_{1}){}\<[E]%
\\
\>[30]{}\mathrel{{>}\hspace{-0.32em}{=}\hspace{-0.32em}{>}}{}\<[30E]%
\>[35]{}\varid{lkMTM}\;(\conid{A}\;\varid{bve}\;\varid{e}_{2}){}\<[E]%
\\
\>[5]{}\hsindent{2}{}\<[7]%
\>[7]{}\conid{Lam}\;\varid{x}\;\varid{e}{}\<[18]%
\>[18]{}\to \varid{mm\char95 lam}{}\<[30]%
\>[30]{}\triangleright{}\<[30E]%
\>[35]{}\varid{lkMTM}\;(\conid{A}\;(\varid{extendDBE}\;\varid{x}\;\varid{bve})\;\varid{e}){}\<[E]%
\\[\blanklineskip]%
\>[3]{}\hsindent{2}{}\<[5]%
\>[5]{}\varid{flexi}\mathrel{=}\varid{mm\char95 pvar}\triangleright\varid{\conid{IntMap}.toList}\triangleright\varid{map}\;\varid{match\char95 one}\triangleright\varid{msum}{}\<[E]%
\\[\blanklineskip]%
\>[3]{}\hsindent{2}{}\<[5]%
\>[5]{}\varid{match\char95 one}\mathbin{::}(\conid{PatVar},\varid{v})\to \conid{MatchME}\;\varid{v}{}\<[E]%
\\
\>[3]{}\hsindent{2}{}\<[5]%
\>[5]{}\varid{match\char95 one}\;(\varid{pv},\varid{v})\mathrel{=}\varid{matchPatVarE}\;\varid{pv}\;\varid{ae}\mathrel{{>}\hspace{-0.4em}{>}}\varid{pure}\;\varid{v}{}\<[E]%
\ColumnHook
\end{hscode}\resethooks
Matching lookup on a trie matches the target
expression against \emph{all patterns the trie represents}.
The \ensuremath{\varid{rigid}} case is no different from exact lookup; compare the
code for \ensuremath{\varid{lkEM}} in \Cref{sec:binders}.  The only difference is that we need
\ensuremath{\varid{liftMaybe}} (from \Cref{sec:matching-monad}) to
take the \ensuremath{\conid{Maybe}} returned by \ensuremath{\varid{\conid{Map}.lookup}} and lift it into the \ensuremath{\conid{MatchME}} monad.

The \ensuremath{\varid{flexi}} case handles the triemap entries whose pattern is simply one of
the quantified pattern variables; these entries are stored in the new \ensuremath{\varid{mm\char95 pvar}} field.
We enumerate these entries with \ensuremath{\varid{toList}}, to get a list of \ensuremath{(\conid{PatVar},\varid{v})} pairs,
match each such pair against the target with \ensuremath{\varid{match\char95 one}}, and finally accumulate
all the results with \ensuremath{\varid{msum}}.  In turn \ensuremath{\varid{match\char95 one}} uses \ensuremath{\varid{matchParVarE}} to match
the pattern variable with the target and, if successful, returns the corresponding
value \ensuremath{\varid{v}}.

The \ensuremath{\varid{matchPatVarE}} function does the heavy lifting, using some
simple auxiliary functions whose types are given below:
\begin{hscode}\SaveRestoreHook
\column{B}{@{}>{\hspre}l<{\hspost}@{}}%
\column{3}{@{}>{\hspre}l<{\hspost}@{}}%
\column{5}{@{}>{\hspre}l<{\hspost}@{}}%
\column{7}{@{}>{\hspre}c<{\hspost}@{}}%
\column{7E}{@{}l@{}}%
\column{10}{@{}>{\hspre}l<{\hspost}@{}}%
\column{13}{@{}>{\hspre}l<{\hspost}@{}}%
\column{15}{@{}>{\hspre}l<{\hspost}@{}}%
\column{28}{@{}>{\hspre}l<{\hspost}@{}}%
\column{E}{@{}>{\hspre}l<{\hspost}@{}}%
\>[B]{}\varid{matchPatVarE}\mathbin{::}\conid{PatKey}\to \conid{AlphaExpr}\to \conid{MatchME}\;(){}\<[E]%
\\
\>[B]{}\varid{matchPatVarE}\;\varid{pv}\;(\conid{A}\;\varid{bve}\;\varid{e})\mathrel{=}\varid{refineMatch}\mathbin{\$}\lambda \varid{ms}\to {}\<[E]%
\\
\>[B]{}\hsindent{3}{}\<[3]%
\>[3]{}\keyword{case}\;\varid{\conid{Map}.lookup}\;\varid{pv}\;\varid{ms}\;\keyword{of}{}\<[E]%
\\
\>[3]{}\hsindent{2}{}\<[5]%
\>[5]{}\conid{Nothing}{}\<[15]%
\>[15]{}\mbox{\onelinecomment  pv is not bound}{}\<[E]%
\\
\>[5]{}\hsindent{2}{}\<[7]%
\>[7]{}\mid {}\<[7E]%
\>[10]{}\varid{noCaptured}\;\varid{bve}\;\varid{e}{}\<[28]%
\>[28]{}\to \conid{Just}\;(\varid{\conid{Map}.insert}\;\varid{pv}\;\varid{e}\;\varid{ms}){}\<[E]%
\\
\>[5]{}\hsindent{2}{}\<[7]%
\>[7]{}\mid {}\<[7E]%
\>[10]{}\varid{otherwise}{}\<[28]%
\>[28]{}\to \conid{Nothing}{}\<[E]%
\\
\>[3]{}\hsindent{2}{}\<[5]%
\>[5]{}\conid{Just}\;\varid{sol}{}\<[15]%
\>[15]{}\mbox{\onelinecomment  pv is already bound}{}\<[E]%
\\
\>[5]{}\hsindent{2}{}\<[7]%
\>[7]{}\mid {}\<[7E]%
\>[10]{}\varid{noCaptured}\;\varid{bve}\;\varid{e}{}\<[E]%
\\
\>[5]{}\hsindent{2}{}\<[7]%
\>[7]{},{}\<[7E]%
\>[10]{}\varid{eqExpr}\;\varid{e}\;\varid{sol}{}\<[28]%
\>[28]{}\to \conid{Just}\;\varid{ms}{}\<[E]%
\\
\>[5]{}\hsindent{2}{}\<[7]%
\>[7]{}\mid {}\<[7E]%
\>[10]{}\varid{otherwise}{}\<[28]%
\>[28]{}\to \conid{Nothing}{}\<[E]%
\\[\blanklineskip]%
\>[B]{}\varid{eqExpr}{}\<[13]%
\>[13]{}\mathbin{::}\conid{Expr}\to \conid{Expr}\to \conid{Bool}{}\<[E]%
\\
\>[B]{}\varid{noCaptured}{}\<[13]%
\>[13]{}\mathbin{::}\conid{DBEnv}\to \conid{Expr}\to \conid{Bool}{}\<[E]%
\ColumnHook
\end{hscode}\resethooks
To match a pattern variable \ensuremath{\varid{pv}} against an expression \ensuremath{(\conid{A}\;\varid{bve}\;\varid{e})},
we first look up \ensuremath{\varid{pv}} in the current substitution (obtained from the
\ensuremath{\conid{MatchME}} monad).  If \ensuremath{\varid{pv}} is not bound we simply extend the
substitution.

But wait!  Consider matching the pattern \ensuremath{([\mskip1.5mu \varid{p}\mskip1.5mu],\lambda \varid{x}\to \varid{p})}
against the target \ensuremath{(\lambda \varid{y}\to \mathrm{3})}.  That's fine: we should succeed, binding
\ensuremath{\varid{p}} to \ensuremath{\mathrm{3}}.  But suppose we match that same pattern against target \ensuremath{(\lambda \varid{y}\to \varid{y})}.
It would be nonsense to ``succeed'' binding \ensuremath{\varid{p}} to \ensuremath{\varid{y}}, because \ensuremath{\varid{y}} is
locally bound within the target.  Hence the \ensuremath{\varid{noCaptured}} test, which
returns \ensuremath{\conid{True}} iff the expression does not mention any of the locally-bound
variables.

If \ensuremath{\varid{pv}} is already bound in the substitution, we have a repeated pattern
variable (see \Cref{sec:matching-spec}), and we must check that
the target expression is equal (using \ensuremath{\varid{eqExpr}}) to the one already bound to \ensuremath{\varid{pv}}.
Once again, however, we must check that the target does not contain any locally-bound
variables, hence the \ensuremath{\varid{noCaptured}} check.

It turns out that \ensuremath{\varid{lookupPatMM}}, described above, is the trickiest case.
The code for \ensuremath{\varid{alterPatMM}}, and
the other operations of the class, is very straightforward, and is given
in the supplement~\cite{triemaps-extended,triemaps-github}.

\subsection{The external API} \label{sec:match-api}

The matching tries we have described so far use canonical pattern keys,
a matching monad, and other machinery that should be hidden from the client.
We seek an external API more like this:
\begin{hscode}\SaveRestoreHook
\column{B}{@{}>{\hspre}l<{\hspost}@{}}%
\column{11}{@{}>{\hspre}l<{\hspost}@{}}%
\column{E}{@{}>{\hspre}l<{\hspost}@{}}%
\>[B]{}\keyword{type}\;\conid{PatMap}\mathbin{::}\conid{Type}\to \conid{Type}{}\<[E]%
\\
\>[B]{}\varid{alterPM}{}\<[11]%
\>[11]{}\mathbin{::}([\mskip1.5mu \conid{Var}\mskip1.5mu],\conid{Expr})\to \conid{TF}\;\varid{v}\to \conid{PatMap}\;\varid{v}\to \conid{PatMap}\;\varid{v}{}\<[E]%
\\
\>[B]{}\varid{lookupPM}{}\<[11]%
\>[11]{}\mathbin{::}\conid{Expr}\to \conid{PatMap}\;\varid{v}\to [\mskip1.5mu (\conid{PatSubst},\varid{v})\mskip1.5mu]{}\<[E]%
\\
\>[B]{}\keyword{type}\;\conid{PatSubst}\mathrel{=}[\mskip1.5mu (\conid{Var},\conid{Expr})\mskip1.5mu]{}\<[E]%
\ColumnHook
\end{hscode}\resethooks
When altering a \ensuremath{\conid{PatMap}}, we supply a client-side pattern, which is
just a pair \ensuremath{([\mskip1.5mu \conid{Var}\mskip1.5mu],\conid{Expr})} of the quantified pattern variables and the pattern.
When looking up in a \ensuremath{\conid{PatMap}}, we supply a target expression, and get back
a list of matches, each of which is a pair of the value and the substitution
for those original pattern variables that made the pattern equal to the target.

So \ensuremath{\varid{alterPM}} must canonicalise the client-side pattern variables
before altering the trie; that is easy enough.
But how can \ensuremath{\varid{lookupPM}} recover the client-side \ensuremath{\conid{PatSubst}}?
Somehow we must remember the canonicalisation used when \emph{inserting}
so that we can invert it when \emph{matching}.   For example, suppose we insert the two
(pattern, value pairs)
$$
\ensuremath{(([\mskip1.5mu \varid{p}\mskip1.5mu],\varid{f}\;\varid{p}\;\conid{True}),\varid{v}_{1})} \quad \text{and} \quad \ensuremath{(([\mskip1.5mu \varid{q}\mskip1.5mu],\varid{f}\;\varid{q}\;\conid{False}),\varid{v}_{2})}
$$
Both patterns will canonicalise their (sole) pattern variable to the de Bruin level 1.
So if we look up the target \ensuremath{(\varid{f}\;\varid{e}\;\conid{True})} the \ensuremath{\conid{MatchME}} monad will produce a
final \ensuremath{\conid{Subst}} that maps \ensuremath{[\mskip1.5mu \mathrm{1}\mapsto\varid{e}\mskip1.5mu]}, paired with the value \ensuremath{\varid{v}_{1}}.  But we want to return
\ensuremath{([\mskip1.5mu (\text{\ttfamily \char34 p\char34},\varid{e})\mskip1.5mu],\varid{v}_{1})} to the client, a \ensuremath{\conid{PatSubst}} that uses the client variable \ensuremath{\text{\ttfamily \char34 p\char34}}, not
the internal index 1.

The solution is simple enough: \emph{we store the mapping in the triemap's domain},
along with the values, thus:
\begin{hscode}\SaveRestoreHook
\column{B}{@{}>{\hspre}l<{\hspost}@{}}%
\column{E}{@{}>{\hspre}l<{\hspost}@{}}%
\>[B]{}\keyword{type}\;\conid{PatMap}\;\varid{v}\mathrel{=}\conid{MExprMap}\;(\conid{PatKeys},\varid{v}){}\<[E]%
\ColumnHook
\end{hscode}\resethooks
Now the code writes itself. Here is \ensuremath{\varid{alterPM}}:
\begin{hscode}\SaveRestoreHook
\column{B}{@{}>{\hspre}l<{\hspost}@{}}%
\column{3}{@{}>{\hspre}l<{\hspost}@{}}%
\column{5}{@{}>{\hspre}l<{\hspost}@{}}%
\column{23}{@{}>{\hspre}l<{\hspost}@{}}%
\column{E}{@{}>{\hspre}l<{\hspost}@{}}%
\>[B]{}\varid{alterPM}\mathbin{::}\keyword{$\forall$}\!\! \hsforall \;\varid{v}\hsdot{\circ }{.\,}([\mskip1.5mu \conid{Var}\mskip1.5mu],\conid{Expr})\to \conid{TF}\;\varid{v}\to \conid{PatMap}\;\varid{v}\to \conid{PatMap}\;\varid{v}{}\<[E]%
\\
\>[B]{}\varid{alterPM}\;(\varid{pvars},\varid{e})\;\varid{tf}\;\varid{pm}\mathrel{=}\varid{atMTM}\;\varid{pat}\;\varid{ptf}\;\varid{pm}{}\<[E]%
\\
\>[B]{}\hsindent{3}{}\<[3]%
\>[3]{}\keyword{where}{}\<[E]%
\\
\>[3]{}\hsindent{2}{}\<[5]%
\>[5]{}\varid{pks}\mathbin{::}\conid{PatKeys}\mathrel{=}\varid{canonPatKeys}\;\varid{pvars}\;\varid{e}{}\<[E]%
\\[\blanklineskip]%
\>[3]{}\hsindent{2}{}\<[5]%
\>[5]{}\varid{pat}\mathbin{::}\conid{PatExpr}\mathrel{=}\conid{P}\;\varid{pks}\;(\conid{A}\;\varid{emptyDBE}\;\varid{e}){}\<[E]%
\\[\blanklineskip]%
\>[3]{}\hsindent{2}{}\<[5]%
\>[5]{}\varid{ptf}\mathbin{::}\conid{TF}\;(\conid{PatKeys},\varid{v}){}\<[E]%
\\
\>[3]{}\hsindent{2}{}\<[5]%
\>[5]{}\varid{ptf}\;\conid{Nothing}{}\<[23]%
\>[23]{}\mathrel{=}\varid{fmap}\;(\lambda \varid{v}\to (\varid{pks},\varid{v}))\;(\varid{tf}\;\conid{Nothing}){}\<[E]%
\\
\>[3]{}\hsindent{2}{}\<[5]%
\>[5]{}\varid{ptf}\;(\conid{Just}\;(\anonymous ,\varid{v})){}\<[23]%
\>[23]{}\mathrel{=}\varid{fmap}\;(\lambda \varid{v}\to (\varid{pks},\varid{v}))\;(\varid{tf}\;(\conid{Just}\;\varid{v})){}\<[E]%
\\[\blanklineskip]%
\>[B]{}\varid{canonPatKeys}\mathbin{::}[\mskip1.5mu \conid{Var}\mskip1.5mu]\to \conid{Expr}\to \conid{PatKeys}{}\<[E]%
\ColumnHook
\end{hscode}\resethooks
The auxiliary function \ensuremath{\varid{canonPatKeys}} takes the client-side pattern \ensuremath{(\varid{pvars},\varid{e})},
and returns a \ensuremath{\conid{PatKeys}} (\Cref{sec:patterns}) that maps each pattern variable
to its canonical de Bruijn index. \ensuremath{\varid{canonPatKeys}} is entirely straightforward:
it simply walks the expression, numbering off the pattern variables in
left-to-right order.

Then we can simply call the internal \ensuremath{\varid{atMTM}} function,
passing it a canonical \ensuremath{\varid{pat}\mathbin{::}\conid{PatExpr}} and
a transformer \ensuremath{\varid{ptf}\mathbin{::}\conid{TF}\;(\conid{PatKeys},\varid{v})} that will pair the \ensuremath{\conid{PatKeys}} with
the value supplied by the user via \ensuremath{\varid{tf}\mathbin{::}\conid{TF}\;\varid{v}}.
Lookup is equally easy:
\begin{hscode}\SaveRestoreHook
\column{B}{@{}>{\hspre}l<{\hspost}@{}}%
\column{3}{@{}>{\hspre}l<{\hspost}@{}}%
\column{5}{@{}>{\hspre}l<{\hspost}@{}}%
\column{11}{@{}>{\hspre}l<{\hspost}@{}}%
\column{29}{@{}>{\hspre}l<{\hspost}@{}}%
\column{E}{@{}>{\hspre}l<{\hspost}@{}}%
\>[B]{}\varid{lookupPM}{}\<[11]%
\>[11]{}\mathbin{::}\conid{Expr}\to \conid{PatMap}\;\varid{v}\to [\mskip1.5mu (\conid{PatSubst},\varid{v})\mskip1.5mu]{}\<[E]%
\\
\>[B]{}\varid{lookupPM}\;\varid{e}\;\varid{pm}{}\<[E]%
\\
\>[B]{}\hsindent{3}{}\<[3]%
\>[3]{}\mathrel{=}[\mskip1.5mu (\varid{\conid{Map}.toList}\;(\varid{subst}\mathbin{`\varid{\conid{Map}.compose}`}\varid{pks}),\varid{x}){}\<[E]%
\\
\>[3]{}\hsindent{2}{}\<[5]%
\>[5]{}\mid (\varid{subst},(\varid{pks},\varid{x}))\leftarrow {}\<[29]%
\>[29]{}\varid{runMatchExpr}\mathbin{\$}{}\<[E]%
\\
\>[29]{}\varid{lkMTM}\;(\conid{A}\;\varid{emptyDBE}\;\varid{e})\;\varid{pm}\mskip1.5mu]{}\<[E]%
\ColumnHook
\end{hscode}\resethooks
We use \ensuremath{\varid{runMatchExpr}} to get a list of successful matches, and then pre-compose
(see \Cref{fig:containers}) the internal \ensuremath{\conid{Subst}} with the \ensuremath{\conid{PatKeys}} mapping that
is part of the match result. We turn that into a list to get the client-side
\ensuremath{\conid{PatSubst}}. The only tricky point is what to do with pattern variables that are
not substituted. For example, suppose we insert the pattern \ensuremath{([\mskip1.5mu \varid{p},\varid{q}\mskip1.5mu],\varid{f}\;\varid{p})}. No
lookup will bind \ensuremath{\varid{q}}, because \ensuremath{\varid{q}} simply does not appear in the pattern. One
could reject this on insertion, but here we simply return a \ensuremath{\conid{PatSubst}} with no
binding for \ensuremath{\varid{q}}.

\subsection{Adding type classes} \label{sec:matching-trie-class}

All of this has been specific to matching tries keyed by \ensuremath{\conid{AlphaExpr}}, but as before
we can profitably define a type class of such matching tries, like this:
\begin{hscode}\SaveRestoreHook
\column{B}{@{}>{\hspre}l<{\hspost}@{}}%
\column{3}{@{}>{\hspre}l<{\hspost}@{}}%
\column{15}{@{}>{\hspre}l<{\hspost}@{}}%
\column{17}{@{}>{\hspre}l<{\hspost}@{}}%
\column{21}{@{}>{\hspre}l<{\hspost}@{}}%
\column{E}{@{}>{\hspre}l<{\hspost}@{}}%
\>[B]{}\keyword{class}\;\conid{Matchable}\;(\conid{MKey}\;\varid{tm})\Rightarrow \conid{MTrieMap}\;\varid{tm}\;\keyword{where}{}\<[E]%
\\
\>[B]{}\hsindent{3}{}\<[3]%
\>[3]{}\keyword{type}\;\conid{MKey}\;\varid{tm}{}\<[21]%
\>[21]{}\mathbin{::}\conid{Type}{}\<[E]%
\\
\>[B]{}\hsindent{3}{}\<[3]%
\>[3]{}\varid{emptyMTM}{}\<[17]%
\>[17]{}\mathbin{::}\varid{tm}\;\varid{v}{}\<[E]%
\\
\>[B]{}\hsindent{3}{}\<[3]%
\>[3]{}\varid{lkMTM}{}\<[17]%
\>[17]{}\mathbin{::}\conid{MKey}\;\varid{tm}\to \varid{tm}\;\varid{v}\to \conid{MatchM}\;(\conid{MKey}\;\varid{tm})\;\varid{v}{}\<[E]%
\\
\>[B]{}\hsindent{3}{}\<[3]%
\>[3]{}\varid{atMTM}{}\<[17]%
\>[17]{}\mathbin{::}\conid{Pat}\;(\conid{MKey}\;\varid{tm})\to \conid{TF}\;\varid{v}\to \varid{tm}\;\varid{v}\to \varid{tm}\;\varid{v}{}\<[E]%
\\[\blanklineskip]%
\>[B]{}\keyword{class}\;\conid{Eq}\;(\conid{Pat}\;\varid{k})\Rightarrow \conid{Matchable}\;\varid{k}\;\keyword{where}{}\<[E]%
\\
\>[B]{}\hsindent{3}{}\<[3]%
\>[3]{}\keyword{type}\;\conid{Pat}\;{}\<[15]%
\>[15]{}\varid{k}\mathbin{::}\conid{Type}{}\<[E]%
\\
\>[B]{}\hsindent{3}{}\<[3]%
\>[3]{}\keyword{type}\;\conid{Subst}\;\varid{k}\mathbin{::}\conid{Type}{}\<[E]%
\\
\>[B]{}\hsindent{3}{}\<[3]%
\>[3]{}\varid{match}\mathbin{::}\conid{Pat}\;\varid{k}\to \varid{k}\to \conid{MatchM}\;\varid{k}\;(){}\<[E]%
\\[\blanklineskip]%
\>[B]{}\keyword{newtype}\;\conid{MatchM}\;\varid{k}\;\varid{v}\mathrel{=}\conid{M}\;(\conid{Subst}\;\varid{k}\to [\mskip1.5mu (\conid{Subst}\;\varid{k},\varid{v})\mskip1.5mu]){}\<[E]%
\ColumnHook
\end{hscode}\resethooks
The following points are worth noting:
\begin{itemize}
  \item The old \ensuremath{\conid{TrieMap}} class, in \Cref{sec:class}, required the keys to be comparable for \emph{equality}.
     Unsurprisingly the new \ensuremath{\conid{MTrieMap}} class instead requires the keys to be \emph{\ensuremath{\conid{Matchable}}}.
   What does it mean to be \ensuremath{\conid{Matchable}}?  As you can see from the \ensuremath{\conid{Matchable}} class, it just
   means that we must provide a \ensuremath{\varid{match}} function, like \ensuremath{\varid{matchE}} defined in \Cref{sec:matching-alphaexpr}.
 \item The \ensuremath{\varid{match}} function returns \ensuremath{\conid{MatchM}}, a slight generalisation of \ensuremath{\conid{MatchME}}.  Specifically,
   \ensuremath{\conid{MatchM}} carries a substitution whose type \ensuremath{\conid{Subst}\;\varid{k}} depends on the key type \ensuremath{\varid{k}}. \ensuremath{\conid{Subst}} is an associated
   type of the \ensuremath{\conid{Matchable}} class.
 \item
 The lookup function takes a key of type \ensuremath{\conid{MKey}\;\varid{tm}}, but
 it returns something in the \ensuremath{\conid{MatchM}} monad, rather than the \ensuremath{\conid{Maybe}} monad (as was the case in \Cref{sec:class}).
\item The \ensuremath{\varid{atMTM}} takes a \emph{pattern} (rather than just a key), of type
  \ensuremath{\conid{Pat}\;(\conid{MKey}\;\varid{tm})}, and alters the trie's value at that pattern.%
\footnote{ Remember, a matching trie represents a set of (pattern,value) pairs.}
\end{itemize}
\noindent
Now we can write instances for these two classes to make them work for \ensuremath{\conid{AlphaExpr}} keys:
\begin{hscode}\SaveRestoreHook
\column{B}{@{}>{\hspre}l<{\hspost}@{}}%
\column{3}{@{}>{\hspre}l<{\hspost}@{}}%
\column{15}{@{}>{\hspre}l<{\hspost}@{}}%
\column{E}{@{}>{\hspre}l<{\hspost}@{}}%
\>[B]{}\keyword{instance}\;\conid{MTrieMap}\;\conid{MExprMap}\;\keyword{where}{}\<[E]%
\\
\>[B]{}\hsindent{3}{}\<[3]%
\>[3]{}\keyword{type}\;\conid{MKey}\;\conid{MExprMap}\mathrel{=}\conid{AlphaExpr}{}\<[E]%
\\
\>[B]{}\hsindent{3}{}\<[3]%
\>[3]{}\varid{emptyMTM}\mathrel{=}\conid{EmptyMEM}{}\<[E]%
\\
\>[B]{}\hsindent{3}{}\<[3]%
\>[3]{}\varid{lkMTM}\mathrel{=}\varid{lookupPatMM}{}\<[E]%
\\
\>[B]{}\hsindent{3}{}\<[3]%
\>[3]{}\mathinner{\ldotp\ldotp}{}\<[E]%
\\[\blanklineskip]%
\>[B]{}\keyword{instance}\;\conid{Matchable}\;\conid{AlphaExpr}\;\keyword{where}{}\<[E]%
\\
\>[B]{}\hsindent{3}{}\<[3]%
\>[3]{}\keyword{type}\;\conid{Pat}\;{}\<[15]%
\>[15]{}\conid{AlphaExpr}\mathrel{=}\conid{PatExpr}{}\<[E]%
\\
\>[B]{}\hsindent{3}{}\<[3]%
\>[3]{}\keyword{type}\;\conid{Subst}\;{}\<[15]%
\>[15]{}\conid{AlphaExpr}\mathrel{=}\conid{SubstE}{}\<[E]%
\\
\>[B]{}\hsindent{3}{}\<[3]%
\>[3]{}\varid{match}\mathrel{=}\varid{matchE}{}\<[E]%
\ColumnHook
\end{hscode}\resethooks
We can also generalise \ensuremath{\conid{SEMap}} (\Cref{sec:singleton}) in a similar way:
\begin{hscode}\SaveRestoreHook
\column{B}{@{}>{\hspre}l<{\hspost}@{}}%
\column{3}{@{}>{\hspre}l<{\hspost}@{}}%
\column{17}{@{}>{\hspre}l<{\hspost}@{}}%
\column{19}{@{}>{\hspre}c<{\hspost}@{}}%
\column{19E}{@{}l@{}}%
\column{22}{@{}>{\hspre}l<{\hspost}@{}}%
\column{33}{@{}>{\hspre}l<{\hspost}@{}}%
\column{E}{@{}>{\hspre}l<{\hspost}@{}}%
\>[B]{}\keyword{data}\;\conid{MSEMap}\;\varid{tm}\;\varid{v}{}\<[19]%
\>[19]{}\mathrel{=}{}\<[19E]%
\>[22]{}\conid{EmptyMSEM}{}\<[E]%
\\
\>[19]{}\mid {}\<[19E]%
\>[22]{}\conid{SingleMSEM}\;(\conid{Pat}\;(\conid{MKey}\;\varid{tm}))\;\varid{v}{}\<[E]%
\\
\>[19]{}\mid {}\<[19E]%
\>[22]{}\conid{MultiMSEM}\;{}\<[33]%
\>[33]{}(\varid{tm}\;\varid{v}){}\<[E]%
\\[\blanklineskip]%
\>[B]{}\keyword{instance}\;\conid{MTrieMap}\;\varid{tm}\Rightarrow \conid{MTrieMap}\;(\conid{MSEMap}\;\varid{tm})\;\keyword{where}{}\<[E]%
\\
\>[B]{}\hsindent{3}{}\<[3]%
\>[3]{}\keyword{type}\;\conid{MKey}\;(\conid{MSEMap}\;\varid{tm})\mathrel{=}\conid{MKey}\;\varid{tm}{}\<[E]%
\\
\>[B]{}\hsindent{3}{}\<[3]%
\>[3]{}\varid{emptyMTM}{}\<[17]%
\>[17]{}\mathrel{=}\conid{EmptyMSEM}{}\<[E]%
\\
\>[B]{}\hsindent{3}{}\<[3]%
\>[3]{}\varid{lkMTM}{}\<[17]%
\>[17]{}\mathrel{=}\varid{lkMSEM}{}\<[E]%
\\
\>[B]{}\hsindent{3}{}\<[3]%
\>[3]{}\varid{atMTM}{}\<[17]%
\>[17]{}\mathrel{=}\varid{atMSEM}{}\<[E]%
\ColumnHook
\end{hscode}\resethooks
Notice that \ensuremath{\conid{SingleMSEM}} contains a \emph{pattern},
not merely a \emph{key}, unlike \ensuremath{\conid{SingleSEM}} in \Cref{sec:singleton}.
The code for \ensuremath{\varid{lkMSEM}} and \ensuremath{\varid{atMSEM}} is straightforward;
we give the former here, leaving the latter for the supplement~\cite{triemaps-extended,triemaps-github}
\begin{hscode}\SaveRestoreHook
\column{B}{@{}>{\hspre}l<{\hspost}@{}}%
\column{18}{@{}>{\hspre}l<{\hspost}@{}}%
\column{31}{@{}>{\hspre}l<{\hspost}@{}}%
\column{38}{@{}>{\hspre}l<{\hspost}@{}}%
\column{E}{@{}>{\hspre}l<{\hspost}@{}}%
\>[B]{}\varid{lkMSEM}\mathbin{::}\conid{MTrieMap}\;\varid{tm}{}\<[31]%
\>[31]{}\Rightarrow \conid{MKey}\;\varid{tm}\to \conid{MSEMap}\;\varid{tm}\;\varid{a}{}\<[E]%
\\
\>[31]{}\to \conid{Match}\;(\conid{MKey}\;\varid{tm})\;\varid{a}{}\<[E]%
\\
\>[B]{}\varid{lkMSEM}\;\varid{k}\;{}\<[18]%
\>[18]{}\conid{EmptyMSEM}{}\<[38]%
\>[38]{}\mathrel{=}\varid{mzero}{}\<[E]%
\\
\>[B]{}\varid{lkMSEM}\;\varid{k}\;{}\<[18]%
\>[18]{}(\conid{MultiMSEM}\;\varid{m}){}\<[38]%
\>[38]{}\mathrel{=}\varid{lkMTM}\;\varid{k}\;\varid{m}{}\<[E]%
\\
\>[B]{}\varid{lkMSEM}\;\varid{k}\;{}\<[18]%
\>[18]{}(\conid{SingleMSEM}\;\varid{pat}\;\varid{v}){}\<[38]%
\>[38]{}\mathrel{=}\varid{match}\;\varid{pat}\;\varid{k}\mathrel{{>}\hspace{-0.4em}{>}}\varid{pure}\;\varid{v}{}\<[E]%
\ColumnHook
\end{hscode}\resethooks
Notice the call to \ensuremath{\varid{mzero}} to make the lookup fail if the map is empty; and, in the
\ensuremath{\conid{SingleMSEM}} case, the call \ensuremath{\varid{match}} to match the pattern against the key.

\subsection{Most specific match and unification} \label{sec:most-specific}

It is tempting to ask: can we build a lookup that returns only the \emph{most
specific} matches? And can we build a lookup that returns all values whose
patterns \emph{unify} with the target. Both would have useful applications, in
GHC at least.

However, both seem difficult to achieve. All our attempts became mired in
complexity, and we leave this for further work, and as a challenge for the
reader. We outline some of the difficulties of unifying lookup in Appendix B of
the extended version of this paper~\cite{triemaps-extended}.

\section{Evaluation} \label{sec:eval}

\pgfplotstableread{bench-plot.txt}\benchdata

\begin{figure}[h]
\begin{tikzpicture}
\begin{axis}[
  ybar,
  bar width=7pt,
  height=6.5cm,
  width=8cm,
  %
  ymin=0,
  ymax=2,
  ylabel={Relative time (lower is better)},
  ytick={0.5,1.5,2.0},
  extra y ticks=1, %
  extra y tick style={grid=major, grid style={solid,black}},
  yticklabel={\pgfmathprintnumber{\tick}x},
  %
  xmin=0.5,
  xmax=4.5,
  xtick=data,
  xticklabels from table={\benchdata}{name},
  x tick label style={font=\small},
  major x tick style={opacity=0},    
  x tick label style={yshift=0.5em}, 
  %
  visualization depends on=rawy \as \rawy,
  nodes near coords={%
    \pgfmathfloatparsenumber{\rawy}%
    \let\rawy\pgfmathresult%
    \pgfmathfloatparsenumber{\pgfkeysvalueof{/pgfplots/ymax}}%
    \pgfmathfloatlessthan{\pgfmathresult}{\rawy}%
    \ifpgfmathfloatcomparison%
      $\cdots$%
    \fi%
    \pgfmathprintnumber[precision=2,fixed zerofill]{\rawy}
  },
  restrict y to domain*={
    \pgfkeysvalueof{/pgfplots/ymin}:\pgfkeysvalueof{/pgfplots/ymax}
  },
  every node near coord/.append style={
    font=\small,
    anchor=west,
    rotate=90,
  },
]

\addplot table[x=id,y=ExprMap]{\benchdata};
\addplot table[x=id,y=Map]{\benchdata};
\addplot table[x=id,y=HashMap]{\benchdata};

\legend{TM,OM,HM}
\end{axis}
\end{tikzpicture}
\caption{Benchmarks comparing our trie map (TM)
  to ordered maps (OM) and hash maps (HM)}
\label{fig:plot}
\end{figure}

So far, we have seen that trie maps offer a significant advantage over other
kinds of maps like ordered maps or hash maps: the ability to do a matching
lookup (in \Cref{sec:matching}). In this section, we will see that query
performance is another advantage. Our implementation of trie maps in Haskell
can generally compete in performance with other map data structures, while
significantly outperforming traditional map implementations on some operations.
Not bad for a data structure that we can also extend to support matching lookup!

We took runtime measurements of the (non-matching) \ensuremath{\conid{ExprMap}} data
structure on a selection of workloads, conducted using the \hackage{criterion}
benchmarking library.%
\footnote{The benchmark machine runs Ubuntu 18.04 on an Intel Core i5-8500 with
16GB RAM. All programs were compiled on GHC 9.0.2 with \texttt{-O2
-fproc-alignment=64} to eliminate code layout flukes and run with \texttt{+RTS
-A128M -RTS} for 128MB space in generation 0 in order to prevent major GCs from
skewing the results.}
\Cref{fig:plot} presents a quick overview of the results.

Appendix A of the extended version of this paper~\cite{triemaps-extended}
featurs a more in-depth analysis and finer runtime as well as space measurements
and indicators for statistical significance.

\subsubsection*{Setup}
All benchmarks except \benchname{fromList} are handed a pre-built
map containing 10000 expressions, each consisting of roughly 100 \ensuremath{\conid{Expr}} data
constructors drawn from a pseudo-random source with a fixed (and thus
deterministic) seed.

We compare three different non-matching map implementations, simply because we
were not aware of other map data structures with matching lookup modulo
$\alpha$-equivalence and we wanted to compare apples to apples.
The \ensuremath{\conid{ExprMap}} forms the baseline. Asymptotics are given with respect to map
size $n$ and key expression size $k$:

\begin{itemize}
  \item \ensuremath{\conid{ExprMap}} (designated ``TM'' in \Cref{fig:plot}) is the trie map
        implementation from this paper. Insertion and lookup perform at most
        one full traversal of the key, so performance should scale with
        $\mathcal{O}(k)$.
  \item \ensuremath{\conid{Map}\;\conid{Expr}} (designated ``OM'') is the ordered map implementation from
        the mature, well-optimised \hackage{containers} library. It uses size
        balanced trees under the hood \cite{adams}. Thus, lookup and insert
        operations incur an additional log factor in the map size $n$, for a
        total of $\mathcal{O}(k \log n)$ factor compared to both other maps.
  \item \ensuremath{\conid{HashMap}\;\conid{Expr}} (designated ``HM'') is an implementation of hash array
        mapped tries \cite{hamt} from the \hackage{unordered-containers}
        library. Like \ensuremath{\conid{ExprMap}}, map access incurs a full traversal of the key
        to compute a hash and then a $\mathcal{O}(\log_{32} n)$ lookup in the
        array mapped trie, as well as an expected constant number of key
        comparisons to resolve collisions. Consequently, lookup and insert
        are both in $\mathcal{O}(k + \log_{32} n)$, where the log summand can
        effectively be treated like a constant for all intents and purposes.
\end{itemize}

Some clarification as to what our benchmarks measure:

\begin{itemize}
  \item The \benchname{lookup} benchmark looks up every
        expression that is part of the map. So for a map of size 10000, we
        perform 10000 lookups of expressions each of which have approximately size
        100.
  \item \benchname{lookup\_lam} is like \benchname{lookup}, but wraps a shared
        prefix of 100 layers of \ensuremath{(\conid{Lam}\;\text{\ttfamily \char34 \$\char34})} around each expression.
  \item \benchname{fromList} benchmarks a naïve \ensuremath{\varid{fromList}}
        implementation on \ensuremath{\conid{ExprMap}} against the tuned \ensuremath{\varid{fromList}} implementations
        of the other maps, measuring map creation performance from batches.
\end{itemize}

\subsubsection*{Querying}
The results suggest that \ensuremath{\conid{ExprMap}} is about as fast as \ensuremath{\conid{Map}\;\conid{Expr}} for
completely random expressions in \benchname{lookup}. But in a more realistic
scenario, at least some expressions share a common prefix, which is what
\benchname{lookup\_lam} embodies. There we can see that \ensuremath{\conid{ExprMap}} wins against
\ensuremath{\conid{Map}\;\conid{Expr}} by a huge margin: \ensuremath{\conid{ExprMap}} looks at the shared prefix exactly
once on lookup, while \ensuremath{\conid{Map}} has to traverse the shared prefix of length
$\mathcal{O}(k)$ on each of its $\mathcal{O}(\log n)$ comparisons.

Although \ensuremath{\conid{HashMap}} loses on most benchmarks compared to \ensuremath{\conid{ExprMap}} and \ensuremath{\conid{Map}}, most
measurements were consistently at most a factor of two slower than \ensuremath{\conid{ExprMap}}.
We believe that is due to the fact that it is enough to traverse the
\ensuremath{\conid{Expr}} twice during lookup barring any collisions (hash and then equate with the
match), thus it is expected to scale similarly as \ensuremath{\conid{ExprMap}}. Thus, both \ensuremath{\conid{ExprMap}}
and \ensuremath{\conid{HashMap}} perform much more consistently than \ensuremath{\conid{Map}}.

\subsubsection*{Modification}
While \ensuremath{\conid{ExprMap}} consistently wins in query performance, the edge is melting into
insignificance for \benchname{fromList} and \benchname{union}. One reason is
the uniform distribution of expressions in these benchmarks, which favors \ensuremath{\conid{Map}}.
Still, it is a surprise that the naïve \ensuremath{\varid{fromList}} implementations of \ensuremath{\conid{ExprMap}} and
\ensuremath{\conid{Map}} as list folds beat the one of \ensuremath{\conid{HashMap}}, although the latter has a tricky,
performance-optimised implementation using transient mutability.

What would a non-naïve version of \ensuremath{\varid{fromList}} for \ensuremath{\conid{ExprMap}} look like? Perhaps
the process could be sped up considerably by partitioning the input list
according to the different fields of \ensuremath{\conid{ExprMap}} and then calling
the \ensuremath{\varid{fromList}} implementations of the individual fields in turn. The process
would be very similar to discrimination sort~\cite{discr-sort}, which is a
generalisation of radix sort to tree-like data and very close to tries.
Indeed, the \hackage{discrimination} library provides such an optimised
$\mathcal{O}(n)$ \ensuremath{\varid{toMap}} implementation for \ensuremath{\conid{Map}}.

\section{Related work} \label{sec:related}

\subsection{Matching triemaps in automated reasoning} \label{sec:discrim-trees}

Matching triemaps have been used in the automated reasoning community for
decades, where they are recognised as a kind of \emph{term
index}~\cite{handbook:2001}, a data structure that allows efficient lookup and
matching of terms such as logic formulas or program expressions.
An automated reasoning system has hundreds or thousands of axioms, each of
which is quantified over some variables (just like the RULEs described in
\Cref{sec:matching-intro}). Each of these axioms might apply at any sub-tree of
the term under consideration, so efficient matching of many axioms is absolutely
central to the performance of these systems.

This led to a great deal of work on so-called \emph{discrimination trees}, starting
in the late 1980's, which is beautifully surveyed in the Handbook of Automated Reasoning
\cite[Chapter 26]{handbook:2001}.
All of this work typically assumes a single, fixed, data type of ``first order terms''
like this\footnote{Binders in terms do not seem to be important
in these works, although they could be handled fairly easily by a de Bruijn pre-pass.}
\begin{hscode}\SaveRestoreHook
\column{B}{@{}>{\hspre}l<{\hspost}@{}}%
\column{3}{@{}>{\hspre}l<{\hspost}@{}}%
\column{E}{@{}>{\hspre}l<{\hspost}@{}}%
\>[3]{}\keyword{data}\;\conid{MKey}\mathrel{=}\conid{Node}\;\conid{Fun}\;[\mskip1.5mu \conid{MKey}\mskip1.5mu]{}\<[E]%
\ColumnHook
\end{hscode}\resethooks
where \ensuremath{\conid{Fun}} is a function symbol, and each such function symbol has a fixed arity.
Discrimination trees are described by imagining
a pre-order traversal that (uniquely, since function symbols have fixed arity)
converts the \ensuremath{\conid{MKey}} to a list of type \ensuremath{[\mskip1.5mu \conid{Fun}\mskip1.5mu]}, and treating that as the key.
The map is implemented like this:
\begin{hscode}\SaveRestoreHook
\column{B}{@{}>{\hspre}l<{\hspost}@{}}%
\column{3}{@{}>{\hspre}l<{\hspost}@{}}%
\column{5}{@{}>{\hspre}l<{\hspost}@{}}%
\column{20}{@{}>{\hspre}l<{\hspost}@{}}%
\column{31}{@{}>{\hspre}l<{\hspost}@{}}%
\column{E}{@{}>{\hspre}l<{\hspost}@{}}%
\>[3]{}\keyword{data}\;\conid{DTree}\;\varid{v}\mathrel{=}\conid{DVal}\;\varid{v}\mid \conid{DNode}\;(\conid{Map}\;\conid{Fun}\;\conid{DTree}){}\<[E]%
\\[\blanklineskip]%
\>[3]{}\varid{lookupDT}\mathbin{::}[\mskip1.5mu \conid{Fun}\mskip1.5mu]\to \conid{DTree}\;\varid{v}\to \conid{Maybe}\;\varid{v}{}\<[E]%
\\
\>[3]{}\varid{lookupDT}\;[\mskip1.5mu \mskip1.5mu]\;{}\<[20]%
\>[20]{}(\conid{DVal}\;\varid{v}){}\<[31]%
\>[31]{}\mathrel{=}\conid{Just}\;\varid{v}{}\<[E]%
\\
\>[3]{}\varid{lookupDT}\;(\varid{f}\mathbin{:}\varid{fs})\;{}\<[20]%
\>[20]{}(\conid{DNode}\;\varid{m}){}\<[31]%
\>[31]{}\mathrel{=}\keyword{case}\;\varid{\conid{Map}.lookup}\;\varid{f}\;\varid{m}\;\keyword{of}{}\<[E]%
\\
\>[3]{}\hsindent{2}{}\<[5]%
\>[5]{}\conid{Just}\;\varid{dt}\to \varid{lookupDT}\;\varid{fs}\;\varid{dt}{}\<[E]%
\\
\>[3]{}\hsindent{2}{}\<[5]%
\>[5]{}\conid{Nothing}\to \conid{Nothing}{}\<[E]%
\\
\>[3]{}\varid{lookupDT}\;\anonymous \;{}\<[20]%
\>[20]{}\anonymous {}\<[31]%
\>[31]{}\mathrel{=}\conid{Nothing}{}\<[E]%
\ColumnHook
\end{hscode}\resethooks
Each layer of the tree branches on the first \ensuremath{\conid{Fun}}, and looks up
the rest of the \ensuremath{[\mskip1.5mu \conid{Fun}\mskip1.5mu]} in the appropriate child.
Extending this basic setup with matching is done by some kind of backtracking.

Discrimination trees are heavily used by theorem provers, such as Coq, Isabelle, and Lean.
Moreover, discrimination trees have been further developed in a number of ways.
Vampire uses \emph{code trees} which are a compiled form of discrimination
tree that stores abstract machine instructions, rather than a data structure
at each node of the tree \cite{voronkov:vampire}.
Spass \cite{spass} uses \emph{substitution trees} \cite{substitution-trees},
a refinement of discrimination trees that can share common \emph{sub-trees}
not just common \emph{prefixes}. (It is not clear whether the extra complexity of
substitution trees pays its way.)  Z3 uses \emph{E-matching code trees}, which solve
for matching modulo an ever-growing equality relation, useful in saturation-based
theorem provers.  All of these techniques except E-matching are surveyed in
\citet{handbook:2001}.

If we applied our ideas to \ensuremath{\conid{MKey}} we would get a single-field triemap which
(just like \ensuremath{\varid{lookupDT}}) would initially branch on \ensuremath{\conid{Fun}}, and then go though
a chain of \ensuremath{\conid{ListMap}} constructors (which correspond to the \ensuremath{\conid{DNode}} above).
You have to squint pretty hard  --- for example, we do the pre-order traversal on the fly
--- but the net result is very similar, although
it is arrived at by entirely different thought process.
%
%
%

Many of the insights of the term indexing world re-appear, in different guise,
in our triemaps.   For example, when a variable is repeated in a pattern we can
eagerly check for equality during the match, or instead gather an equality constraint
and check those constraints at the end \cite[Section 26.14]{handbook:2001}.

A related application of matching tries appear in
\cite[Section 2.2]{linked-visualisations}, where \emph{eliminators} express
both parameter-binding and pattern-matching in a single Core language construct,
with a semantics not unlike GHC's own \text{\ttfamily \char45{}XLambdaCase} extension.
They realise that their big-step interpreter implements eliminators via special
generalised tries that can express variable matching -- which corresponds to our
triemaps applied to linear patterns.

\subsection{Haskell triemaps}

Trie data structures have found their way into numerous Haskell packages over time.
There are trie data structures that are specific to \ensuremath{\conid{String}}, like the
\hackage{StringMap} package, or polymorphically, requiring just a type class for
trie key extraction, like the \hackage{TrieMap} package. None of these
libraries describe how to index on expression data structures modulo
$\alpha$-equivalence or how to perform matching lookup.

Memoisation has been a prominent application of tries in Haskell
\cite{hinze:memo,conal:blog1,conal:blog2}.
Given a function \ensuremath{\varid{f}}, the idea is to build an \emph{infinite},
lazily-evaluated trie, that maps every possible argument \ensuremath{\varid{x}} to (a thunk for)
$\ensuremath{(\varid{f}\;\varid{x})}$.  Now, a function call becomes a lookup in the trie.
These ideas are implemented in the \hackage{MemoTrie} library, and have been generalised
to representable functors in the \hackage{representable-tries} library.
For memo tries, operations like alter, insert, union, and fold are all
irrelevant: the infinite trie is built once, and then used only for lookup.

A second strand of work concerns data type generic, or polytypic, approaches to
generating tries, which nicely complements the design-pattern approach
of this paper (\Cref{sec:generic}).
\citet{hinze:generalized} describes the polytypic approach,
for possibly parameterised and nested data types in some detail, including the
realisation that we need \ensuremath{\varid{atEM}} and \ensuremath{\varid{unionWithEM}} in order to define \ensuremath{\varid{insertEM}} and
\ensuremath{\varid{unionEM}}.
The aforementioned \varid{MemoTrie}, \varid{representable-tries} and
\hackage{generic-trie} libraries generate trie implementations polytypically.
For our approach to do the same, we would need to find a good way to specify
binding structure (as in \ensuremath{\conid{Lam}}) in a polytypic setting, which thus far has been
elusive.

The \hackage{twee-lib} library defines a simple term index data structure based
on discrimination trees for the \varid{twee} equation theorem prover. We would
arrive at a similar data structure in this paper had we started from an
expression data type
\begin{hscode}\SaveRestoreHook
\column{B}{@{}>{\hspre}l<{\hspost}@{}}%
\column{E}{@{}>{\hspre}l<{\hspost}@{}}%
\>[B]{}\keyword{data}\;\conid{Expr}\mathrel{=}\conid{App}\;\conid{Con}\;[\mskip1.5mu \conid{Expr}\mskip1.5mu]\mid \conid{Var}\;\conid{Var}{}\<[E]%
\ColumnHook
\end{hscode}\resethooks
In contrast to our \ensuremath{\conid{ExprMap}}, \varid{twee}'s \ensuremath{\conid{Index}} does path compression not
only for paths ending in leaves (as we do) but also for internal paths, as is
common for radix trees.

It is however unclear how to extend \varid{twee}'s \ensuremath{\conid{Index}} to support
$\alpha$-equivalence, hence we did not consider it for our benchmarks in
\Cref{sec:eval}.

\begin{acks}
We warmly thank Leonardo de Moura and Edward Yang for their very helpful feedback.
\end{acks}

%
%

\section{Conclusion}

We presented trie maps as an efficient data structure for representing a set of
expressions modulo $\alpha$-equivalence, re-discovering polytypic deriving
mechanisms described by~\citet{hinze:generalized}. Subsequently, we showed how to
extend this data structure to make it aware of pattern variables in order to
interpret stored expressions as patterns. The key innovation is that the
resulting trie map allows efficient matching lookup of a target expression
against stored patterns. This pattern store is quite close to discrimination
trees~\cite{handbook:2001}, drawing a nice connection to term indexing problems
in the automated theorem proving community.

\bibliography{refs}

\appendix
\let\Bbbk\relax



\let\restriction\relax

\section{Evaluation} \label{sec:eval-extended}

So far, we have seen that trie maps offer a significant advantage over other
kinds of maps like ordered maps or hash maps: the ability to do a matching
lookup (in \Cref{sec:matching}). In this section, we will see that query
performance is another advantage. Our implementation of trie maps in Haskell
can generally compete in performance with other map data structures, while
significantly outperforming traditional map implementations on some operations.
Not bad for a data structure that we can also extend to support matching lookup!

\subsection{Runtime} \label{sec:runtime}

\begin{table*}

  \caption{Benchmarks of different operations over our trie map \ensuremath{\conid{ExprMap}} (TM),
  ordered maps \ensuremath{\conid{Map}\;\conid{Expr}} (OM) and hash maps \ensuremath{\conid{HashMap}\;\conid{Expr}} (HM), varying the
  size parameter $N$.  Each map is of size $N$ (so $M=N$) and the expressions
  it contains are also each of size $N$ (so $E=N$).
  We give the measurements of OM and HM relative to absolute runtime
  measurements for TM. Lower is better. Digits whose order of magnitude is
  no larger than that of twice the standard deviation are marked by squiggly
  lines.}
  \begin{tabular}{l rrr rrr rrr}
  \toprule
  $N$  & \multicolumn{3}{c}{\textbf{10}} & \multicolumn{3}{c}{\textbf{100}} & \multicolumn{3}{c}{\textbf{1000}} \\
  \cmidrule(lr{.5em}){2-4} \cmidrule(lr{.5em}){5-7} \cmidrule(lr{.5em}){8-10}
  Data structure & TM & OM & HM
                 & TM & OM & HM
                 & TM & OM & HM \\
  \midrule
  \benchname{lookup} & \textbf{3.4\insigdig{9}\textmu{}s} & 1.0\insigdig{7} & 1.4\insigdig{9} & \textbf{213\textmu{}s} & 1.08 & 1.80 & \textbf{30.\insigdig{4}ms} & 1.0\insigdig{0} & 1.6\insigdig{8} \\
\benchname{lookup\_app1} & \textbf{6.2\insigdig{3}\textmu{}s} & 1.76 & 1.5\insigdig{1} & \textbf{481\textmu{}s} & 3.42 & 1.6\insigdig{4} & \textbf{64.4ms} & 4.82 & 1.65 \\
\benchname{lookup\_app2} & \textbf{6.6\insigdig{9}\textmu{}s} & 1.1\insigdig{2} & 1.4\insigdig{3} & \textbf{604\textmu{}s} & 1.42 & 1.4\insigdig{2} & \textbf{100ms} & 3.1\insigdig{2} & 1.30 \\
\benchname{lookup\_lam} & \textbf{7.5\insigdig{6}\textmu{}s} & 2.12 & 1.83 & \textbf{539\textmu{}s} & 4.72 & 2.09 & \textbf{72.6ms} & 6.6\insigdig{3} & 1.94 \\
\benchname{lookup\_one} & \textbf{323ns} & 1.0\insigdig{0} & 1.6\insigdig{4} & \textbf{1.55\textmu{}s} & 1.01 & 1.7\insigdig{0} & 23.0\textmu{}s & \textbf{1.00} & 1.73 \\
\benchname{fold} & 47\insigdig{2}ns & 0.37 & \textbf{0.30} & 6.\insigdig{0}\insigdig{1}\textmu{}s & \textbf{0.30} & 0.33 & 9\insigdig{1}.\insigdig{9}\textmu{}s & \textbf{0.30} & 0.3\insigdig{6} \\
\benchname{insert\_lookup\_one} & \textbf{64.3ns} & 2.19 & 1.4\insigdig{5} & \textbf{2.46\textmu{}s} & 1.15 & 1.54 & \textbf{26.2\textmu{}s} & 1.04 & 1.75 \\
\benchname{fromList} & 3.7\insigdig{1}\textmu{}s & \textbf{0.67} & 0.9\insigdig{8} & 129\textmu{}s & \textbf{0.91} & 1.98 & \textbf{13.\insigdig{1}ms} & 1.0\insigdig{1} & 2.8\insigdig{0} \\
\benchname{fromList\_app1} & 10.4\textmu{}s & 0.97 & \textbf{0.6\insigdig{4}} & 73\insigdig{0}\textmu{}s & 2.03 & \textbf{0.7\insigdig{5}} & \textbf{99.\insigdig{5}ms} & 3.33 & 1.19 \\
\benchname{union} & 4.9\insigdig{0}\textmu{}s & 0.87 & \textbf{0.79} & 306\textmu{}s & 0.94 & \textbf{0.89} & \textbf{43.\insigdig{7}ms} & 1.0\insigdig{7} & 1.0\insigdig{7} \\
\benchname{union\_app1} & \textbf{4.1\insigdig{9}\textmu{}s} & 1.9\insigdig{4} & 1.1\insigdig{3} & \textbf{190\textmu{}s} & 2.39 & 1.\insigdig{9}\insigdig{8} & \textbf{23.\insigdig{6}ms} & 4.34 & 5.26 \\

  \bottomrule
  \end{tabular}

  \label{fig:runtime}
\end{table*}

We measured the runtime performance of the (non-matching) \ensuremath{\conid{ExprMap}} data
structure on a selection of workloads, conducted using the \hackage{criterion}
benchmarking library%
\footnote{The benchmark machine runs Ubuntu 18.04 on an Intel Core i5-8500 with
16GB RAM. All programs were compiled with \texttt{-O2 -fproc-alignment=64} to
eliminate code layout flukes and run with \texttt{+RTS -A128M -RTS} for 128MB
space in generation 0 in order to prevent major GCs from skewing the results.}.
\Cref{fig:runtime} presents an overview of the results, while
\Cref{fig:runtime-finer} goes into more detail on some configurations.

\subsubsection*{Setup}
All benchmarks except the \benchname{fromList*} variants are handed a pre-built
map containing $N$ expressions, each consisting of roughly $N$ \ensuremath{\conid{Expr}} data
constructors, and drawn from a pseudo-random source with a fixed (and thus
deterministic) seed. $N$ is varied between 10 and 1000.

We compare three different non-matching map implementations, simply because we
were not aware of other map data structures with matching lookup modulo
$\alpha$-equivalence and we wanted to compare apples to apples.
The \ensuremath{\conid{ExprMap}} forms the baseline. Asymptotics are given with respect to map
size $n$ and key expression size $k$:

\begin{itemize}
  \item \ensuremath{\conid{ExprMap}} (designated ``TM'' in \Cref{fig:runtime}) is the trie map
        implementation from this paper. Insertion and lookup and have to perform
        a full traversal of the key, so performance should scale with
        $\mathcal{O}(k)$, where $k$ is the key \ensuremath{\conid{Expr}} that is accessed.
  \item \ensuremath{\conid{Map}\;\conid{Expr}} (designated ``OM'') is the ordered map implementation from
        the mature, well-optimised \hackage{containers} library. It uses size
        balanced trees under the hood \cite{adams}. Thus, lookup and insert
        operations incur an additional log factor in the map size $n$, for a
        total of $\mathcal{O}(k \log n)$ factor compared to both other maps.
  \item \ensuremath{\conid{HashMap}\;\conid{Expr}} (designated ``HM'') is an implementation of hash array
        mapped tries \cite{hamt} from the \hackage{unordered-containers}
        library. Like \ensuremath{\conid{ExprMap}}, map access incurs a full traversal of the key
        to compute a hash and then a $\mathcal{O}(\log_{32} n)$ lookup in the
        array mapped trie. The log factor can be treated like a constant for all
        intents and purposes, so lookup and insert is effectively in
        $\mathcal{O}(k)$.
\end{itemize}
Benchmarks ending in \benchname{\_lam}, \benchname{\_app1}, \benchname{\_app2}
add a shared prefix to each of the expressions before building the initial
map:
\begin{itemize}
  \item \benchname{\_lam} wraps $N$ layers of \ensuremath{(\conid{Lam}\;\text{\ttfamily \char34 \$\char34})} around each expression
  \item \benchname{\_app1} wraps $N$ layers of \ensuremath{(\conid{Var}\;\text{\ttfamily \char34 \$\char34}\,\mathbin{`\conid{App}`})} around each expression
  \item \benchname{\_app2} wraps $N$ layers of \ensuremath{(\mathbin{`\conid{App}`}\,\conid{Var}\;\text{\ttfamily \char34 \$\char34})} around each expression
\end{itemize}
where \ensuremath{\text{\ttfamily \char34 \$\char34}} is a name that doesn't otherwise occur in the generated expressions.

\begin{itemize}
  \item The \benchname{lookup*} family of benchmarks looks up every
        expression that is part of the map. So for a map of size 100, we
        perform 100 lookups of expressions each of which have approximately size
        100. \benchname{lookup\_one} looks up just one expression that is
        part of the map.
  \item \benchname{insert\_lookup\_one} inserts a random expression into the
        initial map and immediately looks it up afterwards. The lookup is to
        ensure that any work delayed by laziness is indeed forced.
  \item The \benchname{fromList*} family benchmarks a naïve \ensuremath{\varid{fromList}}
        implementation on \ensuremath{\conid{ExprMap}} against the tuned \ensuremath{\varid{fromList}} implementations
        of the other maps, measuring map creation performance from batches.
  \item \benchname{fold} simply sums up all values that are stored in the map
        (which stores \ensuremath{\conid{Int}}s).
\end{itemize}

\subsubsection*{Querying}
The results show that lookup in \ensuremath{\conid{ExprMap}} often wins against \ensuremath{\conid{Map}\;\conid{Expr}} and
\ensuremath{\conid{HashMap}\;\conid{Expr}}. The margin is small on the completely random \ensuremath{\conid{Expr}}s of
\benchname{lookup}, but realistic applications of \ensuremath{\conid{ExprMap}} often store
\ensuremath{\conid{Expr}}s with some kind of shared structure. The \benchname{\_lam} and
\benchname{\_app1} variants show that \ensuremath{\conid{ExprMap}} can win substantially against
an ordered map representation: \ensuremath{\conid{ExprMap}} looks at the shared prefix exactly
once one lookup, while \ensuremath{\conid{Map}} has to traverse the shared prefix of length
$\mathcal{O}(N)$ on each of its $\mathcal{O}(\log N)$ comparisons. As
a result, the gap between \ensuremath{\conid{ExprMap}} and \ensuremath{\conid{Map}} widens as $N$ increases,
confirming an asymptotic difference. The advantage is less pronounced in
the \benchname{\_app2} variant, presumably because \ensuremath{\conid{ExprMap}} can't share
the common prefix here: it turns into an unsharable suffix in the pre-order
serialisation, blowing up the trie map representation compared to its sibling
\benchname{\_app1}.

Although \ensuremath{\conid{HashMap}} loses on most benchmarks compared to \ensuremath{\conid{ExprMap}} and \ensuremath{\conid{Map}}, most
measurements were consistently at most a factor of two slower than \ensuremath{\conid{ExprMap}}.
We believe that is due to the fact that it is enough to traverse the \ensuremath{\conid{Expr}} once
to compute the hash, thus it is expected to scale similarly as \ensuremath{\conid{ExprMap}}.

Comparing the \benchname{lookup*} measurements of the same map data
structure on different size parameters $N$ reveals a roughly quadratic correlation
throughout all implementations, give or take a logarithmic factor.
That seems plausible given that $N$ linearly affects expression size and map
size (and thus, number of lookups). But realistic workloads tend to have much
larger map sizes than expression sizes!

\begin{table*}
  \centering
  \caption{Varying expression size $E$ and map size $M$ independently on benchmarks
  \benchname{lookup} and \benchname{insert\_lookup\_one}.}
  \resizebox{\textwidth}{!}{%
    \begin{tabular}{cr rrr rrr rrr rrr}
    \toprule
    \multicolumn{2}{c}{\multirow{2}{*}{\diagbox{$E$}{$M$}}} & \multicolumn{3}{c}{\textbf{10}}
                                        & \multicolumn{3}{c}{\textbf{100}}
                                        & \multicolumn{3}{c}{\textbf{1000}}
                                        & \multicolumn{3}{c}{\textbf{10000}} \\
    \cmidrule(lr{.5em}){3-5} \cmidrule(lr{.5em}){6-8} \cmidrule(lr{.5em}){9-12} \cmidrule(lr{.5em}){12-14}
                       & & TM & OM & HM
                         & TM & OM & HM
                         & TM & OM & HM
                         & TM & OM & HM \\
    \midrule
    \multirow{4}{*}{\rotatebox{90}{\benchname{lookup}}}
    & \textbf{10} & \textbf{3.4\insigdig{9}\textmu{}s} & 1.0\insigdig{7} & 1.4\insigdig{9} & \textbf{28.7\textmu{}s} & 1.42 & 1.5\insigdig{3} & \textbf{34\insigdig{3}\textmu{}s} & 1.8\insigdig{6} & 1.35 & \textbf{4.2\insigdig{8}ms} & 2.2\insigdig{9} & 1.15 \\
& \textbf{100} & \textbf{19.5\textmu{}s} & 1.02 & 1.75 & \textbf{213\textmu{}s} & 1.08 & 1.80 & \textbf{2.25ms} & 1.22 & 1.7\insigdig{2} & \textbf{27.\insigdig{3}ms} & 1.4\insigdig{4} & 1.5\insigdig{3} \\
& \textbf{1000} & 249\textmu{}s & \textbf{0.98} & 1.8\insigdig{0} & 2.6\insigdig{6}ms & \textbf{0.99} & 1.78 & \textbf{30.\insigdig{4}ms} & 1.0\insigdig{0} & 1.6\insigdig{8} & 35\insigdig{3}ms & \textbf{0.9\insigdig{9}} & 1.50 \\
& \textbf{10000} & 2.78ms & \textbf{0.98} & 1.76 & 31.\insigdig{7}ms & \textbf{0.9\insigdig{9}} & 1.7\insigdig{1} & \textbf{331ms} & 1.00 & 1.65 & \textbf{3.31s} & 1.01 & 1.64 \\

    \midrule
    \multirow{4}{*}{\rotatebox{90}{\benchname{lo\_a\_app1}}}
    & \textbf{10} & \textbf{6.2\insigdig{3}\textmu{}s} & 1.76 & 1.5\insigdig{1} & \textbf{56.4\textmu{}s} & 3.32 & 1.5\insigdig{4} & \textbf{620\textmu{}s} & 4.66 & 1.4\insigdig{4} & \textbf{7.0\insigdig{7}ms} & 5.6\insigdig{7} & 1.3\insigdig{1} \\
& \textbf{100} & \textbf{46.1\textmu{}s} & 1.9\insigdig{3} & 1.62 & \textbf{481\textmu{}s} & 3.42 & 1.6\insigdig{4} & \textbf{4.8\insigdig{9}ms} & 5.0\insigdig{4} & 1.6\insigdig{3} & \textbf{62.\insigdig{7}ms} & 6.84 & 1.4\insigdig{1} \\
& \textbf{1000} & \textbf{527\textmu{}s} & 1.7\insigdig{6} & 1.7\insigdig{3} & \textbf{5.3\insigdig{3}ms} & 3.1\insigdig{5} & 1.7\insigdig{8} & \textbf{64.4ms} & 4.82 & 1.65 & \textbf{981ms} & 5.08 & 1.09 \\
& \textbf{10000} & \textbf{5.4\insigdig{8}ms} & 1.7\insigdig{4} & 1.7\insigdig{7} & \textbf{73.\insigdig{3}ms} & 3.5\insigdig{0} & 1.69 & \textbf{85\insigdig{0}ms} & 5.12 & 1.47 & \textbf{7.79s} & 8.16 & 1.57 \\

    \midrule
    \multirow{4}{*}{\rotatebox{90}{\benchname{insert\_o\_l}}}
    & \textbf{10} & \textbf{64.3ns} & 2.19 & 1.4\insigdig{5} & 153ns & 2.4\insigdig{9} & \textbf{0.80} & \textbf{16\insigdig{5}ns} & 3.5\insigdig{6} & 1.06 & \textbf{105ns} & 6.\insigdig{4}\insigdig{9} & 1.86 \\
& \textbf{100} & \textbf{2.46\textmu{}s} & 1.05 & 1.5\insigdig{2} & \textbf{2.46\textmu{}s} & 1.15 & 1.54 & \textbf{2.4\insigdig{8}\textmu{}s} & 1.50 & 1.5\insigdig{4} & \textbf{2.5\insigdig{7}\textmu{}s} & 2.0\insigdig{2} & 1.4\insigdig{9} \\
& \textbf{1000} & \textbf{26.1\textmu{}s} & 1.01 & 1.76 & \textbf{26.2\textmu{}s} & 1.02 & 1.7\insigdig{5} & \textbf{26.2\textmu{}s} & 1.04 & 1.75 & \textbf{26.5\textmu{}s} & 1.10 & 1.74 \\
& \textbf{10000} & 253\textmu{}s & \textbf{0.99} & 1.8\insigdig{1} & 255\textmu{}s & \textbf{0.99} & 1.80 & 255\textmu{}s & \textbf{0.99} & 1.8\insigdig{0} & \textbf{255\textmu{}s} & 1.00 & 1.8\insigdig{0} \\

    \midrule
    \multirow{4}{*}{\rotatebox{90}{\benchname{fromList }}}
    & \textbf{10} & 3.7\insigdig{1}\textmu{}s & \textbf{0.67} & 0.9\insigdig{8} & 39.3\textmu{}s & 0.99 & \textbf{0.83} & 58\insigdig{8}\textmu{}s & 1.2\insigdig{6} & \textbf{0.67} & 8.\insigdig{5}\insigdig{4}ms & 1.39 & \textbf{0.5\insigdig{2}} \\
& \textbf{100} & 8.4\insigdig{2}\textmu{}s & \textbf{0.8\insigdig{1}} & 2.52 & 129\textmu{}s & \textbf{0.91} & 1.98 & \textbf{1.55ms} & 1.13 & 1.71 & \textbf{26.\insigdig{8}ms} & 1.3\insigdig{1} & 1.3\insigdig{9} \\
& \textbf{1000} & 8\insigdig{2}.\insigdig{5}\textmu{}s & \textbf{0.96} & 3.37 & 9\insigdig{1}\insigdig{6}\textmu{}s & \textbf{0.9\insigdig{8}} & 3.21 & \textbf{13.\insigdig{1}ms} & 1.0\insigdig{1} & 2.8\insigdig{0} & \textbf{14\insigdig{6}ms} & 1.13 & 2.6\insigdig{5} \\
& \textbf{10000} & 73\insigdig{1}\textmu{}s & \textbf{1.00} & 3.86 & 9.\insigdig{3}\insigdig{9}ms & \textbf{0.\insigdig{9}\insigdig{9}} & 3.6\insigdig{7} & 10\insigdig{8}ms & \textbf{1.0\insigdig{0}} & 3.3\insigdig{8} & \textbf{1.1\insigdig{0}s} & 1.02 & 3.30 \\

    \midrule
    \multirow{4}{*}{\rotatebox{90}{\benchname{union}}}
    & \textbf{10} & 4.9\insigdig{0}\textmu{}s & 0.87 & \textbf{0.79} & 37.\insigdig{6}\textmu{}s & 1.02 & \textbf{0.83} & 38\insigdig{9}\textmu{}s & 1.1\insigdig{0} & \textbf{0.9\insigdig{6}} & \textbf{3.6\insigdig{7}ms} & 1.2\insigdig{6} & 1.0\insigdig{6} \\
& \textbf{100} & 25.4\textmu{}s & 1.02 & \textbf{1.00} & 306\textmu{}s & 0.94 & \textbf{0.89} & 2.8\insigdig{5}ms & 1.0\insigdig{3} & \textbf{0.9\insigdig{6}} & \textbf{37.\insigdig{0}ms} & 1.3\insigdig{1} & 1.0\insigdig{9} \\
& \textbf{1000} & 35\insigdig{3}\textmu{}s & 0.96 & \textbf{0.9\insigdig{2}} & 3.5\insigdig{4}ms & 0.97 & \textbf{0.9\insigdig{6}} & \textbf{43.\insigdig{7}ms} & 1.0\insigdig{7} & 1.0\insigdig{7} & \textbf{427ms} & 1.1\insigdig{7} & 1.1\insigdig{6} \\
& \textbf{10000} & \textbf{3.3\insigdig{8}ms} & 1.05 & 1.0\insigdig{6} & \textbf{45.\insigdig{1}ms} & 1.0\insigdig{1} & 1.0\insigdig{0} & \textbf{450ms} & 1.0\insigdig{3} & 1.0\insigdig{2} & \textbf{4.42s} & 1.0\insigdig{4} & 1.04 \\

    \bottomrule
    \end{tabular}
  }

  \label{fig:runtime-finer}
\end{table*}

Let us see what happens if we vary map size $M$ and expression
size $E$ independently for \benchname{lookup}. The results in
\Cref{fig:runtime-finer} show that \ensuremath{\conid{ExprMap}} scales better than \ensuremath{\conid{Map}} when we
increase $M$ and leave $E$ constant. The difference is even more pronounced than
in \Cref{fig:runtime}, in which $N = M = E$.

The time measurements for \ensuremath{\conid{ExprMap}} appear to grow almost linearly with $M$.
Considering that the number of lookups also increases $M$-fold, it seems the
cost of a single lookup remained almost constant, despite the fact that we store
varying numbers of expressions in the trie map. That is exactly the strength
of a trie implementation: Time for the lookup is in $\mathcal{O}(E)$, i.e.,
linear in $E$ but constant in $M$. The same does not hold for search trees,
where lookup time is in $\mathcal{O}(P \log M)$. $P \in \mathcal{O}(E)$ here and
captures the common short circuiting semantics of the lexicographic order on
\ensuremath{\conid{Expr}}. It denotes the size of the longest shared prefix of all expressions.

By contrast, fixing $M$ but increasing $E$ makes \ensuremath{\conid{Map}} easily catch up
on lookup performance with \ensuremath{\conid{ExprMap}}, ultimately outpacing it. The shared prefix
factor $P$ for \ensuremath{\conid{Map}} remains essentially constant relative to $E$: larger
expressions still are likely to differ very early because they are random.
Increasing $M$ will introduce more clashes and is actually more likely to
increase $P$ on completely random expressions. As written above, realistic
work loads often have shared prefixes like \benchname{lookup\_app1}, where
we already saw that \ensuremath{\conid{ExprMap}} outperforms \ensuremath{\conid{Map}}. The fact that \ensuremath{\conid{Map}} performance
depends on $P$ makes it an extremely workload dependent pick, leading to
compiler performance that is difficult to predict. \ensuremath{\conid{HashMap}} shows performance
consistent with \ensuremath{\conid{ExprMap}} but is a bit slower, as before. There is no subtle
scaling factor like $P$; just plain predictable $\mathcal{O}(E)$ like \ensuremath{\conid{ExprMap}}.

Returning to \Cref{fig:runtime}, we see that folding over \ensuremath{\conid{ExprMap}}s is
considerably slower than over \ensuremath{\conid{Map}} or \ensuremath{\conid{HashMap}}. The complex tree structure is
difficult to traverse and involves quite a few indirections.
This is in stark contrast to the situation with \ensuremath{\conid{Map}}, where it's just a
textbook in-order traversal over the search tree. Folding over \ensuremath{\conid{HashMap}}
performs similarly to \ensuremath{\conid{Map}}.

We think that \ensuremath{\conid{ExprMap}} folding performance dies by a thousand paper cuts: The
lazy fold implementation means that we allocate a lot of thunks for intermediate
results that we end up forcing anyway in the case of our folding operator \ensuremath{(\mathbin{+})}.
That is a price that \ensuremath{\conid{Map}} and \ensuremath{\conid{HashMap}} pay, too, but not nearly as much as the
implementation of \ensuremath{\varid{foldrEM}} does.
Furthermore, there's the polymorphic recursion in the head case of \ensuremath{\varid{em\char95 app}}
with a different folding function \ensuremath{(\varid{foldrTM}\;\varid{f})}, which allocates on each call
and makes it impossible to specialise \ensuremath{\varid{foldrEM}} for a fixed folding function
like \ensuremath{(\mathbin{+})} with the static argument transformation~\cite{santos}. Hence
we tried to single out the issue by ensuring that \ensuremath{\conid{Map}} and \ensuremath{\conid{ExprMap}} in
fact don't specialise for \ensuremath{(\mathbin{+})} when running the benchmarks, by means of a
\texttt{NOINLINE} pragma.
Another possible reason might be that the code generated for \ensuremath{\varid{foldrEM}} is quite
a lot larger than the code for \ensuremath{\conid{Map}}, say, so we are likely measuring caching
effects.
We are positive there are numerous ways in which the performance of \ensuremath{\varid{foldrEM}}
can be improved, but believe it is unlikely to exceed or just reach the
performance of \ensuremath{\conid{Map}}.

\subsubsection*{Building}
The \benchname{insert\_lookup\_one} benchmark demonstrates that \ensuremath{\conid{ExprMap}} also
wins on insert performance, although the defeat against \ensuremath{\conid{Map}} for size
parameters beyond 1000 is looming. Again, \Cref{fig:runtime-finer} decouples
map size $M$ and expression size $E$. The data suggests that in comparison to
\ensuremath{\conid{Map}}, $E$ indeed affects insert performance of \ensuremath{\conid{ExprMap}} linearly. By contrast,
$M$ does not seem to affect insert performance at all.

The small edge that \ensuremath{\conid{ExprMap}} seems to have over \ensuremath{\conid{Map}} and \ensuremath{\conid{HashMap}}
doesn't carry over to its naïve \ensuremath{\varid{fromList}} implementation, though. \ensuremath{\conid{Map}} wins
the \benchname{fromList} benchmark, albeit with \ensuremath{\conid{ExprMap}} as a close second.
That is a bit surprising, given that \ensuremath{\conid{Map}}'s \ensuremath{\varid{fromList}} quickly falls back to a
list fold like \ensuremath{\conid{ExprMap}} on unsorted inputs, while \ensuremath{\conid{HashMap}} has a less naïve
implementation: it makes use of transient mutability and performs destructive
inserts on the map data structure during the call to \ensuremath{\varid{fromList}}, knowing that
such mutability can't be observed by the caller. Yet, it still performs worse
than \ensuremath{\conid{ExprMap}} or \ensuremath{\conid{Map}} for larger $E$, as can be seen in
\Cref{fig:runtime-finer}.


We expected \ensuremath{\conid{ExprMap}} to take the lead in \benchname{fromList\_app1}. And indeed
it does, outperforming \ensuremath{\conid{Map}} for larger $N$ which pays for having to compare the
shared prefix repeatedly. But \ensuremath{\conid{HashMap}} is good for another surprise and keeps
on outperforming \ensuremath{\conid{ExprMap}} for small $N$.

What would a non-naïve version of \ensuremath{\varid{fromList}} for \ensuremath{\conid{ExprMap}} look like? Perhaps
the process could be sped up considerably by partitioning the input list
according to the different fields of \ensuremath{\conid{ExprMap}} like \ensuremath{\varid{em\char95 lam}} and then calling
the \ensuremath{\varid{fromList}} implementations of the individual fields in turn. The process
would be very similar to discrimination sort~\cite{discr-sort}, which is a
generalisation of radix sort to tree-like data and very close to tries.
Indeed, the \hackage{discrimination} library provides such an optimised
$\mathcal{O}(N)$ \ensuremath{\varid{toMap}} implementation for ordered maps.

The \benchname{union*} benchmarks don't reveal anything new; \ensuremath{\conid{Map}} and \ensuremath{\conid{HashMap}}
win for small $N$, but \ensuremath{\conid{ExprMap}} wins in the long run, especially when there's
a sharable prefix involved.

\subsection{Space}

\begin{table*}
  \centering
  \caption{Varying expression size $E$ and map size $M$ while measuring the
  memory footprint of the different map implementions on 4 different expression
  populations. Measurements of \ensuremath{\conid{Map}} (OM) and \ensuremath{\conid{HashMap}} (HM) are displayed as
  relative multiples of the absolute measurements on \ensuremath{\conid{ExprMap}} (TM). Lower is
  better. \dag indicates heap overflow.}
  \resizebox{\textwidth}{!}{%
    \begin{tabular}{rr rrr rrr rrr rrr}
    \toprule
    \multicolumn{2}{c}{\multirow{2}{*}{\diagbox{$E$}{$M$}}} & \multicolumn{3}{c}{\textbf{10}}
                                        & \multicolumn{3}{c}{\textbf{100}}
                                        & \multicolumn{3}{c}{\textbf{1000}}
                                        & \multicolumn{3}{c}{\textbf{10000}} \\
    \cmidrule(lr{.5em}){3-5} \cmidrule(lr{.5em}){6-8} \cmidrule(lr{.5em}){9-12} \cmidrule(lr{.5em}){12-14}
                       & & TM & OM & HM
                         & TM & OM & HM
                         & TM & OM & HM
                         & TM & OM & HM \\
    \midrule
    \multirow{4}{*}{\rotatebox{90}{\benchname{space}}}
    & \textbf{10} & 12.5KB & 0.79 & \textbf{0.78} & 87.8KB & 0.80 & \textbf{0.80} & 766KB & 0.89 & \textbf{0.88} & 6.88MB & 0.96 & \textbf{0.95} \\
& \textbf{100} & 55.5KB & 0.93 & \textbf{0.93} & 531KB & 0.91 & \textbf{0.91} & 4.94MB & 0.91 & \textbf{0.90} & 48.6MB & 0.92 & \textbf{0.92} \\
& \textbf{1000} & 477KB & 0.99 & \textbf{0.99} & 4.77MB & 0.98 & \textbf{0.98} & 46.6MB & 0.98 & \textbf{0.98} & 468MB & 0.98 & \textbf{0.98} \\
& \textbf{10000} & 4.10MB & 1.00 & \textbf{1.00} & 42.1MB & 1.00 & \textbf{1.00} & 421MB & 1.00 & \textbf{1.00} & 4.12GB & 1.00 & \textbf{1.00} \\

    \midrule
    \multirow{4}{*}{\rotatebox{90}{\benchname{space\_app1}}}
    & \textbf{10} & \textbf{15.3KB} & 1.26 & 1.25 & \textbf{90.6KB} & 1.78 & 1.78 & \textbf{769KB} & 1.99 & 1.99 & \textbf{6.88MB} & 2.12 & 2.11 \\
& \textbf{100} & \textbf{83.6KB} & 1.74 & 1.74 & \textbf{560KB} & 2.54 & 2.54 & \textbf{4.97MB} & 2.74 & 2.74 & \textbf{48.6MB} & 2.79 & 2.79 \\
& \textbf{1000} & \textbf{758KB} & 1.86 & 1.86 & \textbf{5.04MB} & 2.74 & 2.74 & \textbf{46.9MB} & 2.92 & 2.92 & \textbf{468MB} & 2.93 & 2.93 \\
& \textbf{10000} & \textbf{6.85MB} & 1.93 & 1.93 & \textbf{44.7MB} & 2.98 & 2.98 & \textbf{423MB} & 3.15 & 3.15 & \textbf{4.13GB} & 3.16 & 3.16 \\

    \midrule
    \multirow{4}{*}{\rotatebox{90}{\benchname{space\_app2}}}
    & \textbf{10} & 29.4KB & 0.66 & \textbf{0.65} & 240KB & 0.67 & \textbf{0.67} & 2.14MB & 0.70 & \textbf{0.70} & 20.2MB & 0.72 & \textbf{0.72} \\
& \textbf{100} & 225KB & 0.65 & \textbf{0.65} & 2.06MB & 0.68 & \textbf{0.67} & 20.2MB & 0.68 & \textbf{0.67} & 200MB & 0.68 & \textbf{0.68} \\
& \textbf{1000} & 2.12MB & 0.65 & \textbf{0.65} & 20.1MB & 0.69 & \textbf{0.69} & 199MB & 0.69 & \textbf{0.69} & 1.95GB & 0.69 & \textbf{0.69} \\
& \textbf{10000} & 20.6MB & 0.64 & \textbf{0.64} & 196MB & 0.68 & \textbf{0.68} & 1.90GB & 0.68 & \textbf{0.68} & \dag & \dag & \dag \\

    \midrule
    \multirow{4}{*}{\rotatebox{90}{\benchname{space\_lam}}}
    & \textbf{10} & 18.3KB & 0.97 & \textbf{0.96} & \textbf{118KB} & 1.24 & 1.23 & \textbf{1.00MB} & 1.36 & 1.35 & \textbf{9.12MB} & 1.45 & 1.45 \\
& \textbf{100} & \textbf{73.3KB} & 1.77 & 1.77 & \textbf{607KB} & 2.08 & 2.08 & \textbf{5.56MB} & 2.18 & 2.17 & \textbf{54.6MB} & 2.20 & 2.20 \\
& \textbf{1000} & \textbf{610KB} & 2.05 & 2.05 & \textbf{4.99MB} & 2.47 & 2.47 & \textbf{47.8MB} & 2.55 & 2.55 & \textbf{478MB} & 2.55 & 2.55 \\
& \textbf{10000} & \textbf{5.33MB} & 2.20 & 2.20 & \textbf{43.4MB} & 2.73 & 2.73 & \textbf{423MB} & 2.79 & 2.79 & \textbf{4.14GB} & 2.79 & 2.79 \\

    \bottomrule
    \end{tabular}
  }

  \label{fig:space}
\end{table*}

We also measured the memory footprint of \ensuremath{\conid{ExprMap}} compared to \ensuremath{\conid{Map}} and
\ensuremath{\conid{HashMap}}. The results are shown in \Cref{fig:space}. All four benchmarks simply
measure the size on the heap in bytes of a map consisting of $M$ expressions of
size $E$. They only differ in whether or not the expressions have a shared
prefix. As before, \benchname{space} is built over completely random expressions,
while the other three benchmarks build maps with common prefixes, as discussed in
\cref{sec:runtime}.

In \benchname{space}, prefix sharing is highly unlikely for reasons discussed
in the last section: Randomness dictates that most expressions diverge quite
early in their prefix. As a result, \ensuremath{\conid{ExprMap}} consumes slightly more space
than both \ensuremath{\conid{Map}} and \ensuremath{\conid{ExprMap}}, the latter of which wins every single instance.
The difference here is ultimately due to the fact that inner nodes in the trie
allocate more space than inner nodes in \ensuremath{\conid{Map}} or \ensuremath{\conid{ExprMap}}.

However, in \benchname{space\_app1} and \benchname{space\_lam}, we can see that
\ensuremath{\conid{ExprMap}} is able to exploit the shared prefixes to great effect: For big
$M$, the memory footprint of \benchname{space\_app1} approaches that of
\benchname{space} because the shared prefix is only stored once. In the other
dimension along $E$, memory footprint still increases by similar factors as
in \benchname{space}. The \benchname{space\_lam} family does need a bit more
bookkeeping for the de Bruijn numbering, so the results aren't quite as close to
\benchname{space\_app1}, but it's still an easy win over \ensuremath{\conid{Map}} and \ensuremath{\conid{HashMap}}.

For \benchname{space\_app2}, \ensuremath{\conid{ExprMap}} can't share any prefixes because the
shared structure turns into a suffix in the pre-order serialisation. As a result,
\ensuremath{\conid{Map}} and \ensuremath{\conid{HashMap}} allocate less space, all consistent constant factors apart
from each other. \ensuremath{\conid{HashMap}} wins here again.

\section{Triemaps that unify?}

In effect, the \ensuremath{\conid{PatVar}}s of the patterns stored in our matching triemaps act
like unification variables. The unification problems we solve are always
particularly simple, because pattern variables only ever match against are
\emph{expression} keys in which no pattern variable can occur.

Another frustrating point is that we had to duplicate the \ensuremath{\conid{TrieMap}} class in
\Cref{sec:matching-trie-class} because the key types for lookup and insertion no
longer match up. If we managed to generalise the lookup key from expressions to
patterns, too, we could continue to extend good old \ensuremath{\conid{TrieMap}}.
All this begs the question: \emph{Can we extend our idiomatic triemaps to facilitate
unifying lookup?}

At first blush, the generalisation seems simple. We already carefully confined
the matching logic to \ensuremath{\conid{Matchable}}. It should be possible to generalise to
\begin{hscode}\SaveRestoreHook
\column{B}{@{}>{\hspre}l<{\hspost}@{}}%
\column{3}{@{}>{\hspre}l<{\hspost}@{}}%
\column{E}{@{}>{\hspre}l<{\hspost}@{}}%
\>[B]{}\keyword{class}\;(\conid{Eq}\;\varid{k},\conid{MonadPlus}\;(\conid{Unify}\;\varid{k}))\Rightarrow \conid{Unifiable}\;\varid{k}\;\keyword{where}{}\<[E]%
\\
\>[B]{}\hsindent{3}{}\<[3]%
\>[3]{}\keyword{type}\;\conid{Unify}\;\varid{k}\mathbin{::}\conid{Type}\to \conid{Type}{}\<[E]%
\\
\>[B]{}\hsindent{3}{}\<[3]%
\>[3]{}\varid{unify}\mathbin{::}\varid{k}\to \varid{k}\to \conid{Unify}\;\varid{k}\;(){}\<[E]%
\\
\>[B]{}\keyword{class}\;(\conid{Unifiable}\;(\conid{Key}\;\varid{tm}),\conid{TrieMap}\;\varid{tm})\Rightarrow \conid{UTrieMap}\;\varid{tm}\;\keyword{where}{}\<[E]%
\\
\>[B]{}\hsindent{3}{}\<[3]%
\>[3]{}\varid{lookupUniUTM}\mathbin{::}\conid{Key}\;\varid{tm}\to \varid{tm}\;\varid{v}\to \conid{Unify}\;(\conid{Key}\;\varid{tm})\;\varid{v}{}\<[E]%
\ColumnHook
\end{hscode}\resethooks
But there are problems:
\begin{itemize}
  \item We would need all unification variables to be globally unique lest we
    open ourselves to numerous shadowing issues when reporting unifiers.
  \item Consider the Unimap for
    $$
      (([a], T\;a\;A), v1) \quad \text{and} \quad (([b], T\;b\;B), v2)
    $$
    After canonicalisation, we get
    $$
      ((T\;\pv{1}\;A), ([(a,\pv{1})], v1)) \quad \text{and} \quad (T\;\pv{1}\;B, ([(b,\pv{1})], v2))
    $$
    and both patterns share a prefix in the trie.
    Suppose now we uni-lookup the pattern $([c,d], T c d)$.
    What should we store in our \ensuremath{\conid{UniState}} when unifying $c$ with $\pv{1}$?
    There simply is no unique pattern variable to \enquote{decanonicalise} to!
    In general, it appears we'd get terms in the range of our substitution that
    mix \ensuremath{\conid{PatVar}}s and \ensuremath{\conid{PatKey}}s. Clearly, the vanilla \ensuremath{\conid{Expr}} datatype doesn't
    accomodate such an extension and we'd have to complicate its definition with
    techniques such as Trees that Grow \cite{ttg}.
\end{itemize}

So while embodying full-blown unification into the lookup algorithm seems
attractive at first, in the end it appears equally complicated to present.

\end{document}